\documentclass[1p]{elsarticle}

\usepackage[pagewise]{lineno}

\biboptions{numbers,sort&compress}

\usepackage{graphicx}

\usepackage[T1]{fontenc}
\usepackage{dsfont}               
\usepackage{mathrsfs}             
\usepackage{slashed}              
\usepackage{amsmath}
\usepackage{amssymb}
\usepackage{amsbsy}
\usepackage{amsfonts}
\usepackage{physics}
\usepackage{bbold}
\usepackage{booktabs}

\usepackage{hyperref}
\usepackage[hyperref]{xcolor}
\hypersetup{colorlinks = true}
\usepackage{orcidlink}

\definecolor{oucrimsonred}{rgb}{0.6, 0.0, 0.0}
\definecolor{DarkGray}{gray}{0.4}
\definecolor{forestgreen}{rgb}{0.13,0.35,0.13}
\definecolor{ocre}{HTML}{F16723}

\usepackage[page,toc,titletoc,title]{appendix}

\numberwithin{equation}{section}
\numberwithin{table}{section}
\numberwithin{figure}{section}

\journal{Progress in Particle and Nuclear Physics}

\usepackage{titlesec}
\usepackage{sectsty}
\titleformat{\section}{\normalfont\Large\bfseries}{\thesection}{1em}{}
\titleformat{\subsection}{\normalfont\large\bfseries}{\thesubsection}{1em}{}
\titleformat{\subsubsection}{\normalfont\normalsize\bfseries}{\thesubsubsection}{1em}{}  

\newcommand{\be}{\begin{equation}}
\newcommand{\ee}{\end{equation}}
\newcommand{\bea}{\begin{eqnarray}}
\newcommand{\eea}{\end{eqnarray}}
\newcommand{\nn}{\nonumber}
\def\Tr{\mbox{Tr}\,}
\def\eq#1{{Eq.~(\ref{#1})}}
\def\eqs#1#2{{Eqs.~(\ref{#1})--(\ref{#2})}}
\def\di{\mbox{d}}

\newcommand{\CC}{C}
\newcommand{\BB}{B}
\newcommand{\hk}{{\bf \hat{k}}}
\newcommand{\hr}{{\bf \hat{r}}}
\newcommand{\hn}{{\bf \hat{n}}}
\newcommand{\hp}{{\bf \hat{p}}}

\newcommand{\lambdap}{\lambda^{\prime}}

\newcommand{\com}[1]{}

\newcommand{\Rchsh}{\mathfrak{m}_{12}}  
\newcommand{\K}{K^{*}(892)^0}

\newcommand{\Rechi}{\Re\!\!\left[\chi(m^2_{\tau\bar{\tau}})\right]}
\newcommand{\Abschi}{\left|\chi(m^2_{\tau\bar{\tau}})\right|}

\newcommand*\xbar[1]{%
  \hbox{\;%
    \vbox{%
      \hrule height 0.5pt 
      \kern0.5ex
      \hbox{%
        \kern-0.25em
        \ensuremath{#1}%
        \kern-0.07em
      }%
    }%
  }%
} 

\newcommand{\lambdaA}{\lambda_{1}}
\newcommand{\lambdaB}{\lambda_{2}}
\newcommand{\lambdaAp}{\lambda_1^{\prime}}
\newcommand{\lambdaBp}{\lambda_2^{\prime}}
\newcommand{\mup}{\mu^{\prime}}
\newcommand{\nup}{\nu^{\prime}}
\newcommand{\ct}{\cos{\Theta}}
\newcommand{\st}{\sin{\Theta}}
\newcommand{\cmb}{\mathscr{C}_2}

\newcommand{\mVV}{m_{\scriptscriptstyle{V\!V}}}
\newcommand{\mWW}{m_{\scriptscriptstyle{W\!W}}}
\newcommand{\mZZ}{m_{\scriptscriptstyle{Z\!Z}}}

\newcommand{\sW}{\sin \theta_W}
\newcommand{\cW}{\cos \theta_W}

\newcommand{\ssW}{s_{\W}}
\newcommand{\ccW}{c_{\W}}

\newcommand{\Ct}{c_{\Theta}}
\newcommand{\St}{s_{\Theta}}
\newcommand{\gAb}{\bar{g}_A^{u}}
\newcommand{\gAbD}{\bar{g}_A^{d}}

\newcommand{\gVb}{\bar{g}_V^{u}}
\newcommand{\gVbD}{\bar{g}_V^{d}}

\newcommand{\gVV}{g_V^{q 2}}
\newcommand{\gAA}{g_A^{q 2}}
\newcommand{\gVVVV}{g_V^{q 4}}
\newcommand{\gAAAA}{g_A^{q 4}}

\newcommand{\fZZ}{f_{\scriptscriptstyle{Z\!Z}}}

\newcommand{\DZ}{{\rm D}_{\Z\!\Z}}

\newcommand{\W}{\scriptscriptstyle{W}}
\newcommand{\Z}{\scriptscriptstyle{Z}}

\newcommand{\betaW}{\beta_{\W}}
\newcommand{\betaZ}{\beta_{\Z}}

\newcommand{\Aqq}{A^{q\bar q}}
\newcommand{\htqq}{\tilde{h}^{q\bar q}}
\newcommand{\ftqq}{\tilde{f}^{q\bar q}}
\newcommand{\gtqq}{\tilde{g}^{q\bar q}}
\newcommand{\gV}{g_V^{q}}
\newcommand{\gA}{g_A^{q}}

\newcommand{\LVV}{a_{V}^{ 2}}
\newcommand{\LV}{a_{V}}
\newcommand{\LtVV}{\widetilde{a}_{V}^{2}}
\newcommand{\LtV}{\widetilde{a}_{V}}

\begin{document}

\hypersetup{citecolor = forestgreen,
linktoc = section, 
linkcolor = forestgreen, 
urlcolor = forestgreen
}
	\begin{frontmatter}
	
		\title{
  {\Large \color{oucrimsonred} \textbf{Quantum entanglement and Bell inequality violation at colliders}}\vskip1cm}
		
		
  \author[alanmainaddress,alanaddress]{Alan J.~Barr\,\orcidlink{0000-0002-3533-3740}}
  \address[alanmainaddress]{{\color{DarkGray} Department of Physics, Keble Road, University of Oxford, OX1 3RH}}
  \address[alanaddress]{{\color{DarkGray}Merton College, Merton Street, Oxford, OX1 4JD}}
		
		\author[marcomainaddress]{Marco Fabbrichesi\,\orcidlink{0000-0003-1937-3854}}
		\address[marcomainaddress]{{\color{DarkGray} INFN, Sezione di Trieste, Via Valerio 2, I-34127 Trieste, Italy}}

\author[marcomainaddress]{Roberto Floreanini\,\orcidlink{0000-0002-0424-2707}}

\author[emidiomainaddress,marcomainaddress,lucamainaddress]{Emidio Gabrielli\corref{mycorrespondingauthor}\,\orcidlink{0000-0002-0637-5124}}
\cortext[mycorrespondingauthor]{Corresponding author}
		\ead{emidio.gabrielli@cern.ch}
\address[emidiomainaddress]{{\color{DarkGray}Physics Department, University of Trieste, Strada Costiera 11,  I-34151 Trieste, Italy}}

\author[lucamainaddress]{Luca Marzola\,\orcidlink{0000-0003-2045-1100}}
\address[lucamainaddress]{{\color{DarkGray}Laboratory of High-Energy and Computational Physics, NICPB, R\"avala pst.~10, 10143 Tallinn, Estonia}}

  
		\begin{abstract}
	The study of entanglement in particle physics  has been gathering pace  in the past few years. It is a new field that is providing important results about the possibility of detecting entanglement and testing Bell inequality at colliders for  final states  as diverse as   top-quark, $\tau$-lepton pairs and $\Lambda$-baryons,  massive gauge bosons and vector mesons.  In this review, after presenting definitions, tools and basic results that are necessary for understanding these developments, we   summarize the main findings---as published by the beginning of year 2024---including   analyses of  experimental data in $B$ meson decays and   top-quark pair production. We include a detailed discussion  of the results for both qubit and qutrits systems, that is, final states containing spin one-half and spin one particles. Entanglement has also been proposed as a new tool to constrain new particles and fields beyond the Standard Model and we introduce the reader to this promising feature as well. \vskip1cm 
		\end{abstract}
		
	\end{frontmatter}
	\newpage

	\tableofcontents
	
	
	\newpage
	\section{Introduction}\label{intro}
	
An unmistakable feature of quantum mechanics is the inseparable nature of states describing physical systems that have interacted in the past. The \textbf{entanglement} among these states  gives rise to correlations that can be stronger than those expected in  classical mechanics and are present even after the systems are separated and can no longer interact, thus introducing a form of nonlocality in our observations which, however, does not imply any violation of relativity.

Entanglement should not be confused with classical correlations, the latter   dealing with intrinsic properties of a system, independently of their measurement.  Consider the simplest situation of two spin-1/2 particles that have been prepared
in a maximally entangled state, then separated by an arbitrary distance and whose spin
is measured with suitable detectors in an arbitrary chosen direction. The result
of the measurement is completely random for both detectors, but if one particle
is found with spin up, then the second is detected with spin down, and vice versa.
As a result, far away though  the two daughter particles  might be, they  must be considered as a single physical entity. This feature represents   the
phenomenon of quantum nonlocality in a nutshell.

The presence of entanglement can lead to the violation of  an inequality---named after J.~S.~Bell, who was first in deriving and discussing it---among the sum of probabilities of the values of certain observables. Whereas  the presence of entanglement  in itself only confirms the existence of  correlations  that must be there because of  quantum mechanics, the observation of the \textbf{violation of Bell inequality}  implies something about the nature of the physical world---namely, its non-separability or, if you prefer,  nonlocality---and it represents therefore  a profound discovery.

Though the study of entangled states has been an ongoing concern in atomic and solid-state physics for many years, it is only recently that the high-energy community has taken up in earnest the study of the subject.\footnote{We are aware, and the reader should too, that the study of quantum entanglement and its many applications is  a broad and ever expanding field of research. The interested readers can look into the review articles and books~\cite{Amico20081,Horodecki:2009zz,Guhne20091,Eisert20101,Laflorencie20161,Nishioka:2018khk,Witten:2018zxz,bruss2019quantum,Benatti20201} for applications beyond particle physics.}  States in quantum field theory are identified  by their mass, momentum and spin (as they  are irreducible representations of the Poincar\`e group) and computations---in the perturbative $S$-matrix framework---are only possible in momentum space; therefore entanglement can only be observed in correlations among the particle spins (or on variables living in the internal flavor space) and it is there that it must be looked for. The investigation of these effects relies on the determination of the full quantum state of the system after the relevant interactions occurred, a process that has been dubbed \textbf{quantum state tomography}.

Collider detectors, while not designed for  probing  entanglement, turn out to be surprisingly good in performing this task, thus ushering in the possibility of many  interesting new measurements  to search for the presence of entanglement as well as to test the violation of Bell inequality. Entanglement also provides  new tools for probing physics beyond the Standard Model (SM) whenever the correlations it affects are experimentally accessible. 

The extension of the physics of entanglement to the realm of particle physics is not just a reformulation at higher energies of the work done within atomic physics. New features come into play, most notably the testing of quantum mechanics beyond electrodynamics, with weak and strong interactions, and the presence of systems possessing more than two possible states, such as massive spin 1 particles with their three polarization states. Other features pertaining to collider physics will become evident as we proceed in our discussion through the following sections. 

Many aspects and peculiarities of quantum physics are taking an increasingly central position in science---from quantum computers to information theory, from theoretical developments to innovative applications.  We  look at the impact of these developments in the  area of high-energy physics.  Our aim  in writing this review is rather circumscribed:
firstly,  we want to present all definitions, tools and basic results that are useful for the study of entanglement and Bell inequality violation at colliders; secondly,  we try to   summarize the main findings reported in the literature up to the beginning of 2024. Our hope is to provide an easily accessible collection of resources to serve as springboard for further study.

\subsection{The ``quantum'' in quantum field theory}

Quantum field theory, coming as it does from the marriage of quantum mechanics and special relativity, inherits the two main features of quantum mechanics:  probabilistic predictions and  amplitude interference. To these two, it adds a quantum feature of its own: radiative corrections arising from closed loops in the propagation of the particles (and their creation out of the vacuum). These quantum effects are part of every computation in quantum field theory. 

Notwithstanding these features, the feel of a computation in \textbf{perturbative $S$-matrix} is distinctively less ``quantum'' than in quantum mechanics proper. There is no wave-function collapse and the variables utilized---momenta and occupation number of the asymptotic in- and out-coming states---commute.  It is so because the $S$-matrix formulation of quantum field theory is part of a shift that has taken place in particle physics (see~\cite{Blum:2017diy} for a nice historical discussion) away from the original framework, which was mostly inspired by atomic physics, and toward the typical setting we find at colliders, in which particles come in and go out and we deduce the interactions they have undergone only (at least for elastic processes) by the change in their momenta or (for inelastic scattering processes) also in  the occupation numbers---with particles being created or destroyed.

The study of entanglement in particle physics goes against this trend. Entanglement is perhaps the quintessential manifestation of quantum mechanical quirkiness: observations on systems retain a form of correlation even after they have been separated and this correlation implies a nonlocal sharing of resources. There is no way to create an entangled state using local operations and classical communication. 
A typical example of study of entanglement in a collider setting sees the spin variables as those that are entangled in the scattering and decay processes.  Spin variables have been  studied until now mostly in the form of classical correlations, which, although sharing some features with entanglement, do not imply entanglement. Quantum tomography, the aim of which is to describe the density matrix of the final state in a scattering process, brings the entanglement among spin variables to center stage.  

The presence of nonlocal effects always brings an ominous note to our relativistic ears. Yet there is no reason for concern, 
for entanglement cannot be used to transfer information between two  separated observers. Any information exchange  can only be carried by local communication in which relativity is not violated, as it should not since it was incorporated in quantum field theory from the very beginning.
Neither is  the cluster decomposition (an essential feature of quantum field theory) violated by entanglement. The decomposition has to take place between initial and final states pertaining to two subsets of the  $S$-matrix which are then assumed to be far away from one another. Entanglement is present only within the two subsets  as long as the relative interactions  take place  independently of each other. 

\subsection{Spin correlations at colliders}

Spin variables of particles and correlations among them  are  accessible at current collider experiments through the study of the distribution of the momenta of the final state into which the original particle decays. These momenta are  commuting variables, but this fact does not prevent entanglement and Bell inequality violation from being accessible at colliders. The measurement takes place (as we shall see in Section~\ref{sec:toolbox}) as the polarized particle  decays (acting as its own polarimeter) and the momenta of the final state only carry the information into the detectors---in the same way as the momenta of the final neutrons carry the information on their spin as their trajectories are separated by the magnetic field in the Stern-Gerlach  experiment.

 The particles created in the collision   first fly through what is (for all practical purposes) a vacuum, going from the collision vertex to hitting the internal surface of the beam pipe and  on  inside the detector. The characteristic time for this flight is given by the radius of the beam pipe---which is of the order of 1 cm, see, for instance~\cite{ATLASCollaboration_2008}---divided by $c$, for a relativistic  particle, that is, $10^{-11}$ s.
On the other hand, spin correlations are measured at the time the particle decays, that is, with a characteristic time given by their lifetimes. These lifetimes  go from $10^{-25}$ s for the top quark and the weak gauge bosons to $10^{-20}$ s for vector mesons and $10^{-13}$ s  for  the $\tau$ leptons. 

 A \textbf{loss in correlation} between the spins of the particles produced at colliders can only take place after they  cross into  the detector, where the particles would necessarily interact with the atoms of which the detector is made. This interaction never happens since the flight-time inside the beam pipe  is  much longer than the lifetime of all the particles we are interested in (except maybe the $\Lambda$ baryons) and they decay before reaching the detector proper. For this reason,  we can safely assume that the spin correlations we measure are not affected, let alone decorrelated, by the presence of the detector.  
 
 The \textbf{hadronization} time scale, a concern only in the case of the top quark, is of the order of $10^{-23}$ s and  takes place well after the spin correlations have been measured as the top quark decays.

\subsection{The story so far}\label{history}

Helicity and polarization amplitudes at colliders are very sensitive probes into the details of the underlying physics  and, for this reason, have been studied for many years. The literature is vast. Older works are reviewed in~\cite{Bourrely:1980mr}. More recent contributions introduce the techniques necessary in computing  polarizations among  fermions~\cite{Kane:1991bg,Dalitz:1991wa,Mahlon:2010gw,Mahlon:1995zn,Mahlon:1996pn,Mahlon:1997uc,Bernreuther:2001rq,Uwer:2004vp,Bernreuther:2010ny}), weak gauge bosons~\cite{Gaemers:1978hg,Hellmund:1981kq,Grau:1982gb,Cortes:1983qn,Tofighi-Niaki:1988ken,Shim:1995ax,Mahlon:1998jd,Bern:2011ie,Stirling:2012zt,Maina:2020rgd,Maina:2021xpe}) or both~\cite{Boudjema:2009fz}. Reconstructing spin-1 polarizations has been well understood since the mid-90s (see~\cite{Dighe:1995pd,Leader2001}) and the framework widely used in  experimental analyses such as in heavy meson decays. All these works look for classical correlations and the possibility of measuring them in cross sections or dedicated observables.

Quantum state tomography falls in the same line of inquiry as the references above except for the twist of using the polarization amplitudes to define no longer the classical but instead the truly quantum correlations. Polarizations are framed in the spin density matrix (as explained in Section~\ref{sec:toolbox}) and made readily accessible to compute entanglement and Bell operators for the processes of interest. 
	
The violation of the Bell inequality has been tested and verified with experiments measuring the polarizations of photons at low energy (that is, in the range of few eVs) in~\cite{Aspect:1982fx,Weihs:1998gy}:
two photons are prepared into a singlet state  and their polarizations  measured along different directions to verify their entanglement and the  violation of  Bell inequality.
More experiments have been performed to further test the inequality~\cite{Clauser:1969ny,Clauser:1974tg} and show that the violation takes place also for separations  of few kilometers~\cite{Tittel:1998ja}. The Bell inequality has also been tested in solid-state physics~\cite{Ansmann2009ViolationOB}. 

No sooner these tests were reported than `loopholes' were put forward --- ways in which, notwithstanding the experimental results, the consequences could be evaded. The presence of these loopholes spurred the experimental community into performing new tests in which the loopholes  were systematically closed  with photons in~\cite{Hensen:2015ccp,Giustina:2015yza},    using superconducting circuits in~\cite{Storz:2023jjx}, and  using atoms in~\cite{PhysRevLett.119.010402}. The reader can find more details and references in the older~\cite{Clauser:1978ng} and the more recent~\cite{Genovese:2005nw} review articles.

In  particle physics, entanglement with low-energy protons has been probed in~\cite{Lamehi-Rachti:1976wey} and  proposed at colliders using charmonium decays in~\cite{Tornqvist:1980af,Tornqvist:1986pe,Baranov:2008zzb,Baranov:2009zza} as well as $\tau$ leptons in~\cite{Privitera:1991nz}. Tests  have been suggested  by means of positronium decays~\cite{Acin:2000cs}  and neutrino oscillations~\cite{Akhmedov:2010ua} (see also~\cite{Naikoo:2017fos,Formaggio:2016cuh}). A Bell inequality is violated in neutral  kaon oscillations due to indirect~\cite{DiDomenico:1995ky} and direct $CP$ violation~\cite{PhysRevD.57.R1332,Benatti2000,Bertlmann:2001ea} (see also~\cite{Banerjee:2014vga}). Flavor oscillations in neutral $B$-mesons have been argued to imply the violation of Bell inequality~\cite{Go:2003tx,Belle:2007ocp} (see also~\cite{Bertlmann:2004cr}). 
 Entanglement among  partons  in scattering processes of nucleons has recently been studied (see, for instance, \cite{Kharzeev:2017qzs,Tu:2019ouv}). A discussion  of entanglement in particle physics  also appears in~\cite{Yongram:2013soa,Cervera-Lierta:2017tdt}.

The interest has been revived recently after entanglement has been convincingly argued~\cite{Afik:2020onf}  to be present in top-quark pair production at the Large Hadron Collider (LHC) and it was  shown that  Bell inequality violation is experimentally accessible in the same system~\cite{Fabbrichesi:2021npl} and in the decay of the Higgs boson into two charged gauge bosons~\cite{Barr:2021zcp}. Entanglement and Bell inequality violation with a significance well in excess of 5$\sigma$ has been shown in LHCb and Belle II data on the decays of the $B$ mesons~\cite{Fabbrichesi:2023idl}. The ATLAS and CMS Collaborations have found~\cite{ATLAS:2023fsd,CMS:2024hgo} that entanglement is present with a significance of more than $5\sigma$ in top-quark pairs produced near threshold at the LHC.

A sizable body of works has been published since. We review it in the Sections~\ref{sec:qubits} and~\ref{sec:qutrits} by organizing it into systems that are qubits and qutrits, that is, entanglement among particles of spin 1/2 and 1. The possibility of using entanglement to probe new physics is reviewed in Section~\ref{sec:newphysics}. 
Before all that,  we introduce in Sections~\ref{sec:EBL}  and~\ref{sec:toolbox}  the definitions and tools necessary in the analysis.

\newpage
\section{Entanglement and Bell locality}	 \label{sec:EBL}

\subsection{Quantum states and observables}\label{sec:oservables}
	
In quantum mechanics, the description of a quantum system $S$, for simplicity taken to be finite dimensional
(\hbox{$n$-level} system), is realized by means of an ($n$-dimensional) Hilbert space $\mathcal{H}_n$, isomorphic
to $\mathbb{C}^n$, where $\mathbb{C}$ is the set of complex numbers, and by the algebra $M_n(\mathbb{C})$
of $n\times n$ complex matrices. The elements $|\psi\rangle$ of $\mathcal{H}_n$, normalized to unity, represent
states of $S$, while the Hermitian matrices in $M_n(\mathbb{C})$, $\mathcal{\hat O}^\dagger = \mathcal{\hat O}$, 
correspond to system observables, whose mean values,
$\langle \mathcal{\hat O} \rangle \equiv \langle\psi| \mathcal{\hat O} |\psi\rangle$,
are statistically linked to measurements of $\mathcal{\hat O}$.

The elements of $\mathcal{H}_n$ are, however, just a particular class of states of $S$, those called {pure states}.
In general, the information on $S$ is incomplete and a set of probabilities $\{p_i\}$, with $\sum_i p_i=1$,
weights the possible (normalized but not necessarily orthogonal) states of the system $|\psi_i\rangle$, $i=1,2,\ldots,m$.
In this case, the mean value of any given system observable $\mathcal{O}$ can be expressed as
the combination of 
the pure state mean values $ \langle\psi_i| \mathcal{\hat O} |\psi_i\rangle$, weighted with the corresponding
probability of occurrence:
\begin{equation}
\langle \mathcal{\hat O} \rangle = \sum_{i=1}^m p_i\, \langle\psi_i| \mathcal{\hat O} |\psi_i\rangle\ .
\label{mean-value}
\end{equation}
It is then natural to describe the statistical mixture $\{p_i, |\psi_i\rangle\}$
by means of the \textbf{density matrix}:
\begin{equation}
\rho = \sum_{i=1}^m p_i\, |\psi_i\rangle \langle\psi_i| \ ,\quad {\rm with}\quad p_i\geq 0 \quad {\rm and}\quad \sum_{i=1}^m p_i= 1\ ,
\label{densitymatrix}
\end{equation}
where the conditions on the set $\{p_i\}$ are those of an ensemble of statistical weights.
The average value of an observable $\mathcal{O}$ can then be most simply  expressed as
\begin{equation}
\langle \mathcal{\hat O} \rangle = {\rm Tr}\big[ \rho\, \mathcal{\hat O} \big]\ ,
\label{trace-mean-value}
\end{equation}
where the trace operation  is explicitly given by
${\rm Tr}[X] \equiv \sum_{i=1}^n \langle \varphi_i| X |\varphi_i\rangle$, with $X\in M_n(\mathbb{C})$ a matrix and the collection of states 
$\{|\varphi_i\rangle\}$ being any orthonormal basis in $\mathcal{H}_n$.

From its definition (\ref{densitymatrix}), any density matrix $\rho$ must satisfy the following three characteristic properties:
\begin{itemize}
    \item $\rho$ is an Hermitian operator, $\rho^\dagger = \rho$,
    \item $\rho$ is normalized, ${\rm Tr}[\rho]=1$,
    \item $\rho$ is a positive semi-definite matrix, {\it i.e.} $\langle\psi| \rho |\psi\rangle\geq 0$;
    for all $|\psi\rangle\in \mathcal{H}_n$,
\end{itemize}
in order to preserve the physically consistent interpretation of $\rho$.

Quantum states are thus positive, normalized operators, with the pure states $|\psi\rangle$
represented by projectors $|\psi\rangle \langle\psi|$ as the statistical mixture in (\ref{densitymatrix})
reduces in this case to a single element.
As a consequence, the eigenvalues of density matrices representing pure states  
are $1$ (non-degenerate) and $0$~($n-1$ times degenerate),
while those, $\{\lambda_i\}$, $i=1,2,\ldots,n$, of a generic density matrix $\rho$ are such that: $0\leq \lambda_i\leq 1$,
with $\sum_i \lambda_i =1$. It follows that in general: ${\rm Tr}[\rho^2]\leq 1$, reaching the upper bound only
when $\rho$ is a pure state. Therefore, a quantum state represented by a density matrix $\rho$ is a \textbf{pure state}
if and only if $\rho$ is a projector:
\begin{equation}
\rho^2 = \rho\ , \qquad {\rm Tr}[\rho^2] =1\ .
\label{pure-state}
\end{equation}
The decomposition of any density matrix $\rho$ in terms of the eigen-projectors 
$|\lambda_i\rangle \langle\lambda_i|$, constructed with
its eigenvectors $|\lambda_i\rangle$, gives its spectral decomposition:
\begin{equation}
\rho = \sum_{i=1}^m \lambda_i\, |\lambda_i\rangle \langle\lambda_i| \ , 
\quad {\rm with}\quad \sum_{i=1}^m \lambda_i= 1 \quad {\rm and}\quad \langle\lambda_i| \lambda_j\rangle=\delta_{ij}\ ;
\label{spectral-decomposition}
\end{equation}
the set $\{\lambda_i\}$ of eigenvalues of $\rho$ constitutes a probability distribution which completely defines the statistical properties of the quantum state. Although the spectral decomposition (\ref{spectral-decomposition}) is unique, 
it should be stressed that there are infinitely many ways of expressing a density matrix as a linear combination 
of projectors as in (\ref{densitymatrix}).

The set of all density matrices describing a quantum system $S$ forms a convex subset of $M_n(\mathbb{C})$, as combining different
density matrices $\eta_i$ into a convex combination $\sum_i r_i \eta_i$, with weights $r_i\geq 0$, and $\sum_i r_i = 1$,
gives again a density matrix. Pure states are extremal elements of this set, that is, they cannot be expressed as
a convex combination of other density matrices; they can be used to decompose non-pure states, see (\ref{densitymatrix}),
and indeed in this way they generate the whole set of density matrices.

Any transformation of the system $S$ can be modeled as a linear map acting on the space of density matrices, $\rho \to \mathbb{E}[\rho]$.
The most general form of such transformations, as
allowed by the statistical interpretation of quantum mechanics outlined above,
is given by the following operator-sum representation:
\begin{equation}
\rho \to \mathbb{E}[\rho]=\sum_i V_i\, \rho\, V_i^\dagger\ ,
\label{operator-sum}
\end{equation}
for some collection of operators $\{V_i\}$. Clearly, the map $\mathbb{E}$ in (\ref{operator-sum}) preserves the hermiticity
and positivity of $\rho$, and, provided $\sum_i V_i^\dagger V_i = \mathbf{1}_n$, with $\mathbf{1}_n\in M_n(\mathbb{C})$ 
the identity matrix,
also its normalization; such a map is called
a \textbf{quantum operation}, or simply a \textbf{quantum channel}.

In particular, the unitary dynamics, $\rho\to \mathbb{U}_t[\rho]$, generated by a system 
Hamiltonian operator $H\in M_n(\mathbb{C})$, is of the form (\ref{operator-sum}), with just one operator $V_i$:
\begin{equation}
\rho \to \mathbb{U}_t[\rho]= e^{-it H}\, \rho\,  e^{it H}\ .
\label{unitary-dynamics}
\end{equation}
The set of transformations $\{\mathbb{U}_t\}$ forms a one-parameter group of linear maps,
$\mathbb{U}_t \circ \mathbb{U}_s = \mathbb{U}_{t+s}$, for all $t,s\in\mathbb{R}$, 
reflecting the reversible character of the unitary Schr\"odinger dynamics;
as such, it preserves the spectrum and the purity of the density matrix:
\begin{equation}
\rho = \rho^2 \implies \big(\mathbb{U}_t[\rho]\big)^2=\mathbb{U}_t[\rho]\ .
\label{purity-preservation}
\end{equation}

Another common transformation affecting quantum states involves \textbf{measurement}. Assuming the system $S$
be initially prepared in a pure state $|\psi\rangle \langle\psi|$, after measuring
a non-degenerate observable $\mathcal{O}=\sum_k \mathcal{O}_k |k\rangle \langle k|$, expressed in its spectral form
with $\mathcal{O}_k$ being its eigenvalues and $|k\rangle$ the corresponding eigenvectors, then
the outcome $\mathcal{O}_k$ occurs with probability $w_k=|\langle k|\psi\rangle|^2$ and,
if the measurement indeed produces $\mathcal{O}_k$, then the post-measurement system state is the projector
$P_k=|k\rangle \langle k|$.
By repeating the measurement operation on copies of the system $S$ equally prepared in the state $|\psi\rangle \langle\psi|$,
the collection of the resulting post-measurement states is described by the statistical mixture
$\{w_k, |k\rangle\}$:
\begin{equation}
|\psi\rangle \langle\psi| \to \sum_k w_k  P_k = 
\sum_k P_k\, \big( |\psi\rangle \langle\psi| \big)\, P_k\ .
\label{}
\end{equation}
This transformation can be extended by linearity to cover any initial density matrix $\rho$ for the system $S$;
as a result, after the given set of measurements the system state is subjected to the transformation:
\begin{equation}
\rho \to \mathbb{P}[\rho]=\sum_k P_k\, \rho \, P_k\ .
\label{measure-map}
\end{equation}
Contrary to the unitary dynamics $\mathbb{U}_t$, the map $\mathbb{P}$ generally transforms pure states into mixtures,
thus involving decoherence effects resulting in the suppression of any initially present phase-interference.
This happens because the quantum operation $\mathbb{P}$ effectively describes $S$ 
as an \textbf{open system}, in this case as a system
interacting with the apparatus used to measure the observable $\mathcal{O}$. Quite in general,
dynamics generating noise and dissipation through decoherence can be modelled as those of systems
in interaction with large external environments; their evolution
must be of the form (\ref{operator-sum}), the only one guaranteeing physical consistency 
in any situation.

\subsection{Quantum correlations} \label{sec:QC}

One of the characteristic properties of quantum mechanics is the possibility of having correlations among constituent quantum
systems, that is, correlations among their observables, that cannot be
accounted for by classical physics.
Initially dismissed as a pure curiosity, the presence of such quantum correlations, that is 
of \textbf{entanglement}~\cite{Horodecki:2009zz,Benatti:2010,Das-doi:https://doi.org/10.1002/9783527805785.ch8}, 
has rapidly become a fundamental resource
in the development of disciplines like quantum information and technology, as it allows
the implementation of protocols and the realization of various apparatus outperforming 
classical ones~\cite{Nielsen:2012yss,bruss2019quantum}.

Many experiments have shown the presence of quantum correlations
in systems involving photons, atoms 
and more recently elementary particles.
Indeed, as entanglement is most likely to emerge as the result of a direct interaction among the constituents
of a quantum system, the interaction among elementary particles as seen
at colliders seems a promising place to study the effects of quantum correlations.

In the following we shall merely deal with bipartite composite quantum systems $S= S_A +S_B$
consisting of two finite-dimensional parties $S_A$ and $S_B$, usually identified with two distant,
well-separated quantum subsystems. An observable $\hat{\mathcal{O}}$ of $S$ can then be expressed in a tensor product form,
$\hat{\mathcal{O}}=\hat{\mathcal{O}}_A\otimes\hat{\mathcal{O}}_B$, where $\hat{\mathcal{O}}_A$, $\hat{\mathcal{O}}_B$ are
observables of $S_A$ and $S_B$, respectively; notice that $\hat{\mathcal{O}}$ is the product of two local
operators, $\hat{\mathcal{O}}_A\otimes {\bf 1}_B$ and ${\bf 1}_A\otimes \hat{\mathcal{O}}_B$.
\begin{quote}
{\sl A state (density matrix) $\rho$ of $S$ is called \textbf{separable} if and only if it can be written
as a linear convex combination of tensor products of density matrices:
\begin{equation}
\rho  = \sum_{ij} p_{ij}\,  \rho^{(A)}_i\otimes \rho^{(B)}_j\ ,\quad \text{with} 
\quad p_{ij}>0 \quad \text{and} \quad \sum_{ij} p_{ij}=1\ , 
\label{separable}
\end{equation}
where $\rho^{(A)}_i$ and $\rho^{(B)}_j$ are density matrices for the subsystems $S_A$ and $S_B$. 
States $\rho$ that cannot be written in the form of (\ref{separable})
are called \textbf{entangled} or \textbf{non-separable}, and exhibit quantum correlations.}
\end{quote}
Notice that, by expressing the density matrices $\rho^{(A)}_i$ and $\rho^{(B)}_j$ 
in terms of their spectral decomposition, that is in terms of their respective eigenprojectors,
separable states as in (\ref{separable}) can always be written as linear convex combinations 
of tensor products of pure states. These states carry statistical correlations, but they are just of classical nature,
reflecting the way the pure states are mixed together. Specifically, a separable state of $S$ of the form
\begin{equation}
\rho  = \sum_{ij} \lambda_{ij}\,  |\psi^{(A)}_i\rangle\langle \psi^{(A)}_i |\otimes 
|\psi^{(B)}_j\rangle\langle \psi^{(B)}_j|\ ,\quad {\rm with}
\quad \lambda_{ij}>0 \quad  {\rm and} \quad \sum_{ij} \lambda_{ij}=1\ , 
\label{classical}
\end{equation}
describes a statistical ensemble that can always be viewed as $N_{ij}$ instances of a system
with state vector $|\psi^{(A)}_i\rangle \otimes |\psi^{(B)}_j\rangle$
coming from a ``source'' that has ``emitted'' a total number $N$ of such systems,
the ratio $N_{ij}/N$ approaching the weight $\lambda_{ij}$ in the large-$N$ limit.
Therefore, in this case the weights $\lambda_{ij}$ just reflect the statistics of the source, 
viewed as a classical stochastic variable.

In addition, due to the structure of (\ref{separable}), the local character of separable states cannot 
be modified by local actions of the form $\hat{\mathcal{O}}_A\otimes\hat{\mathcal{O}}_B$ 
with $\hat{\mathcal{O}}_A$, $\hat{\mathcal{O}}_B$ admissible quantum operations for the subsystems $S_A$, $S_B$.
In other words, in order to change the local character of a separable state into a nonlocal one, 
a nonlocal action involving both parties, for instance a direct interaction, is necessary.

Pure, separable density matrices, such that $\rho^2=\rho$, are projectors onto state vectors in product form,
\hbox{$|\psi\rangle =$} \hbox{$|\psi^{(A)}\rangle \otimes |\psi^{(B)}\rangle$}, $\rho = |\psi\rangle\langle \psi|$,
for some vector states $|\psi^{(A)}\rangle$ of $S_A$ and
$|\psi^{(B)}\rangle$ of $S_B$. Nevertheless, given a generic state vector for the system $S$,
\begin{equation}
|\psi\rangle  = \sum_{ij} \psi_{ij}\,  |i\rangle^{(A)} \otimes |j\rangle^{(B)}\ , 
\label{pure-separable}
\end{equation}
with $\{ | i\rangle^{(A)}\}$, $\{ | i\rangle^{(B)}\}$, orthonormal bases in $S_A$, $S_B$, 
proving that it can or cannot be written in product form is in general a nontrivial
task. Fortunately, the problem can be solved by using
the Schmidt decomposition~\cite{Nielsen:2012yss}; in fact, for any generic state (\ref{pure-separable}), 
one can always find two suitable orthonormal bases for $S_A$ and $S_B$
yielding a diagonal decomposition:
\begin{equation}
|\psi\rangle  = \sum_{k=1}^d \lambda_{k}\,  |k\rangle^{(A)} \otimes |k\rangle^{(B)}\ , 
\label{Schmidt}
\end{equation}
with non-negative Schmidt coefficients $\lambda_k$ and $d$ the lowest dimension among $S_A$ and $S_B$; 
if at least two of these coefficients are nonvanishing,
the state $|\psi\rangle$ is not in product form and thus it is entangled. As a consequence, denoting
with $\rho_A = {\rm Tr}_B[|\psi\rangle\langle \psi|]$, and $\rho_B = {\rm Tr}_A[|\psi\rangle\langle \psi|]$,
the partial traces over $S_B$ and $S_A$ subsystems, respectively, a generic pure state
$|\psi\rangle$ of $S$ is separable if and only if its reduced states $\rho_A$ and $\rho_B$ are pure.

Alternatively, one can focus on the \textbf{von Neumann entropy}, that for a generic density matrix $\rho$
is defined as~\cite{Nielsen:2012yss}
\begin{equation}
{\cal S}[\rho]= -{\rm Tr}[\rho\ln\rho]\ ;
\label{entropyS}
\end{equation}
it is the quantum analogue of the classical Shannon entropy. 
In terms of the reduced density matrices, one
can then define the quantity
\begin{equation}
\mathscr{E}[\rho] \equiv -{\rm Tr}[\rho_A\ln \rho_A] = -{\rm Tr}[\rho_B\ln \rho_B]\ ;
\label{entropy}
\end{equation}
clearly, a pure state $\rho = |\psi\rangle\langle \psi|$ is entangled if and only 
if its reduced density matrices have non-zero entropy. The quantity $\mathscr{E}[\rho]$,
often called in the literature \textbf{entropy of entanglement}, is an entanglement quantifier;
assuming for the two systems $S_A$ and $S_B$ have the same dimension $d$,
one finds $0\leq \mathscr{E}[\rho] \leq \ln d$. The first equality holds if and only if the bipartite pure state is separable, 
while the upper bound is reached by a maximally entangled state,
\begin{equation}
|\Psi_+\rangle=\frac{1}{\sqrt{d}} \sum_{i=1}^d |i\rangle^{(A)}\otimes|i\rangle^{(B)}\ .
\label{max-ent}
\end{equation}

When the state $\rho$ of the compound system $S$ is a generic density matrix,
deciding whether the state is entangled or not, or quantifying its entanglement content, 
is often a hard problem~\cite{gurvits2003classical,Gharibian:2008hgo} and only partial answers are available. In general,
one can only rely on so-called \textbf{entanglement witnesses},
quantities that give sufficient conditions for the presence of entanglement in the system. 

In building such witnesses, a crucial role is played by positive maps $\Lambda$, that is by linear
transformations on the space of matrices, mapping positive matrices, that is, matrices with non-negative eigenvalues,
into positive matrices. Let us assume for simplicity that two systems $S_A$ and $S_B$ have the same dimension $d$;
then the following basic result holds~\cite{Horodecki:19961}:
\begin{quote}
{\sl A state $\rho$ of the bipartite system $S$ is entangled if and only if there exists a positive map $\Lambda_A$
on the subsystem $S_A$, such that the matrix $\rho$ is not left positive by the action of the
compound map $\Lambda_A\otimes {\bf 1}_B$, that is $(\Lambda_A\otimes {\bf 1}_B)[\rho] \ngeq 0$.}
\end{quote}
A well known, easily implementable entanglement test based on this result involves the transposition map, for instance on the
subsystem $S_A$: the compound operation $T_A\otimes {\bf 1}_B$, 
is called partial transposition; then (\textbf{Peres-Horodecki criterion})~\cite{Peres:PhysRevLett.77.1413}:

\begin{quote}
{\sl A state $\rho$ of the bipartite system $S$ is entangled if it does not remain positive under partial transposition,
$(T_A\otimes {\bf 1}_B)[\rho] \ngeq 0$.}
\end{quote}
This entanglement criterion is exhaustive in lower dimensions, for a bipartite system consisting two qubits,
or a qubit and a qutrit~\cite{Woronowicz:1976165}. In addition, quite in general, the absolute sum of the negative eigenvalues
of a partially transposed state, a quantity called \textbf{negativity} and given by
\be
{\cal N}(\rho) = \sum_k \frac{|\lambda_k| - \lambda_k}{2}\, ,
\ee
in which $\lambda_k$ are the eigenvalues of the partially transposed matrix $
\langle i_1,j_2|\rho^{T_2}|\bar i_1 \bar j_2 \rangle $ of the density matrix
$\langle i_1, \bar j_2|\rho|\bar i_1  j_2 \rangle 
$,
can be used to quantify its entanglement content~\cite{Vidal:2002zz}.

The relationship between the entropy of the system and those of its parts
can be used to check whether the state is entangled;  
if the state $\rho$ is separable, than necessarily: ${\cal S}[\rho] \geq {\cal S}[\rho_A]$ and 
${\cal S}[\rho] \geq {\cal S}[\rho_B]$, with $\rho_A$ and $\rho_B$ 
being again the reduced density matrices~\cite{Horodecki:2009zz}.

In applications, entanglement witnesses that can be easily computed are needed: 
a relevant example of such a witness is the \textbf{concurrence}.
Consider again the bipartite quantum system $S$, made of two $d$-dimensional
subsystems $S_A$, $S_B$, described by a normalized pure state $|\psi\rangle$, or equivalently 
by the density matrix $|\psi\rangle\langle\psi|$. The concurrence of the system is 
defined by~\cite{Bennett:PhysRevA.54.3824,Wootters:PhysRevLett.80.2245,Rungta:PhysRevA.64.042315}
\begin{equation}
\mathscr{C}[|\psi\rangle]\equiv\sqrt{2\left( 1-{\rm Tr}\big[(\rho_A)^2\big]\right)}
=\sqrt{2\left( 1-{\rm Tr}\big[(\rho_B)^2\big]\right)}\ .
\label{C_psi}
\end{equation}
As already remarked, any mixed state $\rho$ of the bipartite system can be decomposed 
into a set of pure states $\{|\psi_i\rangle\}$,
\begin{equation}
\rho=\sum_i p_i\, |\psi_i\rangle\langle\psi_i|\ ,\quad {\rm with}\quad p_i\geq0\ ,\quad {\rm and}\quad \sum_i p_i=1\ ;
\end{equation}
its concurrence is then defined by means of the concurrence of the pure states appearing in the decomposition 
through an optimization process:
\begin{equation}
\mathscr{C}[\rho]=\underset{\{|\psi\rangle\}}{\rm inf} \sum_i p_i\, \mathscr{C}[|\psi_i\rangle]\ ,
\label{C_rho}
\end{equation}
where the infimum is taken over all the possible decompositions of $\rho$ into pure states. 
For a pure state the concurrence (\ref{C_psi}) vanishes if and only if the state is separable, 
$|\psi\rangle=|\psi_A\rangle\otimes |\psi_B\rangle$, reaching its maximum value
when $\rho$ is a projection on the pure, maximally entangled state (\ref{max-ent}).
As the same holds for mixed states~\cite{Mintert:PhysRevLett.92.167902}, 
the concurrence appears to be a good entanglement detector. 
Unfortunately, the optimization problem appearing in (\ref{C_rho}) 
makes the evaluation of the concurrence a very hard mathematical task, with 
a simple analytic solution only for two-level systems, $d=2$.

Indeed, in this special low-dimensional case, given a two-qubit, $4\times 4$ density matrix $\rho$ as in (\ref{rho-1/2}), 
its concurrence can be explicitly constructed using the auxiliary matrix
\begin{equation}
R=\rho \,  (\sigma_y \otimes \sigma_y) \, \rho^* \, (\sigma_y \otimes \sigma_y)\, , 
\label{auxiliary-R}
\end{equation}
where $\rho^*$ denotes the matrix with complex conjugated entries. Although non-Hermitian, the matrix $R$
possesses non-negative eigenvalues; denoting with $r_i$, $i=1,2,3,4$, their square roots
and assuming $r_1$ to be the largest,
the concurrence of the state $\rho$ can be expressed as~\cite{Wootters:PhysRevLett.80.2245}
\begin{equation}
\mathscr{C}[\rho] = \max \big( 0, r_1-r_2-r_3-r_4 \big)\ .
\label{concurrence}
\end{equation}
By contrast, for $d>2$, any approximation or numerical computation of (\ref{C_rho}) provides only an upper bound on $\mathscr{C}[\rho]$ and thus cannot serve 
to reliably distinguish between entangled and separable states, or to give an estimate of a state entanglement content.

Fortunately, lower bounds on $\mathscr{C}[\rho]$ for a generic density matrix $\rho$ have been determined and, 
if non-vanishing, unequivocally signal the presence of entanglement. 
One of these bounds is easily computable, yielding~\cite{Mintert:PhysRevLett.98.140505}
\begin{equation}
\big(\mathscr{C}[\rho]\big)^{2} \geq \mathscr{C}_2[\rho]\ ,
\label{C-bound}
\end{equation}
where
\begin{equation}
\mathscr{C}_2[\rho] = 2 \,\text{max}\, \Big( 0,\, \Tr[\rho^{2}] - \Tr[(\rho_A)^2],\, \Tr[\rho^{2}] - \Tr[(\rho_B)^2]  \Big) \ .
\label{C_2}
\end{equation}
A non-vanishing value of $\mathscr{C}_2[\rho]$ implies a concurrence larger than zero, 
and therefore a non-vanishing entanglement content of $\rho$. 
Interestingly, an upper bound for $\mathscr{C}[\rho]$ has also been obtained~\cite{Zhang:PhysRevA.78.042308}; 
explicitly, one finds
\begin{equation}\label{C_2_upper}
\big(\mathscr{C}[\rho]\big)^{2} \leq 2 \,\text{min}\, \Big(1 - \Tr[(\rho_A)^2],\  1 - \Tr[(\rho_B)^2] \Big)\, .
\end{equation}
The easily computable concurrence lower bound (\ref{C_2}) can be used as entanglement witness 
in the study of quantum correlations at colliders.\footnote{For pure states the upper, \eq{C_2_upper}, and lower, \eq{C-bound}, bounds coincide and become a true measure of entanglement.}

Other definitions of non-classical correlations, different from entanglement, have been introduced in the literature,
motivated by the fact that they can be used to enhance selected information tasks
beyond their classical implementation 
(see~\cite{Modi:RevModPhys.84.1655,Adesso_2016,Bera_2018} and references therein).
Rather than nonlocality, these generalized quantum correlations
are the manifestation of non-commutativity and non-invariance under quantum measurements.
Indeed, as disturbance under quantum measurements signals quantumness,
one can characterize classicality through measurement invariance~\cite{Luo:PhysRevA.77.022301}.

Specifically, among separable states,  of the form (\ref{separable}), one can distinguish
the so-called classical-classical states:
\begin{equation}
\rho  = \sum_{ij} p_{ij}\,  \Pi^{(A)}_i \otimes \Pi^{(B)}_j\ ,\quad {\rm with}
\quad p_{ij}>0\ , \quad {\rm and} \quad \sum_{ij} p_{ij}=1\ , 
\label{classical-classical}
\end{equation}
where $\Pi^{(\alpha)}\equiv | i\rangle^{(\alpha)} {}^{(\alpha)}\langle i |$, $\alpha=A,B$, are the projectors on
the elements $\{ | i\rangle^{(\alpha)}\}$ of orthonormal bases in $S_\alpha$. 
There are no non-classical
correlations in these states; indeed, recalling (\ref{measure-map}), they are left undisturbed by any local von Neumann measurement,
performed locally on $S_A$ and $S_B$:
\begin{equation}
\rho  \to  \rho|_{AB} \equiv \sum_{ij}   \Big(\Pi^{(A)}_i \otimes \Pi^{(B)}_j\Big)\, 
\rho\,\Big( \Pi^{(A)}_i \otimes \Pi^{(B)}_j\Big) = \rho\ .
\label{classical-classical-invariance}
\end{equation}
In other terms, the amount of total correlations contained in $\rho$, quantified by its mutual information,
\begin{equation}
I(\rho) =  {\cal S}(\rho_A) +{\cal S}(\rho_B) - {\cal S}(\rho)\ ,
\label{mutual-information}
\end{equation}
where ${\cal S}$ is the von Neumann entropy (\ref{entropyS}),
coincides with the classical Shannon mutual 
information of the joint probability distribution $\{p_{ij}\}$:
the correlations present in $\rho$ are purely classical.

Similarly, one can introduce, separable, quantum-classical states,
\begin{equation}
\rho  = \sum_{i} p_{i}\,  \rho^{(A)}_i \otimes \Pi^{(B)}_j\ ,\quad {\rm with}
\quad p_{i}>0\ , \quad  {\rm and} \quad\sum_{ij} p_{i}=1\ , 
\label{quantum-classical}
\end{equation}
where $\rho^{(A)}_i$ are admissible density matrices for the subsystem $S_A$, 
while, as before, $\Pi^{(B)}_j$ are orthonormal projectors on $S_B$.
These states are left undisturbed under von Neumann measurements  performed on the subsystem $S_B$:
\begin{equation}
\rho  \to  \rho|_B \equiv \sum_{i}   \Big({\bf 1}_A \otimes \Pi^{(B)}_i\Big)\, 
\rho\,\Big( {\bf 1}_A  \otimes \Pi^{(B)}_i\Big) = \rho\ .
\label{quantum-classical-invariance}
\end{equation}
By exchanging the role of $S_A$ and $S_B$, one analogously defines classical-quantum states.

In the case of a more general state, as in (\ref{separable}), in order to obtain its genuine quantum correlation content
one needs to subtract from its quantum mutual information (\ref{mutual-information}) the amount of classical
correlations obtained through local von Neumann measurements.
A possible measure of such classical correlations can be defined as~\cite{Zurek:PhysRevLett.88.017901,Henderson_2001}
\begin{equation}
J^{(B)}(\rho) = \underset{\{\Pi^{(B)}\}}{\rm max}\ I \big( \rho|_B \big)\ ,
\label{mutual-information2}
\end{equation}
where the maximization is over all local von Neumann measurements on $S_B$, 
defined as in (\ref{quantum-classical-invariance}).
One can similarly define $J^{(A)}(\rho)$ by exchanging the roles of $S_A$ and $S_B$, 
or in a symmetric way
\begin{equation}
J(\rho) = \underset{\{\Pi^{(A)}\otimes\Pi^{(B)}\}}{\rm max}\ I \big( \rho|_{AB} \big)\ ,
\label{mutual-information3}
\end{equation}
with the maximization over all local von Neumann measurements $\{\Pi^{(A)}\otimes\Pi^{(B)}\}$
as defined in (\ref{classical-classical-invariance}).
One can then define \textbf{discord} as a measure of the content of non-classical correlations
of a bipartite state $\rho$ as the (non-negative) difference~\cite{Zurek:PhysRevLett.88.017901}:
\begin{equation}
Q^{(B)}(\rho) = I(\rho) - J^{(B)}(\rho)\ .
\label{B-discord}
\end{equation}
One finds that $Q^{(B)}(\rho)=\, 0$ if and only if the state $\rho$ is quantum-classical as in (\ref{quantum-classical}).
A symmetric version of discord can also be introduced~\cite{Wu:PhysRevA.80.032319}:
\begin{equation}
Q(\rho) = I(\rho) - J(\rho)\ ;
\label{discord}
\end{equation}
being the difference between the amount of total correlations and the one of classical correlations,
it vanishes, $Q(\rho)=\,0$, if and only if $\rho$ is classical-classical as in (\ref{classical-classical}).
Extensions of these quantities using generalized quantum Positive Operator-Valued Measures (POVMs)  instead of von Neumann ones
have been discussed in \cite{Henderson_2001}.

In general, discord and entanglement are different measures of the content of quantum correlations in a given bipartite state;
though they coincide for pure states. Classically correlated mixed states are separable,
but the converse is not true: mixed separable state may possess non-zero discord. Additional properties and applications
of discord and other measures of non-classical correlations can be found in~\cite{Modi:RevModPhys.84.1655,Adesso_2016,Bera_2018} and recently discussed in \cite{Afik:2022dgh} for pairs of top quarks.

\subsection{Bell nonlocality}\label{sec:Bell}

One of the most striking and unexpected results of modern physics
is the manifestation of a fundamental indeterminacy in natural phenomena.
Thanks to the advent of quantum mechanics, the use of a statistical language 
became the standard, compelling tool for explaining the behavior of physical phenomena.
Yet, the possibility of recovering a fully deterministic description of natural phenomena is amenable to experimental test, which rests on the presence of classes
of correlations among observables underlying what is 
now known as \textbf{Bell nonlocality}~\cite{Brunner:RevModPhys.86.419,scarani2019bell}.

The simplest situation in which the dichotomy between locality and nonlocality can be appreciated
is that of a bipartite physical system, one part controlled by an agent $A$ (Alice), while that other
by the agent $B$ (Bob), well separated and distinct.%
\footnote{The two parties are usually assumed not to be able to exchange messages, being
in the so-called ``non-signaling settings''.}
Both agents perform measurements
on their respective subsystem parts and by comparing the corresponding results
draw conclusions on the presence of possible correlations.
It is the structure of these correlations that allows distinguishing {\it local} from {\it nonlocal}; indeed,
J.~S.~Bell in 1964~\cite{Bell:1964} was able to introduce a logical formulation, the \textbf{Bell inequalities}, 
allowing a disprovable test for correlations 
being local or nonlocal~\cite{bell2004speakable,redhead1987incompleteness,bell2002quantum,bertlmann2016quantum}. 
A violation of one of these inequalities, as testified in many experiments,  
not only reveals something about the internal structure of quantum physics, 
but strikingly, tells us that correlations in spatially separated systems can exhibit a fundamental nonlocal character.

Bell locality essentially means that the measurement outputs at one party, say $A$, do not depend 
on the outcomes at the remaining one, at $B$; in other terms, all correlations between Alice and Bob
are consequence of shared resources, which, for a quantum system, can even include its wavefunction.
This form of locality can be formalized in full generality.
Let us denote with the (for simplicity, continuous) variable $\lambda$ the set of unspecified common resources, 
shared among Alice and Bob. Further, assume that Alice can choose to measure $M_A$ different
observables $\hat A_1$, $\hat A_2$, \ldots $\hat A_{M_A}$, each one giving rise to $m_A$
different outcomes $a_i=1,2,\ldots, m_A$, $i=1,2,\ldots, M_A$, and similarly for Bob.
Let $P_\lambda(A|a)$ be the probability for Alice of getting the outcome $a$ having chosen to measure the observable $\hat A$
and similarly be $P_\lambda(B|b)$ the probability for Bob of getting $b$ out of the measurement of the observable $\hat B$.
What is important is that $P_\lambda(A|a)$ does not depend on the measurement chosen by Bob
and similarly $P_\lambda(B|b)$ does not depend on the Alice choices; in other terms, the outcome $a$ for Alice and $b$ for Bob
are generated locally, by sampling from the probability distribution $P_\lambda(A|a)$ and $P_\lambda(B|b)$,
respectively.

Within these settings, the probability $P(A,B|a,b)$ of the joint result $(a,b)$, having measured $\hat A$ and $\hat B$,
can be expressed as
\begin{equation}
P(A,B|a,b) = \int {\rm d}\lambda~ \eta(\lambda)\ P_\lambda(A|a)\ P_\lambda(B|b)\ ,
\label{Bell-locality}
\end{equation}
where $\eta(\lambda)$ is the probability distribution of the shared resources.
This is the formal statement of Bell locality; the corresponding statistics of outcomes
is called \textbf{local} if it obeys (\ref{Bell-locality}), 
\textbf{nonlocal} otherwise.
Checking the validity of the hypothesis (\ref{Bell-locality}) is usually done by performing a \textbf{Bell test},
that is, by putting under experimental scrutiny the validity of suitable Bell inequalities that result directly from the hypothesis (\ref{Bell-locality}).

\subsubsection{Qubits}

In order to be more specific, let us study the simplest Bell test, involving two parties, Alice and Bob,
each one having at their disposal two possible observables to measure, ($\hat A_1$, $\hat A_2$), and ($\hat B_1$, $\hat B_2$),
respectively, each giving rise to two possible outcome $(0, 1)$; in the notation introduced above:
$M_A=M_B=m_A=m_B=2$~\cite{Clauser:1969ny,Clauser:1974tg,Clauser:1978ng}.
The test results in checking the following combination of joint expectation values, 
involving an observable of Alice and one of Bob~\cite{Clauser:1969ny}:
\begin{equation}
\mathcal{I}_2=\langle {\hat A}_1 {\hat B}_1\rangle + \langle {\hat A}_1 {\hat B}_2\rangle +\langle {\hat A}_2 {\hat B}_1\rangle
- \langle {\hat A}_2 {\hat B}_2\rangle\ .
\label{CHSH}
\end{equation}
In order to obtain the maximum value of $\mathcal{I}_2$ achieved using only {\it local} resources, 
it is sufficient~\cite{Fine:PhysRevLett.48.291,scarani2019bell}
to see what is the outcome when Alice and Bob share a pre-determined set $(a_1, a_2; b_1, b_2)$
of possible answers to the measurement queries; clearly, as these answers can be either $0$ or $1$,
$\mathcal{I}_2$ can be at most 2, so that Bell locality implies the \textbf{Clauser-Horne-Shimony-Holt (CSHS) inequality}:
\begin{equation}
\mathcal{I}_2\leq 2\ .
\label{CHSH-2}
\end{equation}
If in an actual experiment one finds $\mathcal{I}_2 > 2$, one has to deduce that some sort of nonlocal
resource had been shared between the two parties, and this is precisely what is predicted
in a quantum mechanical setting.%
\footnote{Quantum mechanics predicts for $\mathcal{I}_2$ the maximal value $2\sqrt{2}$~\cite{cirelson}.
Interestingly, hypothetical models ``more nonlocal'' than quantum mechanics have been advocated~\cite{Popescu},
for which the upper value of $\mathcal{I}_2$ may exceed $2\sqrt{2}$.}

A paradigmatic model in which the inequality (\ref{CHSH-2})
can be easily checked is a bipartite system made of two spin-1/2 particles, one belonging to Alice, the other to Bob.
As it will discussed in detail in the following, this physical situation is routinely reproduced at colliders,
where analysis of the spin correlations among products of high-energy collisions
is performed.

A bipartite quantum system, made of two spin-1/2 particles
is described in quantum mechanics in terms of the $4$-dimensional Hilbert space 
$\mathcal{H}_4=\mathcal{H}_2\otimes\mathcal{H}_2\equiv\mathbb{C}^4$, the tensor product of two, \hbox{$2$-dimensional}
Hilbert spaces $\mathcal{H}_2\equiv\mathbb{C}^2$ representing a single spin-1/2 particle.
As already remarked, any observable $\hat{\mathcal{O}}$ of the full system can then be expressed in a tensor product form,
$\hat{\mathcal{O}}=\hat{\mathcal{O}_1}\otimes\hat{\mathcal{O}_2}$, where $\hat{\mathcal{O}_1}$, $\hat{\mathcal{O}_2}$ are
each single-spin observables, for instance they could be spin projections each acting on one of the two particles, and in general in different spatial directions.

The state of the two spin-1/2 system is in general described by a density matrix $\rho$,
that is an operator acting on $\mathcal{H}_4$, that can be represented by a non-negative, $4\times 4$ matrix
of unit trace. As already mentioned,
when the density matrix is a projector operator, $\rho^2=\rho$, than the state of the system
can be represented by a state vector $|\psi\rangle\in\mathcal{H}_4$, such that
$\rho=|\psi\rangle \langle\psi|$.
Knowing $\rho$ allows one to compute the average of any two-spin observable $\hat{\mathcal{O}}$
through its trace with $\rho$, $\langle\hat{\mathcal{O}}\rangle={\rm Tr}[\rho\, \hat{\mathcal{O}}]$;
these expectation values are the quantities measurable in experiments.

The quantum state of a spin-1/2 pair can then be expressed as 
\begin{equation}
\rho=\frac{1}{4}\Big[{\bf 1}_2\otimes{\bf 1}_2 + \sum_{i=1}^3 \BB^{+}_i (\sigma_i\otimes{\bf 1}_2)
+ \sum_{i=1}^3 \BB^-_j ({\bf 1}_2\otimes \sigma_j) + \sum_{i,j=1}^3 \CC_{ij} (\sigma_i\otimes\sigma_j) \Big]\ ,
\label{rho-1/2}
\end{equation}
where $\sigma_i$ are the Pauli matrices, ${\bf 1}_2$ is the unit $2\times 2$ matrix;
the indices $i$, $j$, running over $1$, $2$, $3$, represent any three orthogonal directions in three-dimensional space.
The real coefficients 
\be\label{rho-ab-coefficients}
\BB^+_i = {\rm Tr}[\rho\, (\sigma_i\otimes {\bf 1})] \quad \text{and} \quad
\BB_j^{-}={\rm Tr}[\rho\, ({\bf 1}\otimes\sigma_j)]\ ,
\ee
represent the \textbf{polarization}
of the two particles, while the real matrix 
\begin{equation}\label{rho-c-coefficients}
\CC_{ij} ={\rm Tr}[\rho\, (\sigma_i\otimes\sigma_j)]
\end{equation}
gives their spin \textbf{correlations}. 
The labels `$+$' and `$-$' on the $B$ coefficients simply serve to indicate which particle they refer to; in what follows they are often distinguished by their respective electric charges. 
In the case of a collider setting, $\BB^+_i$, $\BB^-_i$ 
and $\CC_{ij}$ will be functions of the parameters
describing the kinematics of the pair of spin-1/2 production, the total energy $\sqrt s$ in the center of mass
reference frame and the corresponding scattering angle $\theta$.
Note that while the density matrix in (\ref{rho-1/2}) is normalized, ${\rm Tr}[\rho]=1$,
extra constraints on $\BB^+_i$, $\BB^-_i$ 
and $\CC_{ij}$ 
need to be enforced to guarantee its positivity;
these extra conditions are in general non-trivial, as they originate from requiring 
all principal minors of $\rho$ to be non-negative.

The density matrix in \eq{rho-1/2} can be used to re-write the upper bound on the concurrence in \eq{C_2_upper}  as
\be
\big(\mathscr{C}[\rho]\big)^{2} \leq \text{min}\, \Big[1 - \sum_i (\BB_i^+)^2,\;  1 - \sum_j (\BB_j^-)^2 \Big]\,. \label{BC}
\ee
\eq{BC} makes clear that the larger the polarization of each individual particle (as found in the size of the coefficients $B_i^\pm$), the smaller the largest possible value of the polarization entanglement between them, as described by $\mathscr{C}[\rho]$. More precisely, the entanglement in the final state spin correlations is maximal for vanishing polarizations, progressively diminishes as the polarizations increase and vanishes for fully polarized final state particles. 

Let us now express the combination of expectation values appearing in (\ref{CHSH}) in the language of spin,
and choose as observables $\hat{A}_1$ and $\hat{A}_2$, for the first spin-1/2 particle, and 
$\hat{B}_1$, $\hat{B}_2$ for the second one, spin projections along four different unit vectors,
say $\vec{n}_1$, $\vec{n}_3$ for Alice, and $\vec{n}_2$, $\vec{n}_4$ for Bob, so that 
$\hat{A}_1= \vec{n}_1\cdot \vec{\sigma}$ and similarly for the remaining three observables.
Only the correlation matrix $\CC$ is involved in the combinations in (\ref{CHSH}),
that can be conveniently expressed as $\mathcal{I}_2={\rm Tr}[\rho \mathscr{B}]$ where
the quantum \textbf{Bell operator} is given by
\begin{equation}\label{Bchsh}
\mathscr{B} ={\vec n}_1 \cdot {\vec\sigma} \otimes ({\vec n}_2 - {\vec n}_4) \cdot {\vec\sigma}
+ {\vec n}_3 \cdot {\vec\sigma} \otimes ({\vec n}_2 + {\vec n}_4 )\cdot  {\vec\sigma}\ .
\end{equation}
The Bell inequality (\ref{CHSH-2}) then becomes 
\begin{equation}
{\vec n}_1\cdot \CC \cdot \big({\vec n}_2 - {\vec n}_4 \big) +
{\vec n}_3\cdot \CC \cdot \big({\vec n}_2 + {\vec n}_4 \big)\leq 2\ .
\label{algebraic-relation}
\end{equation}
Combining this condition with the analogous one obtained by reversing the direction of  ${\vec n}_1$ and ${\vec n}_3$
one finally gets the following constraint:
\begin{equation}
\Big|{\vec n}_1\cdot \CC \cdot \big({\vec n}_2 - {\vec n}_4 \big) +
{\vec n}_3\cdot \CC \cdot \big({\vec n}_2 + {\vec n}_4 \big)\Big|\leq 2\ .
\label{algebraic-inequality}
\end{equation}
When the spins of the two particle are perfectly anticorrelated, as it happens for a pure singlet state,
\begin{equation}
|\Psi\rangle= \frac{1}{\sqrt{2}}\Big( |\uparrow_{\vec n}\rangle\otimes|\downarrow_{\vec n}\rangle
- |\downarrow_{\vec n}\rangle\otimes|\uparrow_{\vec n}\rangle\Big)\ ,
\label{singlet}
\end{equation}
with $|\uparrow_{\vec n}\rangle$ representing the spin of a particle in the state $\uparrow_{\vec n}$, 
that is with the projection of the spin along the axis determined by the unit vector $\vec n$ pointing in the up direction, one finds
\begin{equation}
\CC_{ij}=-\delta_{ij}\ ,
\label{singlet-correlation}
\end{equation}
and one can easily violate the inequality (\ref{CHSH-2}) by a suitable choice of the four unit vectors
$\vec{n}_1$, $\vec{n}_3$, $\vec{n}_2$, $\vec{n}_4$. In other terms, the nonlocality of quantum mechanics
violates the Bell locality test (\ref{CHSH-2}). 

In order to actually put under experimental test the Bell inequality (\ref{algebraic-inequality}), 
one in principle needs to extract from the collected data the matrix $\CC$
and then choose suitable four independent spatial directions ${\vec n}_1$, ${\vec n}_2$, ${\vec n}_3$ and ${\vec n}_4$
that maximize $\mathcal{I}_2$ in~(\ref{CHSH}).
Fortunately, this maximization process can be performed in full generality 
for a generic spin correlation matrix~\cite{Horodecki:1995340}.
Indeed, consider the matrix $\CC$ and its transpose $\CC^T$ 
and form the symmetric, positive, $3\times 3$ matrix
$M= \CC \CC^T$; its three eigenvalues $m_1$, $m_2$, $m_3$ can be ordered in increasing order:
$m_1\geq m_2\geq m_3$. Then, the following result holds:
\begin{quote}
{\sl The two-spin state $\rho$ in (\ref{rho-1/2}) violates the inequality (\ref{algebraic-inequality}), or
equivalently (\ref{CHSH-2}),
if and only if the sum of the two greatest eigenvalues of $M$ is strictly larger than 1, that is (\textbf{Horodecki condition})}
\begin{equation}
\mathfrak{m}_{12}\equiv m_1 + m_2 >1\, .
\label{eigenvalue-inequality}
\end{equation}
\end{quote}
\noindent
In other terms, given a spin correlation matrix $\CC$ of the state $\rho$ that satisfies (\ref{eigenvalue-inequality}),
then there are choices of the four independent vectors ${\vec n}_1$, ${\vec n}_2$, ${\vec n}_3$, ${\vec n}_4$
for which the left-hand side of (\ref{algebraic-inequality}) is larger than 2.
In the case of the singlet state (\ref{singlet})
the sum of the square of two of its eigenvalue is 2, the condition (\ref{eigenvalue-inequality})
is verified and thus the Bell inequality~(\ref{CHSH-2}) violated, actually at the maximal level~\cite{cirelson}.

\subsubsection{Qudits, mostly qutrits}

The quantum state of a two $d$-level systems, two \textbf{qudits}, can be expressed in a form similar to the one (\ref{rho-1/2}) for 
two qubits, the generalisation being~\cite{kimura2003bloch,Bertlmann-Bloch-2008}:
\begin{equation}
\rho=\frac{1}{d^2}\Big[{\bf 1}_d\otimes{\bf 1}_d + \sum_{i=1}^{d^2-1} \mathcal{A}_i^{(d)} (\tau_i\otimes{\bf 1})
+ \sum_{j=1}^{d^2-1} \mathcal{B}_j^{(d)}({\bf 1}\otimes \tau_j) + \sum_{i,j=1}^{d^2-1} 
\mathcal{C}_{ij}^{(d)} (\tau_i\otimes\tau_j) \Big]\ ,
\label{rho-d}
\end{equation}
where the matrices $\tau_i$, $i=1,2,\ldots, d^2-1$, are the traceless hermitian generators of the fundamental representation 
of the algebra $su(d)$, forming with the normalized identity matrix $\tau_0=\sqrt{2/d}\,{\bf 1}_d$
an orthonormal basis in the space of all $d \times d$ hermitian matrices. 
Recalling that
${\rm Tr}[\tau_i\, \tau_j]= 2\,\delta_{ij}$, one now finds
\be
\mathcal{A}_i^{(d)}=\frac{d}{2}\, {\rm Tr}[\rho\, (\tau_i\otimes {\bf 1}_d)]\quad \text{and} \quad
\mathcal{B}_j^{(d)}=\frac{d}{2}\, {\rm Tr}[\rho\, ({\bf 1}_d\otimes\tau_j)]\ , 
\ee
representing the single qudit polarizations, 
while the real matrix 
\begin{equation}
\mathcal{C}_{ij}^{(d)}=\frac{d^2}{4}{\rm Tr}[\rho\, (\tau_i\otimes\tau_j)]
\end{equation}
gives their correlations.

Given a bipartite setting, sharing a system of two qubits, the Bell test (\ref{CHSH-2})
can be proven to be exhaustive: all other possible Bell tests are just a reformulation
of the basic inequality (\ref{CHSH-2}) (see, for example,  \cite{Brunner:RevModPhys.86.419,scarani2019bell}). 
Extensions to higher dimensions are however possible; 
in order to give one of such generalizations in the case of shared qutrits, that is,  three-level systems,
it is convenient to reformulate the condition (\ref{CHSH-2}) in terms of joint probabilities, by rewriting the
expectation values as:
\begin{equation}
\langle {\hat A}_i {\hat B}_j\rangle = \sum_{m=1}^2 \sum_{n=1}^2 (-1)^{m+n}\ P(A_i, B_j | m, n)\ ,
\label{probabilities}
\end{equation}
where as before $P(A_i, B_j | m, n)$ is the joint probability of finding the outcome $m$ 
in measuring the observable ${\hat A}_i$ by Alice, and the outcome $n$ from the measurement of
${\hat B}_j$ on Bob side. Then, the Bell test (\ref{CHSH-2}) is equivalent to
\begin{equation}
P(A_1=B_1) +P(A_2 \neq B_1) + P(A_2=B_2) + P(A_1=B_2) \leq 3\ ,
\label{Bell-test-2}
\end{equation}
where we have used the shorthand notation
$P(A_1=B_1)$ for the combination $P(A_1, B_1 | 1, 1) + P(A_1, B_1 | 2, 2)$ and similarly for the other terms.

Let us now assume that Alice and Bob share a system made of two qutrits, so that the
outcome of their measurements involve three possible entries, $(0,1,2)$. Let us also denote with
$P(A_i=B_j + k)$ the probability that the measurement outcome of the observables 
${\hat A}_i$ and ${\hat B}_j$ differ by $k$ modulo 3
and rewrite the left-hand side of (\ref{Bell-test-2}) as 
\begin{equation}
P(A_1=B_1) + P(A_2 +1 = B_1) + P(A_2=B_2) + P(A_1=B_2)\ ;
\label{Bell-test-3}
\end{equation}
clearly $P(A_2 +1 = B_1)=P(A_2 \neq B_1)$ in the case of qubits. 

Let us now assume that Alice and Bob share only local resources.
Then consider one possible outcome of their measurements such that
$A_1=B_1$, $A_1=B_2$ and $A_2=B_2$; but then locality would enforce $A_2=B_1$ and the probability $P(A_2 +1 = B_1)$
cannot be one. Clearly, any triple of similar conditions would lead to the same conclusion: for instance,
the choice $A_1=B_1$, $A_1=B_2$ and $A_2 +1 = B_1$ would lead to $A_2+1=B_2$ and thus $P(A_2=B_2)$ cannot be one.
As a result, the combination of probabilities (\ref{Bell-test-3}) cannot exceed 3, exactly as in the case of qubits.
One can prove that under any local deterministic assumptions the maximum of (\ref{Bell-test-3}) is 3
as only three probabilities out of four 
can be satisfied in the sum (\ref{Bell-test-3})~\cite{Fine:PhysRevLett.48.291,scarani2019bell}.

One can further restrict this result by subtracting from the combination (\ref{Bell-test-3}) the conditions
enforced by the four simplest deterministic choices, that is $P(A_2=B_1)$ in the first case discussed above,
$P(A_2 +1 = B_2)$ in the second, and so on. In this way one ends up with the condition:
\begin{align}
\nonumber
&\mathcal{I}_3\equiv P(A_1=B_1) + P(A_2 +1 = B_1) + P(A_2=B_2) + P(A_1=B_2)\\
&- P(A_2=B_1) - P(A_2 = B_2-1) - P(A_1 = B_1-1)- P(B_2 = A_1-1) \leq 2\ .
\label{CGLMP}
\end{align}
This is the Bell inequality introduced in~\cite{Collins:PhysRevLett.88.040404,Kaszlikowski:PhysRevA.65.032118}; 
one can prove that, as in the case of qubits for the
inequality (\ref{CHSH-2}), this inequality is optimal, in the sense that any other Bell inequality
involving two shared qutrits is equivalent to (\ref{CGLMP}).

Similarly to the case (\ref{CHSH}) for qubits, 
the combination of probabilities in
$\mathcal{I}_3$ can be expressed in quantum mechanics as an expectation value 
of a suitable Bell operator $\mathscr{B}$ as 
\begin{equation}
{\cal I}_{3}={\rm Tr}\big[\rho\, \mathscr{B}\big] \ ,
\label{eq:I3}
\end{equation}
where $\rho$ is the $9\times 9$ density matrix representing the state of the two qutrits. Following the current convention\footnote{While some authors maintain the overall $1/d^2$ factor in~\eq{rho-d} in their computation, others directly use the rescaled coefficients. In the following, we adopt the first convention for qubits and the second when dealing with qutrits.}, we denote
\be
f_i=\frac{1}{9} \, \mathcal{A}_i^{(3)}\, ,\quad  g_j=\frac{1}{9} \, \mathcal{B}_j^{(3)}\quad \text{and} \quad h_{ij}=\frac{1}{9} \, \mathcal{C}_{ij}^{(3)}\, .
\ee
The density operator in \eq{rho-d} can thus be written
\be
\rho = \frac{1}{9}\left[\mathbb{1}\otimes
    \mathbb{1}\right]+
    \sum_{a=1}^8 f_a \left[T^a\otimes \mathbb{1}\right]+\sum_{a=1}^8 g_a \left[\mathbb{1}\otimes T^a\right] 
    +\sum_{a,b=1}^8 h_{ab}  \left[T^a\otimes T^b\right]\, ,
\label{eq:rho-qutrit}
\ee
in the form of \eqref{rho-d}, specialised to $d=3$, where now the generators are the standard Gell-Mann matrices $T^a$.

The explicit form of $\mathscr{B}$ depends on the choice of the four measured operators $\hat{A}_i$ and $\hat{B}_i$. 
For the case of the maximally correlated qutrit state, analogous to the qubit state in (\ref{singlet}),
the problem of finding an optimal choice of measurements has been solved~\cite{Collins:PhysRevLett.88.040404},
and the Bell operator takes a particular simple form~\cite{Latorre:PhysRevA.65.052325}:
\be
\small
\mathscr{B} =  \begin{pmatrix} 
  0 & 0 & 0 & 0 & 0 & 0 & 0 & 0 & 0  \\
  0 & 0 & 0 & -\dfrac{2}{\sqrt{3}} & 0 & 0& 0 & 0 & 0  \\
  0 & 0 &0 & 0 & -\dfrac{2}{\sqrt{3}} & 0 &2 & 0 & 0  \\
  0 &  -\dfrac{2}{\sqrt{3}} & 0 & 0 & 0 & 0 & 0 &0 & 0  \\
  0& 0 & -\dfrac{2}{\sqrt{3}} & 0 & 0 & 0 & -\dfrac{2}{\sqrt{3}} & 0 &0  \\
  0 & 0 & 0 & 0 & 0 & 0 & 0 &  -\dfrac{2}{\sqrt{3}} & 0  \\
  0 & 0 & 2 & 0 & -\dfrac{2}{\sqrt{3}} & 0 &0& 0 & 0  \\
  0 & 0 & 0 & 0 & 0 &  -\dfrac{2}{\sqrt{3}} & 0 & 0 & 0  \\
  0 & 0 & 0 & 0 & 0& 0 & 0 & 0 & 0  \\
\end{pmatrix} \ .
\label{B}
\ee

The observable ${\cal I}_3$ defined in 
\eq{eq:I3}, which parametrizes the violations of Bell inequalities for two qutrits systems, then can be written in terms of the  coefficients $h_{ab}$  as 
\be
{\cal I}_3 = 4 \Big(h_{44} + h_{55} \Big)
 - \frac{4 \sqrt{3} }{3} \Big[ h_{61} + h_{66} + h_{72} +  h_{77} +  h_{11} + h_{16} + h_{22} +  h_{27} \Big]\, . \label{ii3}
\ee

Within the choice of measurements leading to the Bell operator (\ref{B}), 
there is still the freedom of modifying the measured observables through local unitary transformations, 
which effectively corresponds to local changes of basis, separately at Alice's and Bob's sites. 
Correspondingly, the Bell operator undergoes the change:
\begin{equation}
\mathscr{B} \to (U\otimes V)^{\dag} \cdot \mathscr{B}\cdot (U\otimes V)\ , \label{uni_rot}
\end{equation}
where $U$ and $V$ are independent $3\times 3$ unitary matrices.
One can use this additional freedom in order to maximize the value of ${\cal I}_3$
for any given qutrit state $\rho$.

Concerning two qutrits\footnote{For qubits, one finds $$\mathscr{C}_2= \frac{1}{2} \max \Big[ -1 +\sum_i (B_i^+)^2 -\sum_j (B_j^-)^2+ \sum_{i,j} C_{ij}^2, \; -1 + \sum_j (B_j^-)^2 - \sum_i (B_i^+)^2 + \sum_{i,j} C_{ij}^2,\; 0\Big]\, .$$} entanglement, it is also useful to collect the explicit form of $\mathscr{C}_2$ in (\ref{C_2}),
giving a lowest bound on concurrence in terms of the coefficients appearing 
in the decomposition (\ref{eq:rho-qutrit}):
\bea
\cmb &=& 2\max \Big[ -\frac{2}{9}-12 \sum_a f_{a}^{2} +6 \sum_a g_{a}^{2} + 4 \sum_{ab} h_{ab}^{2}\, ;\nn \Big.\\
  & & \hskip 2cm \Big. -\frac{2}{9}-12  \sum_a g_{a}^{2} +6 \sum_a f_{a}^{2} + 4 \sum_{ab} h_{ab}^{2},\; 0 \Big]\,.
\label{eq:cmb-qutrit}
\eea
Moreover, the same inverse proportionality  between entanglement and polarizations in the final state, as given in \eq{BC}, holds  for quatrits, the necessary changes having been made.

The Bell test in (\ref{CGLMP}) can be extended to the case in which Alice and Bob
share two $d$-dimensional systems, with $d>3$; also, Bell tests involving more than two parties
have been proposed (see, for example, \cite{Brunner:RevModPhys.86.419,scarani2019bell}). 
A classification of these generalized Bell inequalities is 
quite intricate~\cite{Peres:1998sf,10.1063/1.1928727,PhysRevA.108.062404}.

\subsection{Quantum correlations and relativity}\label{sec:relativity}

As particles at colliders are created at relativistic velocities, one may wonder what is the fate of quantum correlations,
and entanglement in particular, under the action of a Lorentz transformation. One should keep in mind that these transformations
are implemented on the Hilbert space of particle states by means of unitary operators that always act separately on each particle
created in a high-energy collision. As local quantum operations cannot change the amount of quantum correlation of a state,
its entanglement remains unchanged by the action of any Lorentz transformation.

Nevertheless, when the change of reference frame is implemented by a transformation involving different degrees of freedom, for instance 
 momentum and spin, then the entanglement encoded in the purely spin part of the 
multi-party state might change~\cite{Peres-Terno-2004,Alsing-2012,Castro-Ruiz-2015}.
Indeed, it is known that the von Neumann entropy of the reduced spin state is not, in general, relativistically invariant~\cite{Peres-Scudo-Terno}.
Yet, violations of Bell inequalities is assured in any reference frame by a careful choice of the directions along which
particle spin is measured~\cite{Lee-2004,Streiter-2021}. 
In this respect, observables as (\ref{eigenvalue-inequality}) that optimize this choice are indeed
of most valuable practical utility.

In addition, it should be stressed that the violation of Bell inequalities is pervasive in relativistic quantum field theory: 
if we take a bipartite system, each party living in space-like separated space-time regions, there always exists a state for 
which the inequality (\ref{algebraic-inequality}) is maximally 
violated~\cite{Summers-Werner-1,Summers-Werner-2,Summers-Werner-3,Landau-1987,Summers-1990,PhysRevD.108.L081701}.


\section{The toolbox} \label{sec:toolbox}

\subsection{A Cartesian basis for bipartite systems at colliders}\label{sec:basis}

When discussing the production of pairs of entangled particles, a natural coordinate system is that formed by a right-handed orthonormal basis $\{\hn, \hr, \hk\}$, introduced in~\cite{Bernreuther:2001rq} and defined in the particle-pair center of mass (CM) frame as follows.

\begin{figure}[h!]
\begin{center}
\includegraphics[width=4.5in]{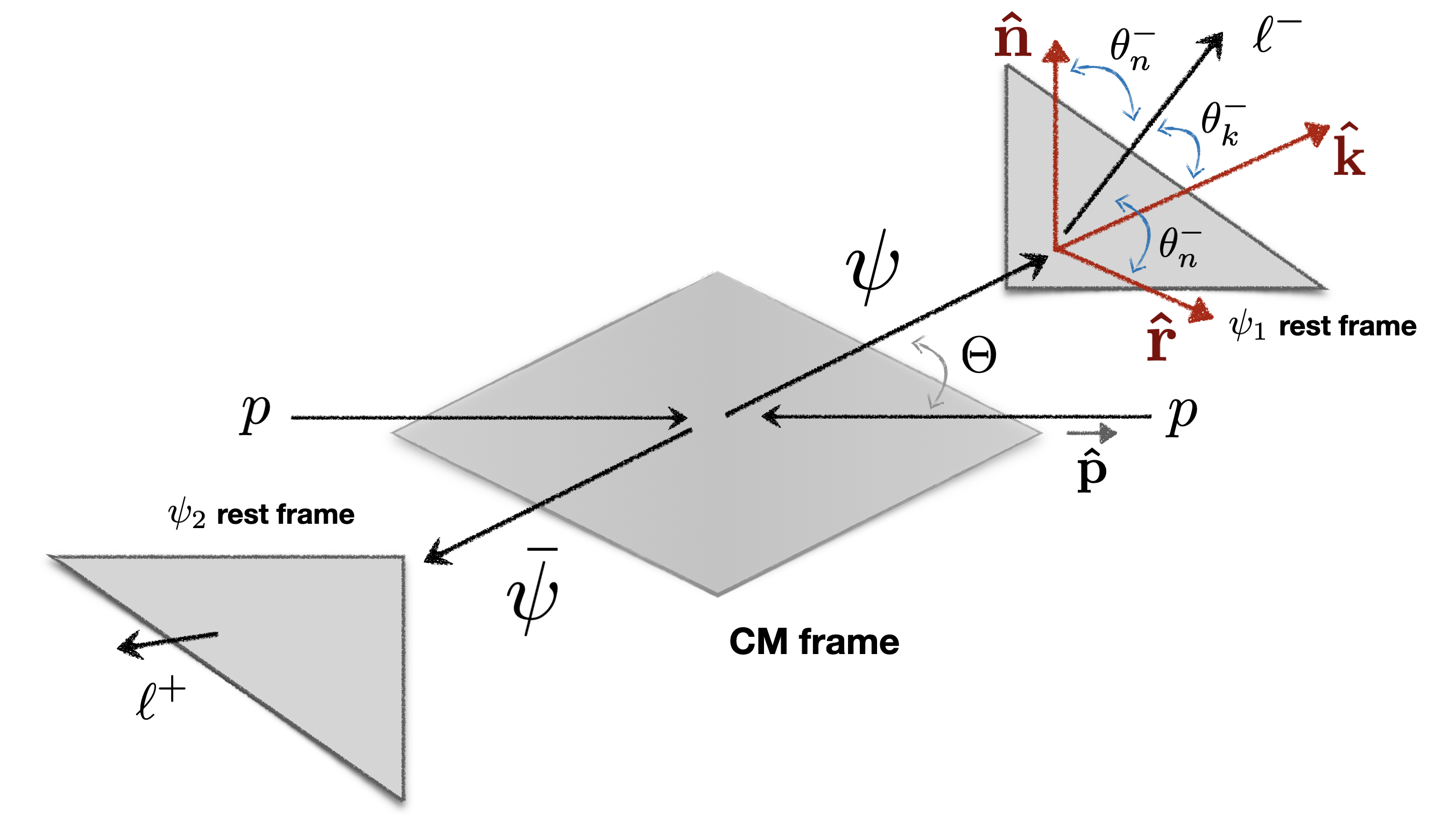}
\caption{\small Unit vectors and momenta in the CM system~\cite{Bernreuther:2001rq}, here specified for the production $p\, p \to \psi \bar \psi$. The angles $\theta^-_i$ define the directions of the final lepton in the rest frame of the fermion $\psi$ with respect to the quantization axis. The same holds for $\bar \psi$.
\label{fig:coordinates} 
}
\end{center}
\end{figure}

Let $\hp$ be the unit vector along the direction of one of the incoming beams in the CM frame and $\hk$ the direction of the momentum of one of the produced particles in the same frame. Then the remaining unit vectors of the basis can be defined as 
\begin{equation}
 \hn =\frac{1}{\st}\left(\hp\times\hk\right)\, ,
 \qquad
 \hr=\frac{1}{\st}\left(\hp-\ct\hk\right)\, , 
 \label{basis}
 \end{equation} 
with  $\Theta$ being the scattering angle satisfying $\hp\cdot\hk = \ct$.
This basis is then used to decompose the spin components of a particle in the corresponding rest frame (reached via a boost along the $\pm \hk$ direction, which leaves the basis vectors unchanged) as illustrated for the case of two particles $V_1$ and $V_2$ in Figure~\ref{fig:coordinates}; it is customary to take the spin quantization axis along $\hk$.

%

\subsection{Polarization density matrices}\label{sec:third:toolbox}

\subsubsection{Qubit polarization matrices: Spin-half fermions}

The density matrix describing the polarization state $\lambda$ of a spin-half fermion  $\psi_\lambda$ can be computed straightforwardly from the amplitude of the underlying production process
	\begin{equation}
		\mathcal{M}(\lambda) = \qty[\bar u_\lambda \mathcal{A}],
	\end{equation}
with polarization $\lambda\in\{-\frac{1}{2},\frac{1}{2}\}$ along a given quantization direction. In the above formula we have indicated with $\mathcal{A}$ the term in the amplitude that multiplies the spinor $\bar u_\lambda$ of the produced fermion and we used square brackets to track the contractions of spinor indices. 

The outgoing particle is then described by a state
	\begin{equation}
		\ket{\psi} = \sum_\lambda \mathcal{M}(\lambda) \ket{u_\lambda}
	\end{equation}
where $\ket{u_\lambda}$ is the Hilbert space representation of the spinor. The spinor-space density matrix is then obtained as
	\begin{equation}
		\tilde{\rho}_\psi = \frac{\dyad{\psi}{\psi}}{\braket{\psi}} 
		= 
		\frac{\sum_{\lambda\lambdap} \qty[\bar u_\lambda\mathcal{A}]\qty[\bar u_{\lambdap}\mathcal{A}]^\dagger\,\dyad{u_\lambda}{\bar u_{\lambdap}}}{{\sum_{\lambda\lambdap}\qty[\bar u_\lambda\mathcal{A}]\qty[\bar u_{\lambdap}\mathcal{A}]^\dagger\,\braket{\bar u_{\lambdap}}{u_{\lambda}}}}.
	\end{equation}
By using the orthogonality relation $\braket{\bar u_{\lambdap}}{u_{\lambda}} \equiv \qty[\bar u_{\lambdap} u_{\lambda}] = 2 m \delta_{\lambdap \lambda}$ the denominator can be rewritten as
	\begin{equation}
		\tilde{\rho}_\psi   
		= 
		\frac{\sum_{\lambda\lambdap}\qty[\bar u_\lambda\mathcal{A}]\qty[\bar u_{\lambdap}\mathcal{A}]^\dagger\,\dyad{u_\lambda}{\bar u_{\lambdap}}}{2m{\sum_{\lambda}\qty[\bar u_\lambda\mathcal{A}]\qty[\bar u_{\lambda}\mathcal{A}]^\dagger}}
		=
		\frac{\sum_{\lambda\lambdap}\qty[\bar u_\lambda\mathcal{A}]\qty[\bar u_{\lambdap}\mathcal{A}]^\dagger\,\dyad{u_\lambda}{\bar u_{\lambdap}}}
		{2m \,\abs{\mathcal{M}}^2}\, ,
	\end{equation}
where $m$ is the mass of the fermion and $\abs{\mathcal{M}}^2$ is the squared amplitude for the production process summed over the spin. 
	
To obtain the polarization density matrix we use the projection operators~\cite{Bouchiat:1958yui,Leader2001}
	\be
		\label{eq:projU}
		\frac{\dyad{u_\lambda}{\bar u_{\lambdap}}}{2m}=\frac{\Pi^u_{\lambda\lambdap}}{2m}= \frac{1}{4m}\qty(\slashed{p}+m)\qty(\delta_{\lambda \lambdap}+\gamma_5\sum_i\slashed{n}_i\sigma^i_{\lambdap \lambda})
  \ee
  and
  \be
		\label{eq:projV}
		\frac{\dyad{v_\lambda}{\bar v_{\lambdap}}}{2m}=\frac{\Pi^v_{\lambda\lambdap}}{2m}= \frac{1}{4m}\qty(\slashed{p}-m)\qty(\delta_{\lambda \lambdap}+\gamma_5\sum_i\slashed{n}_i\sigma^i_{\lambda \lambdap})\, ,
	\ee
where $\sigma_i$ are the Pauli matrices and $\{n_i^\mu\}$ is a triad of space-like four-vectors, each satisfying $n_i^\mu p_\mu =0$, obtained by boosting the canonical basis of the spin four-vector $n$
to the frame\footnote{In the rest frame of the fermion we have $n=(0, \vec n)$ and
	\begin{equation}
	n_1 = \begin{pmatrix}
	0 \\ 1 \\ 0 \\ 0    
	\end{pmatrix}, \qquad 
	n_2 = \begin{pmatrix}
	0 \\ 0 \\ 1 \\ 0    
	\end{pmatrix}, \qquad
	n_3 = \begin{pmatrix}
	0 \\ 0 \\ 0 \\ 1    
	\end{pmatrix}.  
	\end{equation}} 
 where the fermion has four-momentum $p$. By means of the projector operators we then obtain   
	\begin{equation}\label{project-spin-half}
		\rho_{\lambda \lambdap}=\qty[\frac{\Pi^u_{\lambda\lambdap}}{2m} \tilde{\rho}_\psi] = \frac{\qty[\mathcal{A}\mathcal{A}^\dagger \Pi^u_{\lambda\lambdap}]}
		{\abs{\mathcal{M}}^2}
		\equiv\frac{1}{2}\left(\mathbb{1}+\sum_{i=1}^3 s^i\sigma_i\right)_{\lambda \lambdap}\, ,
	\end{equation} 
where $s^i$ are the components of the fermion polarization vector that generally depend on the kinematic variables of the underlying production process. The generalization to processes yielding more than one spin-half fermion in the final state is straightforward and the resulting density matrices can be decomposed on the basis of the tensor products of Pauli and unit matrices. For the case of two fermions, this yields the bipartite density matrix \eq{rho-1/2}, the parameters of which are given in terms of expectation values in \eq{rho-ab-coefficients} and \eq{rho-c-coefficients}.

\subsubsection{Qubit polarization matrices: Photons}  
The production of massless spin-1 particles (photons) is the other instance  of a system whose polarizations are qubits. Let us consider the amplitude for the production of a photon with helicity $\lambda\in\{+1,-1\}$ and momentum $k$
\begin{equation}
    \mathcal{M}(\lambda,k) = \mathcal{A}_\mu \varepsilon^{\mu*}_\lambda(k)
    \label{eq:amp-photon} 
\end{equation}
where $\mathcal{A}_\mu$ denotes the coefficient multiplying the (conjugated) polarization vector $\varepsilon^{\mu}_\lambda$ of the produced photon. In the following we will remove the momentum dependence in  $\mathcal{M}$.

The polarization four-vectors $\varepsilon_{\lambda}^\mu$, $\lambda \in\{1,2\}$ obey the conditions $\varepsilon_{\lambda} \cdot \varepsilon_{\lambda^\prime}=-\delta_{\lambda\, \lambda^\prime}$,  $\varepsilon_{\lambda} \cdot k=0$, where the contractions of Lorentz indices is left implied, and provide a basis for the linear polarizations.  The polarization state $\ket{V^\mu}$ of the photon $V$ is consequently determined as
\begin{equation}
    \ket{V^\nu} = \sum_\lambda \mathcal{M}(\lambda) \ket{\varepsilon^{\nu}_\lambda}\,,
    \label{eq:state-photon}
\end{equation}
where $\ket{\varepsilon^{\nu}_\lambda} $ is a representation of the polarization vector $\varepsilon^{\mu}_\lambda$ in the Hilbert space. The covariant density matrix describing the state is then obtained as
\begin{equation} 
    \tilde\rho^{\mu\nu} = - \frac{\dyad{V^\mu}{V^\nu}}{\braket{V^\mu}{V_\mu}} 
    \label{eq:rcov-photon} 
\end{equation}
after the normalization of the state vector and having inserted a factor of $(-1)$ to account for the signature $(1,\, -1,\, -1,\, -1)$ of the Minkowski metric. The polarization density matrix $\rho_{\lambda\lambda'}$ is then obtained by contracting the density matrix in \eq{eq:rcov-photon} with the projector $\mathcal{P}^{\mu\nu}_{\lambda\lambda'}(k)=\varepsilon^{\mu*}_\lambda(k) \, \varepsilon^{\nu}_{\lambda'}(k) \, $ as
\begin{equation}
    \rho_{\lambda\lambda'}=\mathcal{P}^{\mu\nu}_{\lambda\lambda'}\, \tilde\rho_{\mu\nu}.
    \label{eq:rpol-photon}
\end{equation}
From the orthonormality relation $\varepsilon_{\lambda} \cdot \varepsilon_{\lambda^\prime}=-\delta_{\lambda\, \lambda^\prime}$ and Eqs.~\eqref{eq:amp-photon}-\eqref{eq:rpol-photon} it follows that
\begin{equation}
    \rho_{\lambda\lambda'}=\frac{\mathcal{M}(\lambda)\mathcal{M}^\dagger(\lambda')}{\sum_{\lambda''} \mathcal{M}^\dagger(\lambda'')\mathcal{M}(\lambda'')}
    =\frac{\mathcal{A}_\mu \mathcal{A}^\dagger_\nu \mathcal{P}^{\mu\nu}_{\lambda\lambda'}}{\abs{\mathcal{M}}^2}.
    \label{eq:rpol3}
\end{equation}

The covariant density matrix in \eq{eq:rcov-photon} can  be decomposed in terms of the \textbf{Stokes parameters} $\xi^i$~\cite{Landau1982} as:
\begin{align}
\tilde\rho_{\mu\nu}&=
\frac{1}{2}\sum_{\lambda\lambdap} \varepsilon^{\lambda}_\mu \Big( \mathbb{1}  + \sum_{i=1}^3\xi^i  \sigma_i \Big)_{\lambda \lambda'} \varepsilon^{\lambda'}_\nu \nn \\
 & = \frac{1}{2} 
 \Big( \varepsilon_\mu^{(1)} \varepsilon_\nu^{(1)} + \varepsilon_\mu^{(2)} \varepsilon_\nu^{(2)} \Big) +\frac{\xi_1}{2} 
 \Big( \varepsilon_\mu^{(1)} \varepsilon_\nu^{(2)} + \varepsilon_\mu^{(2)} \varepsilon_\nu^{(1)} \Big)\nn \\
 & \quad  - \frac{i\xi_2}{2} 
 \Big( \varepsilon_\mu^{(1)} \varepsilon_\nu^{(2)} - \varepsilon_\mu^{(2)} \varepsilon_\nu^{(1)} \Big) 
 + \frac{\xi_3}{2} 
 \Big( \varepsilon_\mu^{(1)} \varepsilon_\nu^{(1)} - \varepsilon_\mu^{(2)} \varepsilon_\nu^{(2)} \Big)\, .
 \label{densitymat}
\end{align}  
In matrix form, the density matrix on the helicity basis in \eq{eq:rpol3} is given by
\begin{align}
 \rho_{\lambda\lambdap}=\frac{1}{2}\Big(\mathbb{1}  + \sum_{i=1}^3\xi^i  \sigma_i \Big)_{\lambda\lambdap}
 \label{dm-compact}
\end{align}  
and the Stokes coefficients $\xi^i$ can be obtained by taking the traces, namely
$\xi^i ={\rm Tr}[\rho \sigma_i]$.

 In the case of a two-photon system, the corresponding density matrix will depend on the Stokes parameters $\vec{\xi}^{(a)}$ and $\vec{\xi}^{(b)}$ of the photons $a$ and $b$. The generalization is straightforward and the resulting density matrix can be decomposed on the basis of the tensor products of Pauli and unit matrices as in \eq{rho-1/2}.
 
 \subsubsection{X states} 
 A great deal of simplification  occurs if the
 matrix  $\CC$, written on the basis of (\ref{basis}), 
 \be
\CC
=  \begin{pmatrix} C_{nn}&C_{nr}&C_{nk}\\
C_{rn}&C_{rr}&C_{rk} \\
C_{kn}&C_{kr}&C_{kk} \label{matrixC}
\end{pmatrix} \,,
 \ee
only has a pair of non-vanishing off-diagonal terms,  for instance  $C_{kr}=C_{rk}$.  The eigenvalues of 
$\mathfrak{m}_{12}$ are given in this case by
\be
 C_{nn}^2 , \quad
 \frac{1}{4} \Big[ C_{kk} +  C_{rr} + \sqrt{(C_{kk}^2 - C_{rr})^2+ 4 C_{kr}^2}  \Big]^2 , \quad
  \frac{1}{4} \Big[ C_{kk} +  C_{rr}- \sqrt{(C_{kk}^2 - C_{rr})^2+ 4 C_{kr}^2} \Big]^2 \,. \label{eq:eigenC}
\ee
The  result in \eq{eq:eigenC} is an example of the simplification that occurs for  a class of states, dubbed X states~\cite{Quesada:2012} because their density matrix takes the form
\be
\rho_X= \begin{pmatrix} a&0&0&w\\
0&b&z&0 \\
0&z^*&c&0 \\
w^*&0&0&d \label{rhoX}
\end{pmatrix} \, .
\ee
All  matrices $\CC$ with only one non-vanishing coefficient off diagonal give rise to a density matrix that falls into this class.  

The eigenvalues of the  matrix $R$ in \eq{auxiliary-R} in the case of $\rho_X$ can be readily written  and 
the concurrence  $\mathscr{C}[\rho]$   computed by means of a particularly simple formula;
when $B_i^\pm=0$ and the only off-diagonal non-vanshing element of $C$ is $C_{rk}=C_{kr}$, one has
\be
\mathscr{C}[\rho] = \frac{1}{2} \max \Big[0, \, |\CC_{rr}+\CC_{kk}| -(1+\CC_{nn})\,,\,  \sqrt{(\CC_{rr}-\CC_{kk})^2 + 4\, \CC_{rk}^2} - |1 - \CC_{nn}| \Big] \, .\label{Crho}
\ee

\subsubsection{Qutrit polarisation matrices}

Massive spin-1 particles provide an instance of a system whose polarizations implement qutrits. 
Let us consider the amplitude for the production of a massive gauge boson with helicity $\lambda\in\{+1, 0, -1\}$ and momentum $p$
\begin{equation}
    \mathcal{M}(\lambda, p) = \mathcal{A}_\mu \varepsilon^{\mu*}_\lambda(p)
    \label{eq:amp} 
\end{equation}
where $\mathcal{A}_\mu$ denotes the coefficient multiplying the (conjugated) polarization vector $\varepsilon^{\mu}_\lambda$ of the produced boson.   The polarization state $\ket{V^\mu}$ of the boson $V$ is consequently determined as
\begin{equation}
    \ket{V^\nu} = \sum_\lambda \mathcal{M}(\lambda) \ket{\varepsilon^{\nu}_\lambda}\,,
    \label{eq:state}
\end{equation}
where $\ket{\varepsilon^{\nu}_\lambda} $ is a representation of the polarization vector in the Hilbert space. The covariant density matrix describing the state is then obtained as
\begin{equation}
    \tilde\rho^{\mu\nu} = - \frac{\dyad{V^\mu}{V^\nu}}{\braket{V^\mu}{V_\mu}} 
    \label{eq:rcov}
\end{equation}
after the normalization of the state vector and having inserted a factor of $(-1)$ to account for the signature $(1,\, -1,\, -1,\, -1)$ of the Minkowski metric $g_{\mu\nu}$. The polarization density matrix is then obtained through the projector $\mathcal{P}^{\mu\nu}_{\lambda\lambda'}(p)=\varepsilon^{\mu*}_\lambda(p) \, \varepsilon^{\nu}_{\lambda'}(p) \, $:
\begin{equation}
    \rho_{\lambda\lambda'}=\mathcal{P}^{\mu\nu}_{\lambda\lambda'}\, \tilde\rho_{\mu\nu}.
    \label{eq:rpol}
\end{equation}
From the orthonormality relation $g_{\mu\nu}\,\varepsilon^\mu_\lambda(p) \,\varepsilon^\nu_{\lambda'}(p)=-\delta_{\lambda\lambda'}$ and Eqs.~\eqref{eq:amp}-\eqref{eq:rpol} it follows that
\begin{equation}
    \rho_{\lambda\lambda'}=\frac{\mathcal{M}(\lambda)\mathcal{M}^\dagger(\lambda')}{\sum_{\lambda''} \mathcal{M}^\dagger(\lambda'')\mathcal{M}(\lambda'')}
    =\frac{\mathcal{A}_\mu \mathcal{A}^\dagger_\nu \mathcal{P}^{\mu\nu}_{\lambda\lambda'}}{\abs{\mathcal{M}}^2}.
    \label{eq:rpol2}
\end{equation}  
In order to obtain an expression for the projector $\mathcal{P}$, consider the explicit form of the wave vector of a massive gauge boson with helicity $\lambda$
\begin{equation}
    \varepsilon^{\mu}_\lambda (p)=-\frac{1}{\sqrt{2}}|\lambda|\left(\lambda \, n_1^{\mu}+i \, n_2^{\mu}\right)
+\Big(1-|\lambda| \Big)n_3^{\mu} ,
\label{eps}
\end{equation}
where the four-vectors $n_i=n_i(p)$, $i\in\{1,2,3\}$, form a right-handed triad and are obtained by boosting the linear polarization vectors defined in the frame where the boson is at rest to a frame where it has momentum $p$. With the above expression one finds~\cite{Kim1980,Choi1989,Fabbrichesi:2023cev}
\begin{equation}
\mathcal{P}^{\mu\nu}_{\lambda \lambda '}(p) =
\frac{1}{3}\left(-g^{\mu\nu}+\frac{p^{\mu}p^{\nu}}{m_V^2}\right)
\delta_{\lambda\lambda '}-\frac{i}{2m_V}
\epsilon^{\mu\nu\alpha\beta}p_{\alpha} n_{i\,\beta} \left(S_i\right)_{\lambda\lambda '}-\frac{1}{2}n_i^{\mu}n_j^{\nu} \left(S_{ij}\right)_{\lambda\lambda '},
\label{proj}
\end{equation}
where $m_V$ is the invariant mass of the vector boson $V$, $\epsilon^{\mu\nu\alpha\beta}$ the permutation symbol ($\epsilon^{0123}=1$) and $S_i$, $i\in\{1,2,3\}$, are the $SU(2)$ generators in the spin-1 representation---the eigenvectors of $S_3$, corresponding to the eigenvalues $\lambda\in\{+1, 0, -1\}$, define the helicity basis. The spin matrix combinations appearing in the last term are given by
\begin{equation}
S_{ij}= S_iS_j+S_jS_i-\frac{4}{3} \mathbb{1}\, \delta_{ij},
\label{Sij}
\end{equation}
with $i,j\in\{1,2,3\}$ and $\mathbb{1}$ being the $3\times 3$ unit matrix.

Eqs.~\eqref{eq:rpol2} and \eqref{proj} make it possible to compute the polarization density matrix for an ensemble of $V$ bosons produced in repeated reactions described by the amplitude $\mathcal{M}$. The formalism can be straightforwardly extended to processes yielding a bipartite qutrit state formed by two massive gauge bosons, $V_1$ and $V_2$. In this case we have
\begin{equation}
    \rho=\frac{{\mathcal A}_{\mu\nu}{\mathcal A}^{\dagger}_{\mu'\nu'}}{\abs{\mathcal{M}}^2}
    \left( \mathcal{P}^{\mu\mu'}(k_1)\otimes\mathcal{P}^{\nu\nu'}(k_2) \right),
    \label{eq:rhoVV}
\end{equation}
where $k_1$ and $k_2$ denote the momenta of the vector bosons in a given frame.  
The eight components of $f_a$ and $g_a$, as well as the 64 elements of $h_{ab}$, can be obtained by projecting the density matrix (\ref{eq:rho-qutrit}) on the desired subspace basis using the orthogonality relations, yielding 
\bea
f_a=\frac{1}{6}\,\Tr\left[\rho \left(T^a \otimes \mathbb{1}\right)\right]\, , ~~
g_a=\frac{1}{6}\,\Tr\left[\rho \left(\mathbb{1}\otimes T^a\right)\right]\, , ~~
h_{ab}=\frac{1}{4}\,\Tr\left[\rho \left(T^a \otimes T^b\right)\right]\, .
\label{fgh}
\eea
All the terms computed via Eq.~(\ref{fgh}) are Lorentz scalars.

	\subsection{Reconstructing density matrices from events}\label{sub:first}

The preceding Section described how to calculate the probability of the directions of the emitted decay products based on the spin density matrix of the parent particle. The experimental analysis must instead provide the spin density matrix from the observable angular distributions. This inverse problem is possible provided that 
(i) the decays depend sufficiently on $\rho$ that the process is invertible in principle and (ii) that the daughter particle angular distributions  can be determined in the rest-frame of the parent particles.

The simplest case of the two-body decay of a scalar state is uninteresting in this regard; the spin density matrix is the one-dimensional identity $\mathbb{1}_1$, and the angular distributions are isotropic. 

\subsubsection{Qubits}

 \begin{figure}[h!]
\begin{center} 
 \includegraphics[width=2.5in]{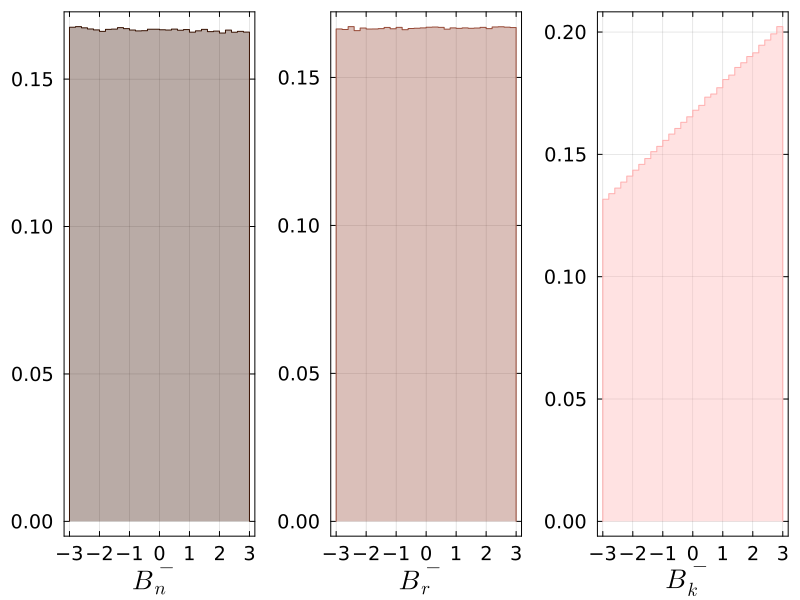}
 \hskip1cm
 \includegraphics[width=2.5in]{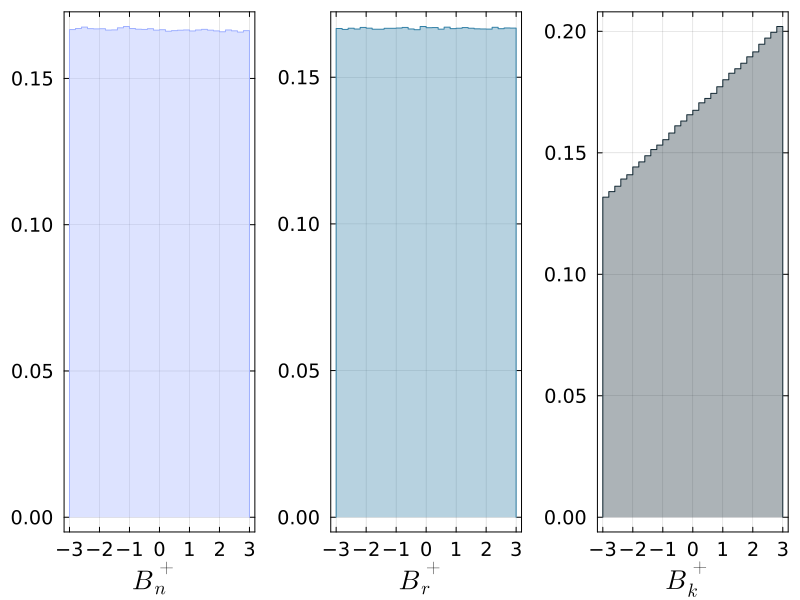}
 \vskip0.5cm
 \includegraphics[width=4.5in]{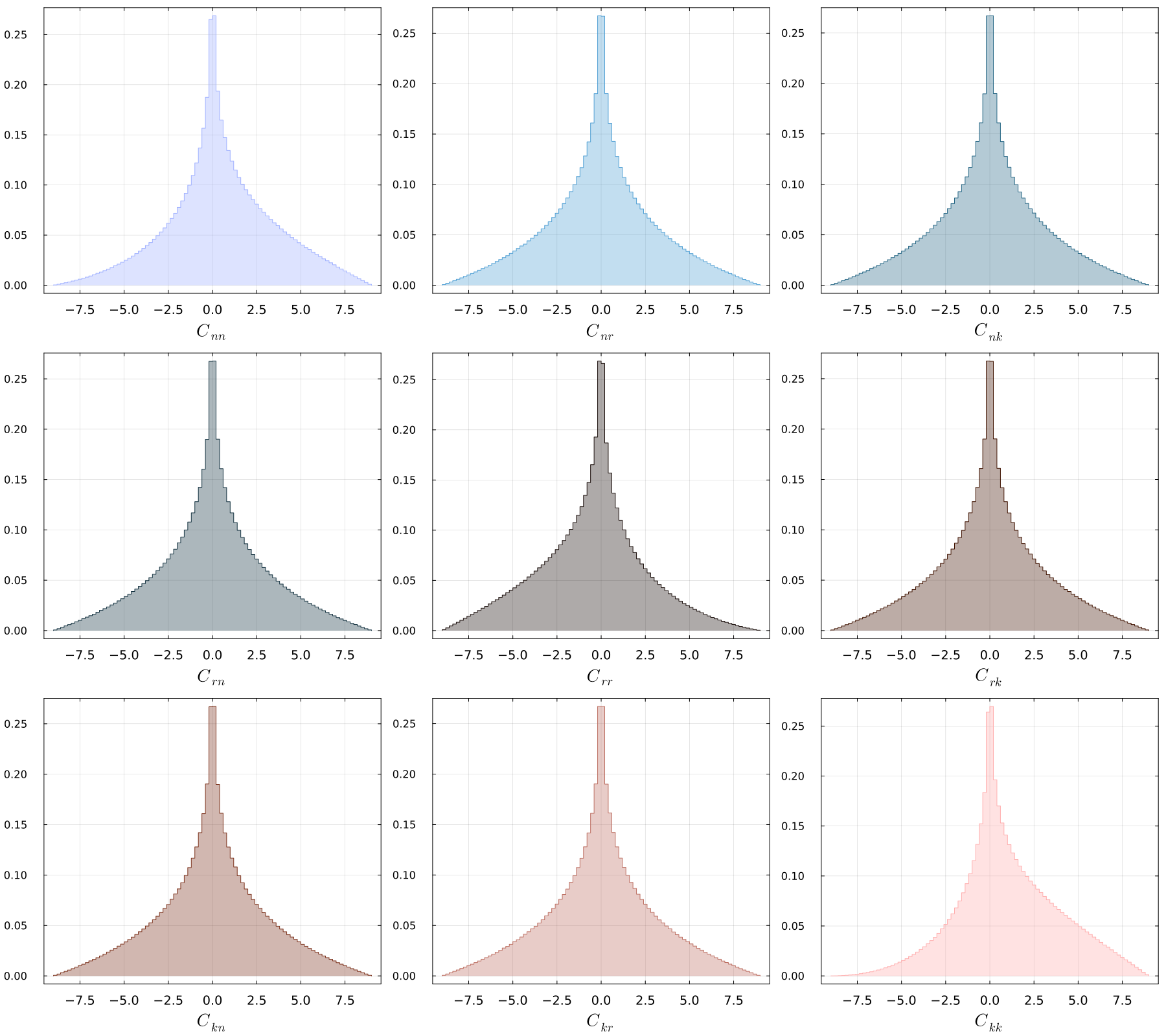}
\caption{\small  Example of distributions of the elements of the matrix  $C_{ij}$ and the  vectors $B_i^\pm$. The ordinate axes represent the respective frequencies. The average (see \eqs{eq:xsec2dA}{eq:xsec2dB}) and standard deviation of these histograms give the mean value and uncertainty of the corresponding coefficient. The non-vanishing values of $B_i^\pm$ or $C_{ij}$ manifest here as asymmetries in the histograms as vanishing values  would be perfectly symmetric. The plots are  from a  simulation of the process $e^+e^-\to \tau^+\tau^-$ at $\sqrt{s}=91.19$ GeV, by two of the authors. 
  \label{fig:distC} 
}
\end{center}
\end{figure}

For the simplest non-trivial case, the decay of a spin-half particle, such as a top quark or antiquark, the density matrix \eq{project-spin-half} can be represented by the polarization vector $\vec{B}\equiv\langle\vec s\rangle$, where the average is taken over the distributions of the kinematic parameters that determine $\vec s$. The role of the projectors in \eq{project-spin-half} is to produce an angular dependence that the probability density function for the decay product lie into the infinitesimal solid angle close to $\vec{n}$:
\begin{equation}\label{pdf-polarisation}
p(\vec{n}; \rho)
               =\frac{1}{4\pi} (1+ \kappa\, {\vec B}\cdot\vec{n}).
\end{equation}
The decay depends only on $\vec B$ and on the so-called `spin-analysing power' $\kappa: -1\leq \kappa\leq 1$ of the daughter particle in the decay. 
Near-maximum values of $|\kappa|\approx 1.0$ are obtained for charged leptons emitted in top-quark decays~\cite{Brandenburg:2002xr}.

The process of measuring $\rho$ from data in this case is equivalent to determining the polarisation $\vec B$ from the angular distribution. This can be achieved by measurement of the angular distributions, except the (not infrequent) special case when $\kappa= 0$ when the decay is isotropic and hence the process non-invertible. 
For $\kappa\ne0$ the polarisation components are given by projecting out the polarisation components of \eq{pdf-polarisation} which can be achieved from the averages of the angular distributions of the polarimetric vector $\vec n$

\begin{equation}
B^\pm_i =  \frac{3}{\kappa_\pm}\frac{1}{\sigma}\int \di\Omega^\pm \frac{\di\sigma}{\di \Omega^\pm} (\vec{n}^\pm\cdot\hat{e}_i),  \label{avXIA}
\end{equation}
where $\{\hat{e}_i\}$, $i=1,2,3$, is an orthonormal basis---usually $\{\hn, \hr, \hk\}$. 
The correlation parameters $\CC_{ij}$ can also be determined by taking the average 
\be
\CC_{ij} = \frac{9}{\kappa_+\kappa_-} \frac{1}{\sigma} \int \di\Omega^+ \di\Omega^-\frac{\di \sigma}{\di\Omega^+ \di\Omega^-} \, (\vec{n}^+\cdot\hat{e}_i)(\vec{n}^-\cdot\hat{e}_j) \label{avXIB}
\ee
weighted again by the differential cross section. For the case of measuring the spin of $t\bar{t}$ or $\tau^-\tau^+$ systems from the final particle angular distributions in their parents' respective rest frames, the spin analysing powers in \eqs{avXIA}{avXIB} are $\kappa_+=+1.0$ and $=\kappa_-=-1.0$ for the positive and negative leptons respectively.

Alternatively, quantum tomography can be performed if the following distributions can be reconstructed
\begin{equation}
  \frac{1}{\sigma} \,\frac{\dd\sigma}{\dd\cos\theta^\pm_i} = \frac{1}{2}\,
  \qty(1\mp\BB^\pm_i\cos\theta^\pm_i) \,,
   \label{eq:xsec2dA}
\end{equation}

\begin{equation}
  \frac{1}{\sigma} \, \frac{\dd\sigma}{\dd\cos\theta^+_{i} \, \dd\cos\theta^-_{j}} = \frac{1}{4} \, \left( 1 + \CC_{ij} \, \cos\theta^+_{i} \, \cos\theta^-_{j} \right) \, ,
  \label{eq:xsec2dB}
\end{equation}
in which $\cos\theta^\pm_{i}$ are the projections of the spin vector (or, equivalently, of the polarimetric vector) on the $\{\hn, \hr, \hk\}$ basis as computed in the rest frame of the qubit of interest. An example of the distributions obtained for the $\BB^\pm_{i}$ and $\CC_{ij}$ coefficients through Monte Carlo simulations of the $e^+ e^- \to \tau^+\tau^-$ process can be found in Figure~\ref{fig:distC} for the case of the $\tau$ leptons. Non-vanishing values of the coefficients are signaled by asymmetric distributions.

\subsubsection{Qutrits}

The spin 1 gauge bosons also act as their own polarimeters. For instance, in the decay $W^{+}\to \ell^{+}  \nu_{\ell}$ the lepton $\ell^{+}$ is produced in the positive helicity state while the neutrino $\nu_{\ell}$ in the negative helicity state. The polarization of the $W^{+}$ is therefore measured to be $+1$ in the direction of the lepton  $\ell^{+}$. The opposite holds for  the decay $W^{-}\to \ell^{-}  \bar \nu_{\ell}$ and the polarization of the $W^{-}$ is therefore  measured to be $-1$ in the direction of the lepton $\ell^{-}$. In both the cases, the momenta of the final leptons (as in Fig.~\ref{fig:coordinates})  provide a measurement of the gauge boson polarizations. The same is true for final jets from $d$ and $s$ quarks.
These momenta are the only information that we need to extract from  the  numerical simulation or the actual data. 

The challenge of reconstructing the correlation coefficients $h_{ab}$, $f_a$ and $g_a$ has of the density matrix of the final leptons has recently been discussed in~\cite{Ashby-Pickering:2022umy}, which we mostly follow in the remainder of this section. 

The cross section we are interested in can be written as~\cite{Rahaman:2021fcz}
\be 
  \frac{1}{\sigma}\frac{\di \sigma}{\di\Omega^{+}\,\di\Omega^{-}}
  = \left( \frac{3}{4 \pi} \right)^2
 \Tr  \Big[ \rho_{V_1V_2} \left(\Pi_+ \otimes \Pi_-\right)\Big]\, , \label{x-sec-leptons}  
\ee
in which the angular volumes $\di \Omega^\pm= \sin \theta^\pm \di \theta^\pm\,\di \phi^\pm$ are written in terms of the spherical coordinates  (with independent polar axes) for the momenta of the final charged leptons in the respective rest frames of the decaying particles. The dependence on the invariant mass $m_{VV}$ and scattering angle $\Theta$ in \eq{x-sec-leptons} is implied. The density matrix $\rho_{V_1V_2}$ in \eq{x-sec-leptons} is that for the production of two gauge bosons given in \eq{eq:rho-qutrit}. 

The density matrices $\Pi_\pm$ describe the polarization of the decaying gauge bosons. The final leptons are taken to be massless---for their masses are negligible with respect to that of the gauge boson. They are projectors  in the case of the $W$-bosons because of their chiral coupling  to leptons. These matrices can be computed by rotating to an arbitrary polar axis the  spin $\pm 1$ states of the weak gauge bosons taken in the $z$ direction and are given,  in the Gell-Mann basis, as 
\be
\Pi_\pm=\frac{1}{3}\,\mathbb{1}
+\frac{1}{2} \sum_{i=a}^8 \mathfrak{q}^a_\pm\, T^a \, , \label{pi_{f}}
\ee
where the Wigner functions $\mathfrak{q}^a_\pm$ can be written in terms of the respective spherical coordinates, as reported in \eq{Q} of  Appendix~\ref{sec:Aqp}, for the decay of $W$-bosons.

We can define another set of functions \label{sec:qp} 
\be
\mathfrak{p}_\pm^n =   \sum_{m} (\mathfrak{m}^{-1}_{\pm})_{m}^{n} \,\mathfrak{q}_\pm^m \label{P_first}
\ee
orthogonal to those in \eq{Q}:
\be
\left(\dfrac{3}{4\, \pi} \right) \int \mathfrak{p}_\pm^n\, \mathfrak{q}_\pm^m \, \di \Omega^\pm = 2\,\delta^{nm} \, .
\ee
In \eq{P_first}, $\mathfrak{m}^{-1}$ is the inverse of the matrix 
\be
(\mathfrak{m}_{\pm})^{nm}= \left(\dfrac{3}{8\, \pi}  \right)\int \mathfrak{q}_\pm^n \, \mathfrak{q}_\pm^m \, \di \Omega^\pm \, ,
\ee
which is assumed to exist.
The explicit form of the functions $\mathfrak{p}_\pm^n$ are given in Appendix~\ref{sec:Aqp}~\eq{P}.

The functions in \eq{P_first}  can be used to extract the correlation coefficients  $h_{ab}$ from the bi-differential cross section in \eq{x-sec-leptons} through the projection 
\bea
h_{ab} &=&  \frac{1}{4\, \sigma} \int \int \frac{\di \sigma}{\di\Omega^{+}\,\di\Omega^{-}} \, \mathfrak{p}_+^a \, \mathfrak{p}_-^b \,\di \Omega^+ \di \Omega^-\, .\label{hh}
\eea
The correlation coefficients  $f_a$ and $g_a$ can be obtained in similar fashion by projecting the single differential cross sections: 
\bea
f_{a} &=&  \frac{1}{2\, \sigma}  \int  \frac{\di \sigma}{\di\Omega^{+}} \, \mathfrak{p}_+^a \, \di \Omega^+ \, ,\nn\\
g_{a} &=&  \frac{1}{2\, \sigma}  \int  \frac{\di \sigma}{\di\Omega^{-}} \, \mathfrak{p}_-^a \, \di \Omega^- \, . \label{ffgg}
\eea

The density matrices $\Pi_\pm$ are not projectors in the case of the $Z$-bosons because  the coupling between $Z$-bosons and  leptons in the Lagrangian,
\be
-i \frac{g}{\cos \theta_{W}} \Big[ g_{L} (1-\gamma^{5}) \gamma_{\mu} + g_{R} (1+\gamma^{5}) \gamma_{\mu}\Big]\, Z^{\mu}\ ,
\ee
contains both right- and  left-handed components, whose strengths are controlled by the coefficients $g_{L} =-1/2 + \sin^2 \theta_{W}$  and $g_{R}= \sin^2 \theta_{W}$. In this case, one must introduce a generalized form of the functions in \eq{Q} which is defined as the following linear combinations
\be
\tilde{\mathfrak{q}}^{n} = \dfrac{1}{g_{R}^2 + g_{L}^{2}} \Big[ g_{R}^2 \, \mathfrak{q}^{n}_{+}+  g_{L}^2 \, \mathfrak{q}_{-}^{n}\Big]\, ,
\ee
and define from these the corresponding orthogonal functions $\tilde{\mathfrak{p}}^{n}$ to be used in \eq{fgh}. They are the same for both the $\pm$ coordinate sets and given by
\be
\tilde{\mathfrak{p}}^{n} = \sum_{m}\mathfrak{a}^{n}_{m} \mathfrak{p}_+^m \, , \label{p_f}
\ee
where the matrix $\mathfrak{a}^{n}_{m}$ is given in \eq{Anm} in Appendix~\ref{sec:Aqp}. The \eqs{hh}{ffgg} can be used after replacing the functions $\mathfrak{p}_\pm^m$ with $\tilde{\mathfrak{p}}^{n} $. 

 \eqs{hh}{ffgg} provide the means to reconstruct the correlation functions of the density matrix from the distribution of the lepton momenta and thus allow to infer the expectation values of the observables ${\cal I}_3$ and  $\cmb$ from the data. 
In a numerical simulation, or working with actual events, one extracts from  each single event the coefficient of the combinations of trigonometric functions indicated in \eq{P} in \ref{sec:Aqp}; that coefficient is the corresponding entry of the correlation matrix in \eqs{hh}{ffgg}. Running this procedure over all events gives an average value and its standard deviation.

\begin{figure}[h!]
\begin{center} 
 \includegraphics[width=4in]{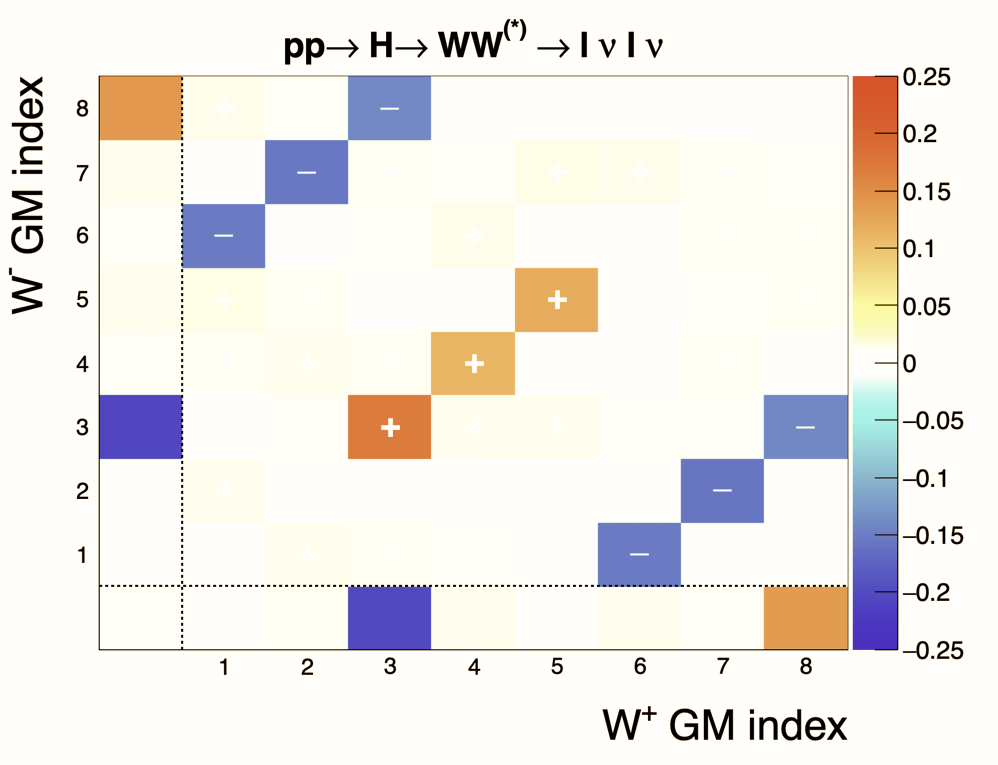}
\caption{\label{Hww_parameters}
Reconstructed Gell-Mann parameters obtained from quantum state tomography of pairs of simulated $W^\pm$ bosons obtained from $H\rightarrow WW^{(*)} \rightarrow \ell^+\nu \ell^- \bar\nu$, (the bottom row of each plot contains the $B^+$ parameters for a $W^+$ boson, the leftmost column the $B^-$ parameters for the $W^-$ boson and the rows and columns 1-8 the $C_{ij}$ parameters.
Bins are marked with ``$+$'' or ``$-$'' to indicate the sign of the reconstructed coefficient. 
The (0,0) element has no meaning. Adapted from~\cite{Ashby-Pickering:2022umy} (\href{https://creativecommons.org/licenses/by/4.0/}{CC BY 4.0}).}

\end{center}
\end{figure}

An example showing the corresponding parameters, after this averaging for the process $H\rightarrow WW^{(*)} \rightarrow \ell^+\nu \ell^- \bar\nu$, assuming that the parental rest frames can be determined is shown in Figure~\ref{Hww_parameters}.

\subsubsection{Tensor representation for qutrits} 

The Gell-Mann representation of the density matrix \eq{eq:rho-qutrit} is only one possible parameterization. An alternative representation of the density matrix is in terms of tensor operator components, which for a single system can be written~\cite{BOURRELY198095,Leader2001,Rahaman:2021fcz,Aguilar-Saavedra:2015yza,Bernal:2023jba}
\begin{equation}\label{eq:rho-tensor}
\rho=\frac{1}{2s+1}\sum_{L,M}(2L+1)(t_M^L)^*T_M^L\, ,
\end{equation}
where $T^L_M$ are the matrices that represent the irreducible spherical tensor operators. 
We note that for the case of a qubit representation of the density matrix the Tensor representation and the Gell-Mann representation are identical, since both are provided by the standard Bloch vector, that is a parameterization based on the Pauli matrices.

For the general tensor representation, the orthogonality relationship
\begin{equation}
\Tr{\left(T_{M'}^{L'} T_M^{L\dag}\right)} = \frac{2s+1}{2L+1}\delta_{LL'}\delta_{MM'}
\end{equation}
allows determination of the coefficients 
\begin{equation}
t^L_M = \Tr{\left(\rho T_M^L\right)}
\end{equation}
from the observables. The procedure for extracting the coefficients from angular distributions in this framework is described in \cite{Bernal:2023jba}, which also includes discussion of the Wigner $\mathfrak{q}$ and $\mathfrak{p}$ functions for the irreducible tensors.
The density matrices for bipartite systems can similarly be parameterized in terms of tensor products of tensor operators for the respective particles
\be
\rho = \frac{1}{9}\Bigg\{ \mathbb{1}\otimes
    \mathbb{1}+
    A^1_{LM} \left[T^L_M\otimes \mathbb{1}\right]+ A^2_{LM}  \left[\mathbb{1}\otimes T^L_M \right] 
    +C_{L_1 M_1 L_2 M_2}  \left[T^{L_1}_{M_1}\otimes T^{L_2}_{M_2}\right]\Bigg\}\, .
\label{eq:rho-tensor2}
\ee

The resultant angular distributions for $W^\pm$ boson decays, in terms of related parameters are given in \cite{Aguilar-Saavedra:2015yza}. The equivalent distributions for the $Z$ boson are provided in \cite{Aguilar-Saavedra:2022wam}.


The analyses outlined in this Section can be experimentally challenging because both the CM frame of the collision  and the rest frame of the parents must be determined in order to compute the various correlation coefficients with reasonable uncertainties. We discuss more details of the experimental aspects of these analyses in Section~\ref{sec:qubits} for qubits and Section~\ref{sec:qutrits} for qutrits.
 
	\newpage
\section{Qubits: $\Lambda$ baryons, top quarks, $\tau$ leptons and photons}\label{sec:qubits}

        Systems of two qubits, such as those arising from the  polarizations of pairs of fermions (or photons), are routinely produced particle colliders such as the LHC, SuperKEKB and BEPC II.
        We consider the production of $\Lambda$ baryons at BEPC II and collected by the BESIII experiment, top-quark pair $t\bar{t}$,  $\tau$-lepton pair $\tau\bar{\tau}$ via the Drell-Yan mechanism and in the resonant Higgs boson decay $h\to \tau \bar\tau$  at the LHC, and in the $e^+e^-\to \tau \bar{\tau}$ at SuperKEKB. We also  include the di-photon system via the resonant Higgs decay  process $h\to \gamma \gamma$, assuming (and it is a significant assumption) that polarizations of the high-energy photons could be determined.  For each of the considered processes, we  provide the analytical predictions for the corresponding Bell inequality violation and quantum entanglement observables. 
        Side by side with the analytical computation, it is  crucial to have access to Monte Carlo  simulations of the same processes in order to have an estimate of the uncertainty and therefore of the significance that can be reached for the values of the observables. The predictions, obtained by the reconstruction of the polarization density matrix by means of  simulations of events, are  provided,  in dedicated sub-sections, for each of the considered processes. 
 \subsection{Entangled $\Lambda$ baryons}\label{sec:qubits:lambda}

The decays of charmonium $\eta_c,\, \chi_c$ and $J/\psi$  produce  pairs of entangled $\Lambda$ baryons. These processes were
suggested in  \cite{Tornqvist:1980af,Tornqvist:1986pe} and studied in \cite{Baranov:2008zzb,Baranov:2009zza,Chen:2013epa} as a promising setting for testing a Bell inequality. 

The  helicity states of the final system in
\be
\eta_c \to \Lambda +  \bar \Lambda
\ee
fall in the singlet representation of the product $\tfrac{1}{2} \otimes \tfrac{1}{2} = 0 \oplus 1$. The same holds for the decays of $\chi_c$. It is constrained by the conservation of the angular momentum to be described by the state
\be
|\psi_0\rangle  \propto w_{ \frac{1}{2}\,- \frac{1}{2}} \, |\tfrac{1}{2}, \tfrac{1}{2} \rangle \otimes |\tfrac{1}{2}, -\tfrac{1}{2} \rangle -
w_{-\frac{1}{2}\,\frac{1}{2} }\, |\tfrac{1}{2}, -\tfrac{1}{2} \rangle \otimes |\tfrac{1}{2}, \tfrac{1}{2} \rangle  \, ,\label{state0}
\ee
in which $w_{ij}$ are the normalized helicity amplitude are given by the Clebsh-Gordon coefficients: $w_{\frac{1}{2}\, -\frac{1}{2}}=w_{-\frac{1}{2}\, \frac{1}{2}} = \frac{1}{\sqrt{2}}$.

 The case of the decays of the $J/\psi$, which is a spin-one particle, is different. The spin state of the pairs of $\Lambda$ depends on the polarization of  the $J/\psi$  and is therefore  in general in a mixed state with less entanglement. Accordingly, this process is less favorable to the observation of large entanglement and a significant violation of Bell inequality, as already noted in~\cite{Baranov:2008zzb,Baranov:2009zza,Chen:2013epa}.

 Data on these processes have been collected by the BESIII Collaboration~\cite{BESIII:2018cnd,BESIII:2021ypr,BESIII:2022qax}, with sufficient numbers of events for an experimental observation of entanglement and Bell inequality violation to be possible.

\subsubsection{Entanglement and Bell inequality violation in $\eta_c \to \Lambda +  \bar \Lambda$}

The  state in \eq{state0}  gives rise to the density matrix
\be
\rho_{\Lambda\,\Lambda}  =  |\psi_0\rangle \langle \psi_0| = \frac{1}{2}\,  \begin{pmatrix} 
0&0&0&0\\
0& 1 & 1 &0\\
0&1&1 &0\\
0&0&0&0 
\end{pmatrix}\,,
\ee
which can only depends on the overall strength of the coupling.  The conservation of the angular momentum forces the final state into the singlet and the helicity amplitudes are completely fixed except for an overall function that is factorized out in the normalization of the density matrix.   
Using the Pauli matrices $\sigma_i$, we can write  the correlation matrix
\be 
C_{ij} =\Tr \rho_{\Lambda \Lambda} \,\sigma_{i} \otimes \sigma_{j} =\begin{pmatrix} 1 &0 &0\\ 0&1&0\\ 0&0&-1 \end{pmatrix}
\ee
from which it is possible to  compute the concurrence $\mathscr{C} =1$ and determine the Horodecki condition $\mathfrak{m}_{12} =2$. These maximum values show that one can expect  maximum entanglement and maximal violation of Bell inequality in this process. This is also the original result of the computation in~\cite{Baranov:2008zzb,Baranov:2009zza}.

        \subsection{Top-quark pair production at the LHC}\label{sec:qubits:top}

\begin{figure}[h!]
\begin{center}
\includegraphics[width=5.5in]{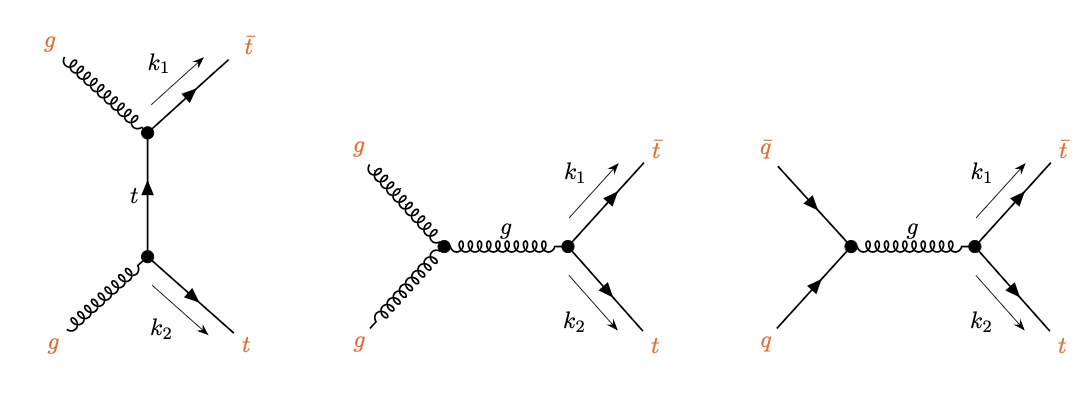}
\caption{\small \label{fig:ttbar}  Feynman diagrams (at partonic tree-level)  for top-antitop ($t\bar t$) production, for  gluon ($gg$) and quark-pair ($q\bar{q}$) initial states.}
\end{center}
\end{figure}

 At the parton level, the production of top-quark pair $t\bar{t}$ at the LHC receives two distinct contributions, namely from quark anti-quark annihilation ($q\bar q \to t \bar{t}$) and gluon-gluon fusion ($gg \to t \bar{t}$) respectively. Corresponding Feynman diagrams in the SM are shown in Fig.~\ref{fig:ttbar}. 
The analysis of the kinematics and polarizations is described for $q\bar q \to f \bar{f}$, where $f$ stands for a generic fermion in Appendix \ref{appendix:kinematics}. The same considerations on the kinematics and polarizations of the final states hold for the top-quark production via gluon-gluon fusion.

The unpolarized differential cross section for the process 
\be
p + p \to t + \bar{t}\, .
\label{eq:pptt}
\ee
 is given in the basis \eq{basis} by 
\cite{Bernreuther:2001rq,Afik:2020onf,Fabbrichesi:2021npl}
\be
\frac{\di \sigma}{\di \Omega\, \di m_{t\bar t}} = \frac{\alpha_s^2 \beta_t}{64 \pi^2  m_{t\bar t}^2} \Big\{ L^{gg} (\tau) \, \tilde A^{gg}[m_{t\bar t},\, \Theta]+L^{qq} (\tau)\, \tilde A^{qq}[m_{t\bar t},\, \Theta]  \Big\}\, ,
\label{eq:x-sec-tt}
\ee
where the combination of the two channels at partonic tree-level (see Fig.~\ref{fig:ttbar}) $g+g\to t +\bar t$ and $q + \bar q \to t +\bar t$ in \eq{eq:x-sec-tt} is weighted by the respective  parton luminosity functions $L^{gg,qq}(\tau)$
\be
L^{gg} (\tau)= \frac{2 \tau}{\sqrt{s}} \int_\tau^{1/\tau} \frac{\di z}{z} q_{g} (\tau z) q_{g} \left( \frac{\tau}{z}\right)\quad \text{and}\quad
L^{qq} (\tau)= \sum_{q=u,d,s}\frac{4 \tau}{\sqrt{s}} \int_\tau^{1/\tau} \frac{\di z}{z} q_{q} (\tau z) q_{\bar q} \left( \frac{\tau}{z}\right)\, ,
\label{eq:lumPDF}
\ee
where the functions $q_j(x)$ are the PDFs, $\alpha_s=g^2/4 \pi$ and $\tau = m_{t\bar t}/\sqrt{s}$, with $m_{t\bar t}$ the invariant mass of the $t\bar{t}$ system. The explicit expressions for $\tilde A^{gg}$ and  $\tilde A^{qq}$  are given in Appendix \ref{appendix:top}.     
Their numerical values can be taken from, for instance,  those provided by a recent sets ({\sc PDF4LHC21}~\cite{PDF4LHCWorkingGroup:2022cjn}) for $\sqrt{s}=13$ TeV and factorization scale $q_0=m_{t\bar t}$.

The correlation coefficients $\CC_{ij}$ in the
polarization density matrix for the $t\bar{t}$ pair production is  given as \cite{Bernreuther:2001rq,Afik:2020onf,Fabbrichesi:2021npl}
\be
\CC_{ij} [m_{t\bar t},\, \Theta]= \frac{L^{gg} (\tau)\, \tilde C_{ij}^{gg}[m_{t\bar t},\, \Theta]+L^{qq} (\tau)\, \tilde C_{ij}^{qq}[m_{t\bar t},\, \Theta]} {L^{gg}(\tau) \, \tilde A^{gg}[m_{t\bar t},\, \Theta]+L^{qq} (\tau)\, \tilde A^{qq}[m_{t\bar t},\, \Theta]} \, .\label{eq:Cij-top}\,
\ee
Notice that in the SM the polarization coefficients  for the quark-pair 
$B_{i}^{qq}=0$ identically vanish---barring higher order electroweak corrections.

The explicit expression for the coefficient $\tilde C_{ij}^{gg}$ and $\tilde C_{ij}^{qq}$ in \eq{eq:Cij-top} for the SM are collected in Appendix~\ref{appendix:top}. They are related to the corresponding correlation coefficients $C^{qq,gg}_{ij}$ for partonic processes by $ \tilde{C}_{ij}^{gg}= C_{ij}^{gg}  A^{gg}$ and $ \tilde{C}_{ij}^{qq}= C_{ij}^{qq}  A^{qq}$. 

\subsubsection{Entanglement in  $t\bar{t}$ production}
\label{sec:qubits:top:entanglement}

  Top-quark pair production is the first  process that has been  considered in the current run of analyses. In \cite{Afik:2020onf} the expected entries of the density matrix were evaluated in the frame proposed in \cite{Bernreuther:2010ny} (in which they were computed for estimating classical correlations) and the concurrence computed. 
  
 The dependence of the entries of the polarization density matrix in \eq{eq:Cij-top} on the kinematic variables $\Theta$, the scattering angle,  and $\beta_t=\sqrt{1-4m_t^2/m_{t\bar t}^2}$, is in general rather involved but
it simplifies  at $\Theta=\pi/2$ for which the top-quark pair is  transversally produced and the entanglement is  maximal. To understand the behaviour in this limit, one can choose the three vectors
$\{\hn,\, \hr, \,\hk\}$ to point in the $\{\hat{x},\hat{y},\hat{z}\}$ directions and  denote by $|0\rangle$ and $|1\rangle$ the eigenvectors of the Pauli matrix $\sigma_z$
with eigenvalues $-1$ and $+1$, respectively; similarly, let $|\small{-}\rangle$ and $|\small{+}\rangle$ be the analogous
eigenvectors of $\sigma_x$ and $|\text{\small L}\rangle$ and $|\text{\small R}\rangle$ those of $\sigma_y$.

A set of quark pair spin density matrices that are relevant to this case are the projectors on pure, maximally entangled Bell states,
\be
\rho^{(\pm)}= |\psi^{(\pm)}\rangle\langle\psi^{(\pm)}|\ ,\qquad 
|\psi^{(\pm)}\rangle=\frac{1}{\sqrt{2}}\big( |01\rangle \pm |10\rangle \big)\ ,
\label{Bell-states}
\ee
together with the mixed, unentangled states,
\begin{eqnarray}
&&\rho^{(1)}_{\rm mix}=\frac{1}{2}\Big( |{\small ++}\rangle\langle {\small ++}| + |\small{--}\rangle\langle {\small --}| \Big)\ ,\\
\label{rho-mix-2}
&&\rho^{(2)}_{\rm mix}=\frac{1}{2}\Big( |\text{\small LR}\rangle\langle \text{\small LR}| + |\text{\small RL}\rangle\langle \text{\small RL}| \Big)\ ,\\
&&\rho^{(3)}_{\rm mix}=\frac{1}{2}\Big( | 01 \rangle\langle 01| + |10\rangle\langle 10| \Big)\ .\\
\nonumber
\end{eqnarray}

Let us  treat separately the quark-antiquark $q \bar{q}$ and 
gluon-gluon $gg$ production channels.
For the $q \bar{q}$ production channel, using the explicit expression   collected in Appendix \ref{appendix:top} for the correlation coefficients $\CC_{ij}$, one obtains that 
the $t \bar{t}$ spin density matrix can be expressed as the following convex combination 
 \cite{Fabbrichesi:2022ovb}~:
\be
\rho_{t\bar t}^{(q\bar{q})}= \lambda \rho^{(+)} + (1-\lambda)\rho^{(1)}_{\rm mix}\ ,\quad \text{with} \quad \lambda=\frac{\beta_t^2}{2-\beta_t^2}\in [0,1]\ , 
\label{rho-qq}
\ee
so that at high transverse momentum, for $\beta_t \rightarrow 1$, the spins of the $t\bar{t}$ pair 
tend to be generated in a maximally entangled state; this quantum correlation is however progressively
diluted for $\beta_t < 1$, vanishing at threshold, $\beta_t=0$, as the two spin state becomes a
totally mixed, separable state.

The situation is different for the $gg$ production channel, as both at threshold and at high momentum
the $t \bar{t}$ spins result maximally entangled, with $\rho_{t\bar t}^{(gg)}=\rho^{(+)}$ for $\beta_t \rightarrow 1$
and $\rho_{t\bar t}^{(gg)}=\rho^{(-)}$ when $\beta_t=\, 0$. For intermediate values of $\beta_t$, the situation becomes more involved,
and the two-spin density matrix can be expressed as the following convex combination:
\be
\rho_{t\bar t}^{(gg)}= a\rho^{(+)} + b \rho^{(-)} + c \rho^{(1)}_{\rm mix} + d \rho^{(2)}_{\rm mix}\ ,
\ee
with non-negative coefficients \cite{Fabbrichesi:2022ovb}
\be
a=\frac{\beta_t^4}{1+2\beta_t^2-2\beta_t^4}\ ,\quad
b=\frac{(1-\beta_t^2)^2}{1+2\beta_t^2-2\beta_t^4}\ ,\quad
c=d=\frac{2\beta_t^2\big(1-\beta_t^2\big)}{1+2\beta_t^2-2\beta_t^4}\ ,\quad
\ee
so that $a+b+c+d=1$, while entanglement is less than maximal.

Including both the $q \bar{q}$- and $gg$-contributions leads to more mixing and therefore in general to additional loss of quantum correlations. 


 \begin{figure}[h!]
\begin{center} 
 \includegraphics[width=3.45in]{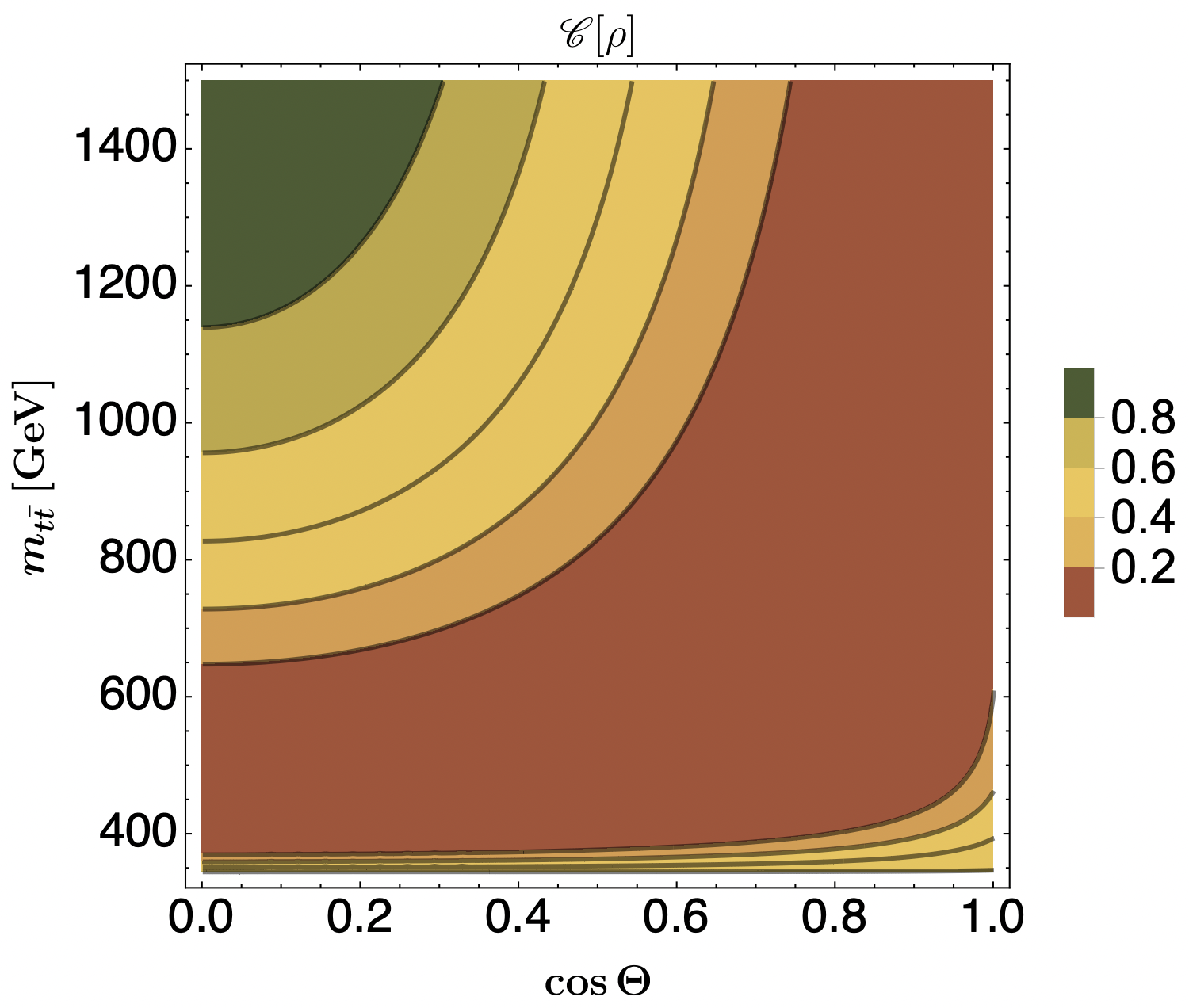}
  \includegraphics[width=3.6in]{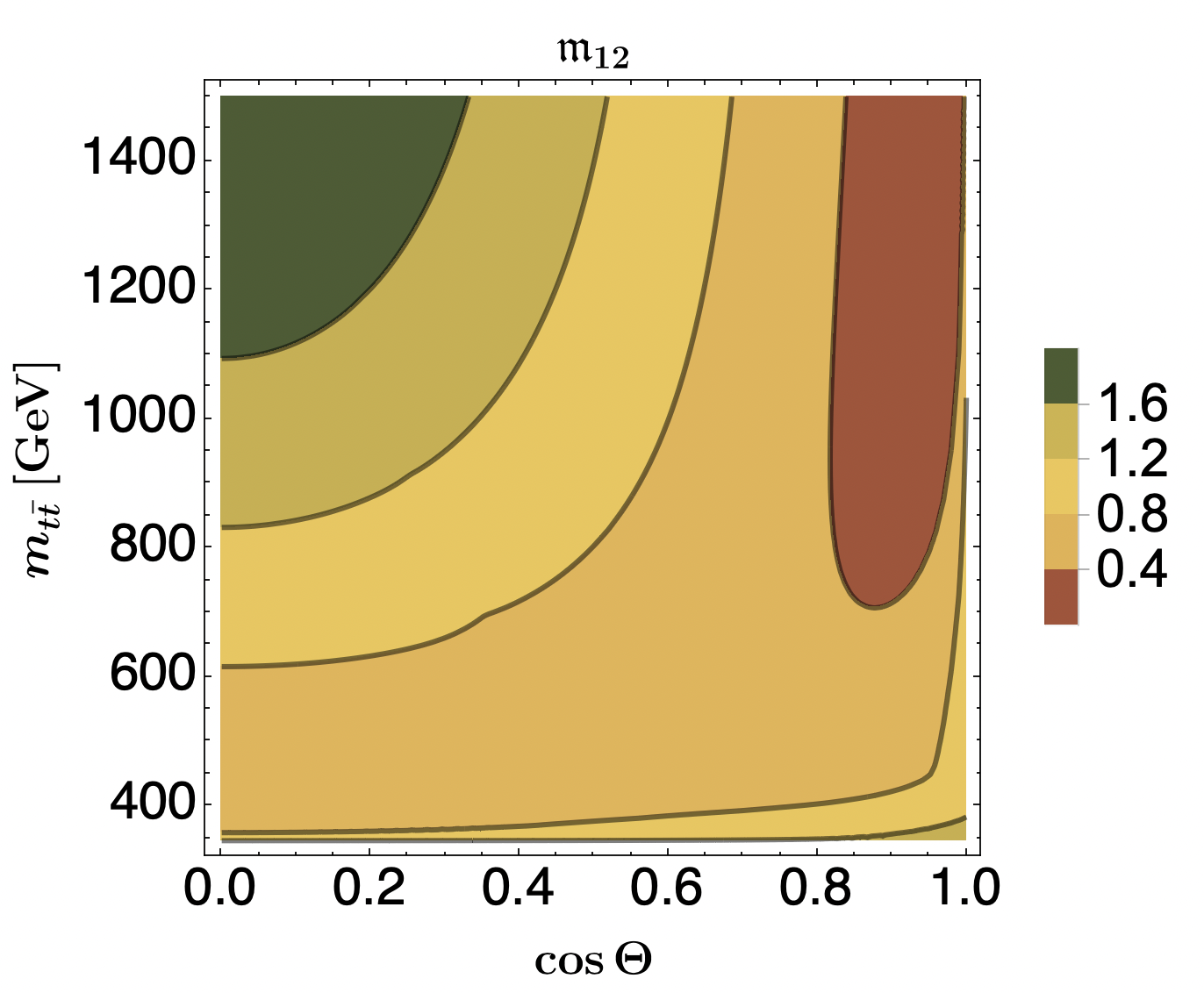}
\caption{\small  The observables $\mathscr{C}[\rho]$ (contour plot on the left) and $\mathfrak{m}_{12}$ (contour plot on the right) for $t\bar t$ production as  functions of the kinematic variables $\Theta$ and $m_{t \bar t}$ across the entire available space (they  are symmetric for $\cos \Theta<0$). Figures revised from \cite{Fabbrichesi:2022ovb} (\href{https://creativecommons.org/licenses/by/4.0/}{CC BY 4.0}). 
  \label{fig:topm1m2} 
}
\end{center}
\end{figure}

All these features are manifest in the plot on the left-side in Fig.~\ref{fig:topm1m2}. There are two regions where entanglement is significant: in a narrow region near threshold; and for boosted tops for scattering angles close to $\pi/2$.

The strong dependence of the entanglement on the kinematic variables was first shown in \cite{Afik:2020onf}. 
That paper calculated the quantity
\be
D= \frac{1}{3} \, \Tr \CC_{ij}
\ee
and showed that close to threshold it is expected to be smaller than $-1/3$. This is a sufficient condition for  entanglement,
as $D$ is directly connected to concurrence by the relation $\mathscr{C}[\rho]=\max [-1-3D,\,0]/2$~\cite{Afik:2020onf}.

The ATLAS Collaboration, applying the method proposed in \cite{Afik:2020onf},  has  recently~\cite{ATLAS:2023fsd} analyzed the $pp$ data and extracted the value of $D$ from the differential cross section
\be
\frac{1}{\sigma} \frac{\di \sigma}{\di \cos \phi} = \frac{1}{2} \Big( 1 - D \cos \phi \Big)\, ,
\ee
where $\phi$ is the angle between the respective leptons as computed in the rest frame of the decaying top and anti-top. 

The analysis selected fully leptonic top pair events with one electron and one muon of opposite signs, and measured $D$ at the particle level in the near-threshold region $340\,{\rm GeV} < m_{t\bar{t}}< 380$\,GeV. After calibrating for  detector acceptance and efficiency they measured~\cite{ATLAS:2023fsd}
\be
D = -0.547 \pm 0.002\; [{\rm stat.}] \pm 0.021\; [{\rm syst.}]\, .
\ee
This value is smaller than $-1/3$ with a significance of more than $5\sigma$, thus provides the first experimental observation of the presence of entanglement between the spins of the top quarks.

The observed entanglement is larger than that predicted by the simulations, suggesting that the simulations might require improved modelling of near-threshold effects in $t\bar{t}$ production.

A preliminary analysis\footnote{This review aimed to survey papers released prior to the beginning of 2024. An exception was made for this recent experimental result.} by CMS~\cite{CMS:2024hgo} in an overlapping near-threshold region, $345\,{\rm GeV} < m_{t\bar{t}} < 400\,{\rm GeV}$, observed $D$ to be
\be
D = -0.478 \pm 0.017\; [{\rm stat.}] ~^{+0.018}_{-0.021}\; [{\rm syst.}]\, 
\ee
in that region, with a statistical significance to be smaller than $-1/3$ of $5.1\sigma$. The Monte Carlo simulations for the CMS analysis included a calculation of the colour-singlet contribution of toponium bound states, the inclusion of which tends to increase the predicted level of entanglement, and to improve agreement between simulation and data.

\subsubsection{Bell inequalities}
\label{sec:qubits:top:bell}

 The violation of the Bell inequality, coming from the  entanglement of the top-quark pair, can be measured~\cite{Fabbrichesi:2021npl} by means of the Horodecki condition~(\ref{eigenvalue-inequality})  
 \be
\mathfrak{m}_{12}\equiv m_1 + m_2 >1
\label{eq:m12}
\ee
as defined in  Section~\ref{sec:Bell}. 
The values of the observable $\mathfrak{m}_{12}$ across the entire kinematic space available  are shown on the right-hand side of Fig.~\ref{fig:topm1m2}.

 Fig.~\ref{fig:topm1m2} shows how the quantum entanglement as well as Bell inequality violation, encoded in the observable $\mathfrak{m}_{12}[C]$,  increases as we consider larger scattering angles and $m_{t\bar t}$ masses. As expected from the qualitative discussion in the previous Section, the kinematic window where the observable $\mathfrak{m}_{12}$ is larger is for $m_{t\bar t} > 900$ GeV and $\cos  \Theta/\pi < 0.2$. The mean value of $\mathfrak{m}_{12}$ in this bin is found to be 1.44 \cite{Fabbrichesi:2022ovb}.

\subsubsection{Monte Carlo simulations and predictions}
\label{sec:qubits:top:simulations}

A number of MC simulations have been performed of quantum observables in top-quark pair production. They consider fully- as well as semi-leptonic decays, and all agree with the analytic results. In addition, they provide an estimate of the uncertainty in both the amount of entanglement and of violation of Bell inequality. All works predict entanglement to be measurable at the LHC while they differ about the possibility of having a significant violation of Bell inequality. This process is now under scrutiny by the experimental Collaborations.

In \cite{Fabbrichesi:2021npl}, the process
\be
p+p \to t+\bar t \to \ell^\pm \ell^\mp + \text{jets}+ E^{\rm miss}_{\rm T}
\ee
is simulated by means of \textsc{MadGraph5\_aMC@NLO}~\cite{Alwall:2014hca} at leading order at parton level and then hadronised and showered using \textsc{Phytia8}~\cite{Sjostrand:2014zea};  the detector reconstruction is simulated within the \textsc{Delphes}~\cite{deFavereau:2013fsa} framework using the ATLAS detector card. 

The operators related to entanglement and Bell inequality violation are computed from the simulated events by looking at the angular correlations of the pairs of charged leptons, as represented by the product of the cosines $\cos \theta^i_+$ and $\cos \theta^j_-$ as in \eq{eq:xsec2dB}. The matrix $\CC_{ij}$ is reconstructed from these by going to the rest frame of the top quark (which requires the reconstruction of the neutrino momenta). 

 \begin{figure}[h!]
\begin{center}
\includegraphics[width=4.5in]{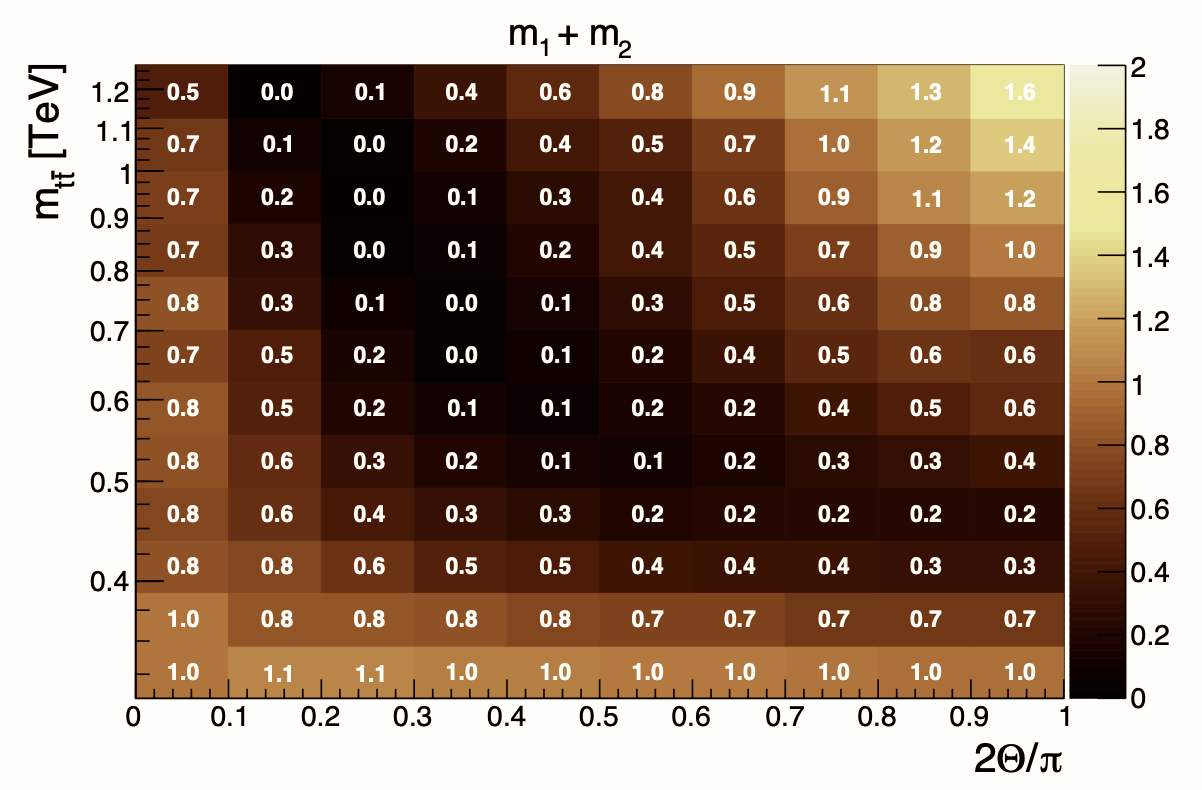}
\caption{\small Simulation of the values of  $\mathfrak{m}_{12}$ (here indicated as $m_1+m_2$) for the top-quark pair production at the LHC in bins, as a function of the invariant mass and the scattering angle. Values greater than 1 indicate violation of Bell inequality. Figure revisited from \cite{Fabbrichesi:2021npl} (\href{https://creativecommons.org/licenses/by/4.0/}{CC BY 4.0})}   
\end{center}
\label{fig:ttbarBINS}
\end{figure}

In~\cite{Fabbrichesi:2021npl}, the authors concentrate on the region of high invariant mass and large scattering angles and estimate the value of  $\mathfrak{m}_{12}$, after correcting for the bias. They predict that the violation can have a significance of $3\sigma$ for the combined Run~1 plus Run~2 at the LHC (with 300 fb$^{-1}$ of luminosity) and $4\sigma$ at the high-luminosity (Hi-Lumi) LHC (with  3 ab$^{-1}$ of luminosity). A smaller significance is found in \cite{Severi:2021cnj} for the same kinematic region: below $1\sigma$ at Run~1 plus Run~2 and only 1.8$\sigma$ at the Hi-Lumi LHC. 
The difference seems to come from a different treatment of the uncertainties in going from the parton level (where the two analyses agree) to the unfolded events. The neutrino weighting technique~\cite{D0:1997pjc} is used in \cite{Fabbrichesi:2021npl} to reconstruct the top quark momenta, while \cite{Severi:2021cnj} uses weighted kinematic reconstruction and
then \texttt{RooUnfold} to unfold detector effects.

A method to enhance the violation of Bell inequality was discussed in \cite{Aguilar-Saavedra:2022uye} for the threshold region, by imposing a cut on the velocity of the $t\bar{t}$ system in the laboratory frame which suppresses $q\bar{q}$  production contributions. A study of the optimal bases in which to define the quantum ensemble was presented in \cite{Cheng:2023qmz}. Different optimal event-by-event defined frames were found for near-threshold production (for which the optimum is close to the lab basis) vs high $m_{tt}$ production (for which it is the helicity basis). 

In \cite{Dong:2023xiw} and \cite{Han:2023fci} the simulation is extended to include the semi-leptonic decays:
\be
p +p \to t+\bar t \to \ell \nu + 2 b +2 j \, . 
\ee
The semi-leptonic channel contains more events, and fewer undetected particles, and could therefore provide a result with less uncertainty than the fully leptonic one. 
However the spin analysis in this channel is more challenging due to the difficulty in determining which jet from the $W$ originated from an up-type quark and which from a down-type quark,  reducing the spin analysing power. 
The same software packages, as described above, are used in the numerical simulations. 
The result is that tagging through the semi-leptonic channel brings more events even though the efficiency is reduced. An increase of a factor 1.6 in significance is expected between the fully leptonic and the semi-leptonic channels.

The combinations, derived from the CSHC inequality in \eq{CHSH-2},
\be
|C_{rr} - C_{nn}| -\sqrt{2} > 0 \quad \text{or} \quad |C_{kk} + C_{rr}| > 0
\ee
are used to mark the violation of the Bell inequality. 
Both works find a significance of $4\sigma$ at Hi-Lumi (with  3 ab$^{-1}$ of luminosity) for the violation of the Bell inequality in the region of large invariant mass and scattering angle.

One would expect the experimental Collaborations eventually to use both the semi- and leptonic channels in the analysis of the actual data.


\subsection{$\tau$-lepton pair production at the LHC and SuperKEKB}
\label{sec:qubit:tau}

 The study of entanglement in $\tau$-lepton pairs  was first proposed for $e^+e^-$ collisions at LEP~\cite{Privitera:1991nz}. It was extended in \cite{Fabbrichesi:2022ovb} for the production at the LHC and in \cite{Ehataht:2023zzt} for that at SuperKEKB. 

\begin{figure}[h!]
\begin{center}
\includegraphics[width=5.5in]{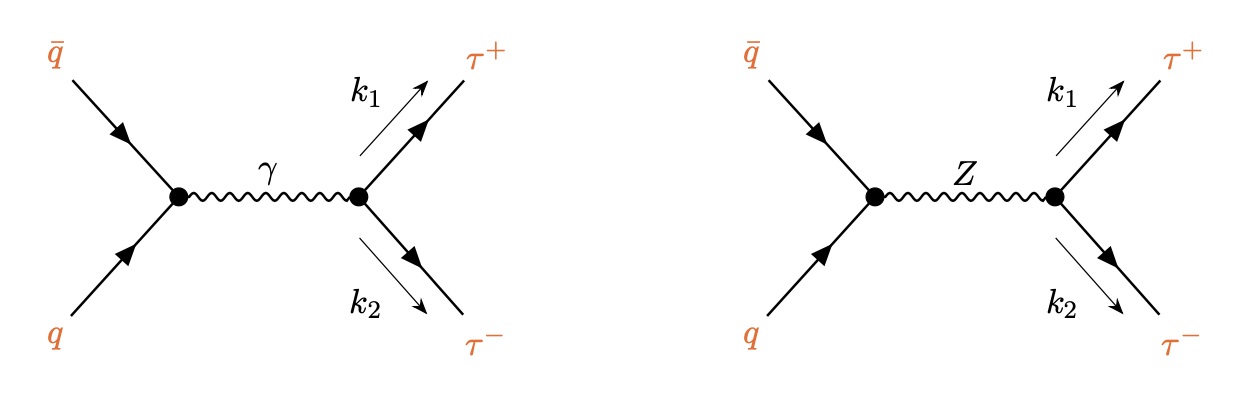}
\caption{\small  Feynman diagrams for $\tau^-\tau^+$ production via Drell-Yan mechanism at hadron collider. 
}
\end{center}
\label{fig:taus_cont}
\end{figure}

 The procedure for computing the polarization density matrix for this process  at the LHC follows the same steps as for the top quarks analyzed in Section~\ref{sec:qubits:top}, except for the main production mechanism.
The dominant process in this case is the Drell-Yan production in which the quarks go  into the $s$-channel either via a photon or a $Z$-boson which, in turn, decay into the $\tau$-lepton pair. The corresponding tree-level relevant Feynman diagrams for the $\tau$-pair production are shown in Fig.\ref{fig:taus_cont}.

In addition to the Drell-Yan mechanism production, in this case we also have the  process in which the $\tau$ leptons originate from the resonant Higgs boson decay channel. 
Here we focus on the Drell-yan production and leave resonant Higgs to Section~\ref{sec:qubit:higgs-tau}, which is devoted to the qubits systems arising from the Higgs boson decay.

The production process of $\tau$-lepton pairs via Drell-Yan mechanism in the SM receives contributions from the diagrams mediated by the $s$-channel photon, the $Z$-boson and their interference. These contributions provide an ideal laboratory for studying quantum entanglement among the qubits pairs of  $\tau$-lepton pairs. Due to the fact that the fewer the contributions, the larger the entanglement  (as mixing among quantum states suppresses quantum correlations), we expect this to be larger at low-energies (where the photon diagram dominates) or around the $Z$-boson pole (where the $Z$-boson diagram dominates).  
At low energies, the cross section is dominated by the photon term which produces entangled $\tau$-lepton pairs, while at high-energies all terms contribute and entanglement is suppressed. Around the $Z$-boson pole the cross section is dominated by the corresponding term with maximal entanglement.
 
The entries of the correlation matrix $\CC_{ij}$ from the process
\be
p+p \to \tau^- + \tau^+  \, .
\label{eq:process}
\ee
are collected in Appendix~\ref{sec:appendix:qubits:tau}.

The cross-sections for up-type and down-type initial state $q\bar{q}$ pairs are then combined by weighting the respective contributions through the parton luminosity functions  $L^{qq}(\tau)$ defined in \eq{eq:lumPDF}. The corresponding unpolarized cross section  is given by \cite{Fabbrichesi:2022ovb}
\be
\frac{\di \sigma}{\di \Omega \, \di m_{\tau\bar\tau}} = \frac{\alpha^2 \beta_\tau}{64 \pi^2 m_{\tau \bar \tau}^2} \Big\{ L^{uu} (\tau) \, \tilde A^{uu}[m_{\tau\bar\tau},\, \Theta]+\big[L^{dd} (\tau)+L^{ss}(\tau) \big]\, \tilde A^{dd}[m_{\tau\bar\tau},\, \Theta]  \Big\}
\ee
where $\tau = m_{\tau^- \tau^+}/\sqrt{s}$  and $\alpha=e^2/4 \pi$.
For the numerical values of $L^{qq}(\tau)$, we can use those provided by {\sc PDF4LHC21}~\cite{PDF4LHCWorkingGroup:2022cjn} for $\sqrt{s}=13$ TeV, as for the top pair before, but with factorization scale $q_0=m_{\tau \bar \tau}$.
The explicit expressions for $\tilde A^{uu,dd}(m_{\tau\bar\tau})$  are given in  Appendix \ref{sec:appendix:qubits:tau}. 

The full correlation matrix $\CC_{ij}$ is obtained by putting together all relevant   contributions from the various $q\bar q$-production channels, weighted by suitable luminosity functions and with appropriate normalization. This effect leads to further mixing and in general to additional loss of entanglement.

For the correlation coefficients $C_{ij}$ we have
\cite{Fabbrichesi:2022ovb}
\be
C_{ij} [m_{t\bar t},\, \Theta]= \frac{L^{uu} (\tau) \, \tilde C_{ij}^{uu}[m_{\tau\bar\tau},\, \Theta]+\big[L^{dd} (\tau)+L^{ss} (\tau) \big]\, \tilde C_{ij}^{dd}[m_{\tau\bar\tau},\, \Theta]} {L^{uu}(\tau) \, \tilde A^{uu}[m_{\tau\bar\tau},\, \Theta]+\big[L^{dd} (\tau)+L^{ss} (\tau)\big] \, \tilde A^{dd}[m_{\tau\bar\tau},\, \Theta]} \label{eq:Cij-tau}\, ,
\ee
where the down-quark luminosity functions can be grouped together because they multiply the same correlation functions.

A much simpler formula holds at lepton colliders for the process
\be
e^+ + e^- \to \tau^- + \tau^+  \, .
\label{eq:ee}
\ee
because there are no PDF luminosity functions,  the CM energy is fixed at $\sqrt{s}=10$ GeV at SuperKEKB and  there is only the photon diagram. 
This process was studied  at SuperKEK in~\cite{Ehataht:2023zzt} to show how promising this setting can be for  a study of Bell inequality violation.
The expected concurrence is  given in this case by a closed formula~\cite{Ehataht:2023zzt}:
\begin{equation}
    \mathscr{C}[\rho]=\frac{\left(s - 4 \,m_{\tau}^2\right) \sin^2\Theta}{4\, m_{\tau}^{2} \sin^2\Theta + s \left(\cos^2\Theta + 1\right)}\, .
\end{equation}

\subsubsection{Entanglement in $\tau\bar \tau$ production}
\label{sec:qubit:tau:entanglement}

The two spin-1/2 state of the
 $\tau$ pairs can be expressed by a density matrix having a general form as in \eq{rho-1/2}, whose entries depend on the kinematic variable $\beta_\tau=\sqrt{1-4m_\tau^2/m_{\tau\bar\tau}^2}$,  with $m_{\tau\bar\tau}$ the $\tau$-pair invariant mass, and on the scattering angle $\Theta$ in the  $\tau\bar{\tau}$ CM frame. 

Following  the same notation and reference frame adopted for the top-pair production in Section \ref{sec:qubits:top}, and focusing on the configuration of transversally produced lepton pairs ($\Theta=\pi/2$), we can distinguish three kinematic regions according to the following energy ranges: the low-energy one, at $m_{\tau\bar\tau}\ll m_Z$, where photon exchange dominates, the intermediate one at $m_{\tau\bar\tau}\simeq m_Z$, which is dominated by the $Z$ exchange, and finally the high-energy one, $m_{\tau\bar\tau} \gg m_Z$. 

 In the low-energy regime ($m_{\tau\bar\tau}\ll m_Z$), by using the results provided in Appendix~\ref{sec:appendix:qubits:tau} for the polarization and correlation coefficients of the density matrix in the $\tau$-pair case, we can see that
the $\tau$-pair spin state can be represented by the convex combination as in (\ref{rho-qq}) for the top-pair \cite{Fabbrichesi:2022ovb}, 
\be
\rho_{\tau\bar\tau}= \lambda \rho^{(+)} + (1-\lambda)\, \rho^{(1)}_{\rm mix} \quad \text{with} \quad  \lambda=\frac{\beta_\tau^2}{2-\beta_\tau^2}\in [0,1]\, ;
\ee
at threshold, $\beta_\tau\simeq 0$, the quantum state is a totally mixed one, with no quantum correlations, while as  
$\beta_\tau \rightarrow 1$ (i.e. when the tau leptons become relativistic, while still satisfying $m_{\tau\bar\tau}\ll m_Z$), the spins of the $\tau$-lepton pair tend to be generated in a maximally entangled state.

In the intermediate energy region, where the $Z$-channel starts to become relevant, this entanglement begins to loose coherence due to the increasing contribution of the interference term
between the photon and $Z$ diagrams. Nevertheless, a revival of entanglement reappears as the $m_{\tau\bar\tau}$ approaches the resonant  $Z$-channel region. In this region, using the notation and conventions introduced in the Appendix \ref{sec:appendix:qubits:tau}, the two-spin density matrix can be described by the following convex combination, for all quark production channels :
\be
\rho_{\tau\bar\tau}= \lambda \tilde\rho^{(+)} + (1-\lambda)\tilde\rho^{(2)}_{\rm mix}\ ,\qquad 
\lambda=\frac{(g_A^\tau)^2 - (g_V^\tau)^2 }{(g_A^\tau)^2 + (g_V^\tau)^2 }\ ,
\label{eq:rho-tau}
\ee
where, 
\be
\tilde\rho^{(2)}_{\rm mix}=\frac{1}{2}\Big( |\text{\small RR}\rangle\langle \text{\small RR}| 
+ |\text{\small LL}\rangle\langle \text{\small LL}| \Big)\ .\\
\ee
while
\be
\tilde\rho^{(+)}= |\tilde\psi^{(+)}\rangle\langle\tilde\psi^{(+)}|\ ,\qquad 
|\tilde\psi^{(+)}\rangle=\frac{1}{\sqrt{2}}\Big( |\small{+-}\rangle + |\small{-+}\rangle \Big)\ ,
\ee
is a projector on a Bell state as in (\ref{Bell-states}), expressed in terms of the eigenvectors of $\sigma_x$. Then, we could see that when $\lambda\to 1$,
the density matrix $\rho_{\tau\bar\tau}$ in \eq{eq:rho-tau} turns out to be very close to the maximally entangled state $\tilde\rho^{(+)}$.

 \begin{figure}[h!]
\begin{center}
\includegraphics[width=3.4in]{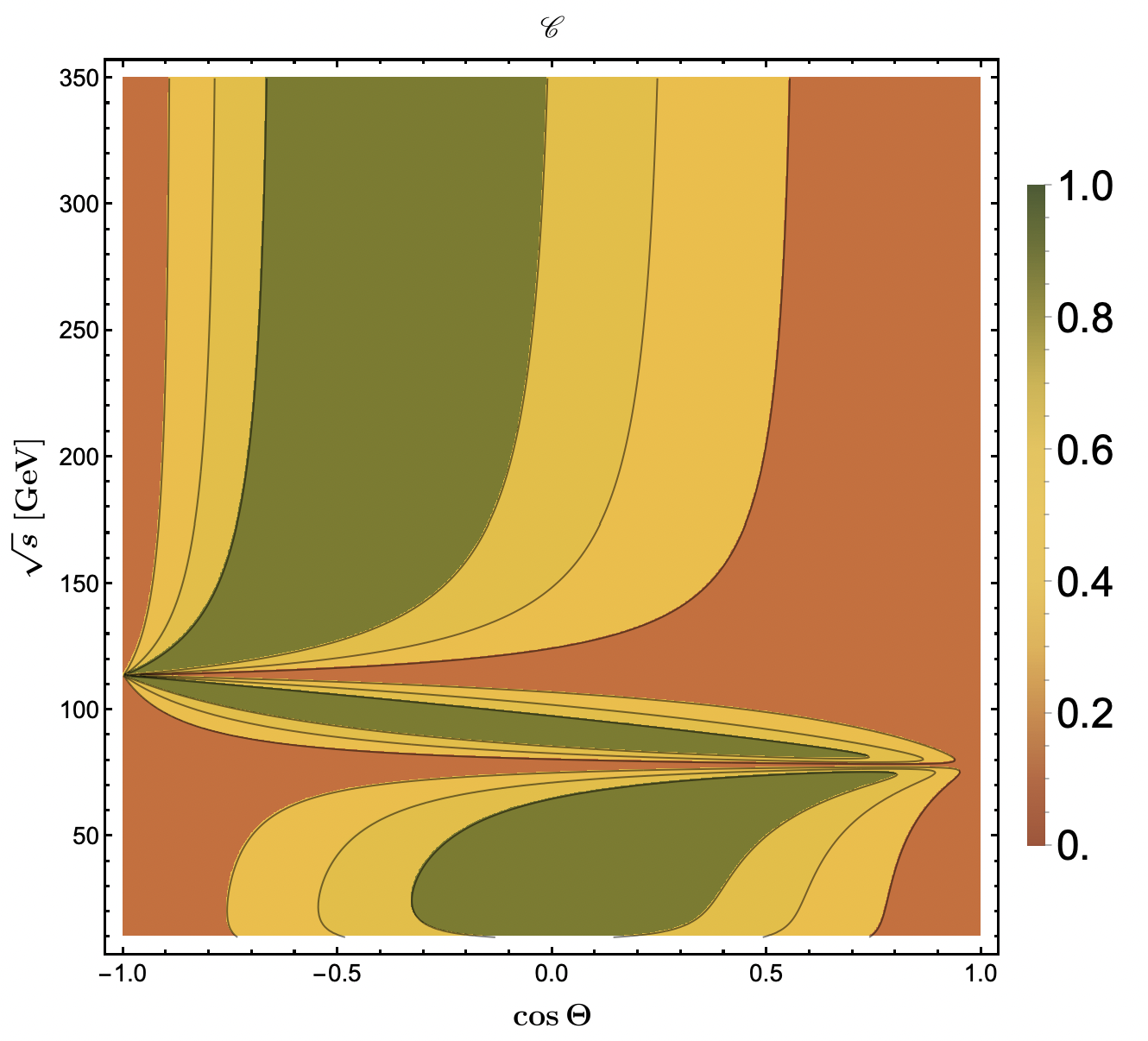}
\includegraphics[width=3.4in]{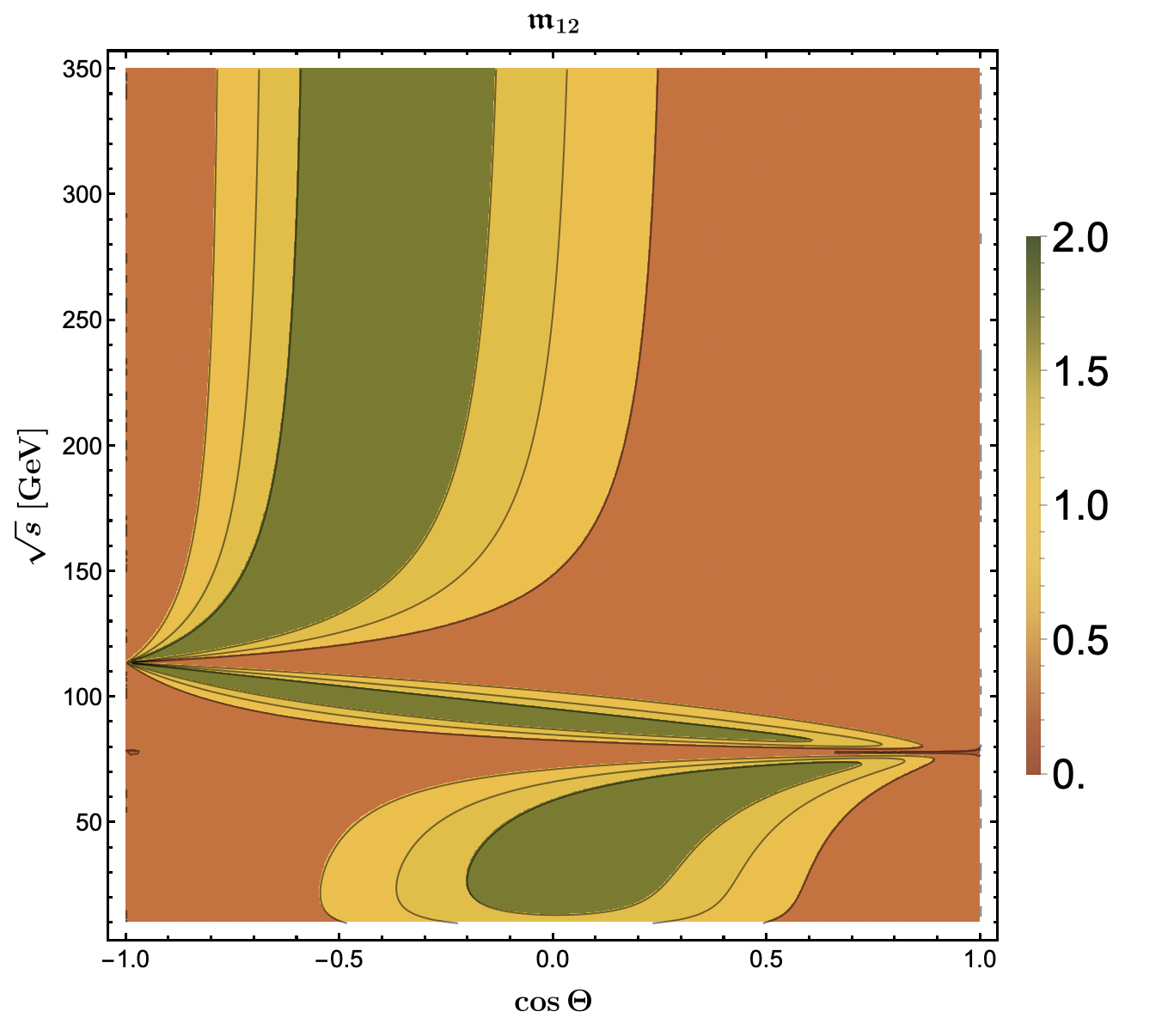}
\caption{\small Concurrence and  $\mathfrak{m}_{12}$ for the $e^+ e^- \to \tau \bar{\tau}$ pair production, as a function of the kinematic variables $\Theta$ and $m_{\tau \bar \tau}$ across the entire available space. Figures revisited from \cite{Fabbrichesi:2022ovb} (\href{https://creativecommons.org/licenses/by/4.0/}{CC BY 4.0}). }

\label{fig:m1m2tau} 

\end{center}
\end{figure}

Finally, in the high energy regime ($m_{\tau\bar\tau}\gg m_Z$)  both photon and $Z$ channel contribute, and, due to their mixing, a rapid depletion of entanglement is induced. In particular, for each $q\bar q$ production channel, the $\tau$-pair spin correlations can be described in terms of the following density matrix \cite{Fabbrichesi:2022ovb}~:
\be
\rho_{\tau\bar\tau}= \lambda^q \rho^{(+)} + (1-\lambda^q)\tilde\rho^{(2)}_{\rm mix}\ ,
\qquad \lambda^q=\frac{1-R_-^q}{1+R_+^q}\ ,
\label{rho-high}
\ee
where $\rho^{(+)}$ is as in (\ref{Bell-states}), while
\be
R^q_\pm = \frac{\chi^2(m_{\tau\bar\tau}^2) \big[ (g_A^q) + (g_V^q) \big] \big[(g_A^\tau) \pm (g_V^\tau)\big]}
{(Q^q)^2 (Q^\tau)^2 + 2\,{\rm Re}\chi(m_{\tau\bar\tau}^2)\, Q^q Q^\tau\,  g_V^q g_V^\tau}\ .
\ee
Namely, in the case of the $u$ quark production channel, we have $\lambda^u\simeq 0.7$,  so that some entanglement is preserved. On the other hand, 
for the $d$ quark production channel, since $\lambda^d\simeq 0.1$, the entanglement is essentially lost.

For completeness, it should be noticed that each $\tau$ lepton
is produced in a partially polarized state, as some of the single-spin polarization coefficient $\BB_i^\pm$ 
in the spin density matrix are non-vanishing (see Appendix~\ref{sec:appendix:qubits:tau}).
This is particularly relevant for the quark $d$ production channel, where the magnitude
of these single particle terms is of the same order of the entries of the correlation matrix $\CC_{ij}$,
while for the $u$ production channel they are about one order of magnitude smaller.
This implies that the full density matrix describing the $\tau$-pair spin state $\rho_{\tau\bar\tau}$ is really in this case
a mixture of (\ref{rho-high}) with additional states
further reducing in general its entanglement content.

\subsubsection{Bell inequalities}
\label{sec:qubit:tau:bell}

The same method presented in Section \ref{sec:qubits:top:bell} for the top-quark pairs can be followed here.  To make the discussion simpler, we focus on  the case in which the $\tau$-lepton pairs are produced at a lepton collider.
The values of the observable $ \mathfrak{m}_{12}$, are shown  in Fig.~\ref{fig:m1m2tau} across the entire kinematic space \cite{Fabbrichesi:2022ovb} for existing and future $e^+$-$e^-$ machines. The results are similar in the case of a hadron collider (with a little modulation because of the parton luminosity functions) and confirm the qualitative analysis of entanglement  in Section
\ref{sec:qubit:tau:entanglement}:   Entanglement is close to maximal (that is, $\mathfrak{m}_{12}$ close to 2)  for large scattering angles
whenever the invariant mass of the $\tau$-lepton pairs selects one of the two possible channels with either the photon or the $Z$-boson exchange dominating.

For the process at the LHC, the authors of \cite{Fabbrichesi:2022ovb}
take the kinematic window where the $\tau$-lepton pair invariant mass is in the range $20{\rm GeV} <m_{\tau \bar \tau} < 45{\rm GeV}$ and $|\cos \Theta| < 0.2$ as the most favorable to test the Bell inequalities and there estimate the mean value of $\mathfrak{m}_{12}$ to be  1.88.

\begin{figure}[h!]
\begin{center}
\includegraphics[width=4in]{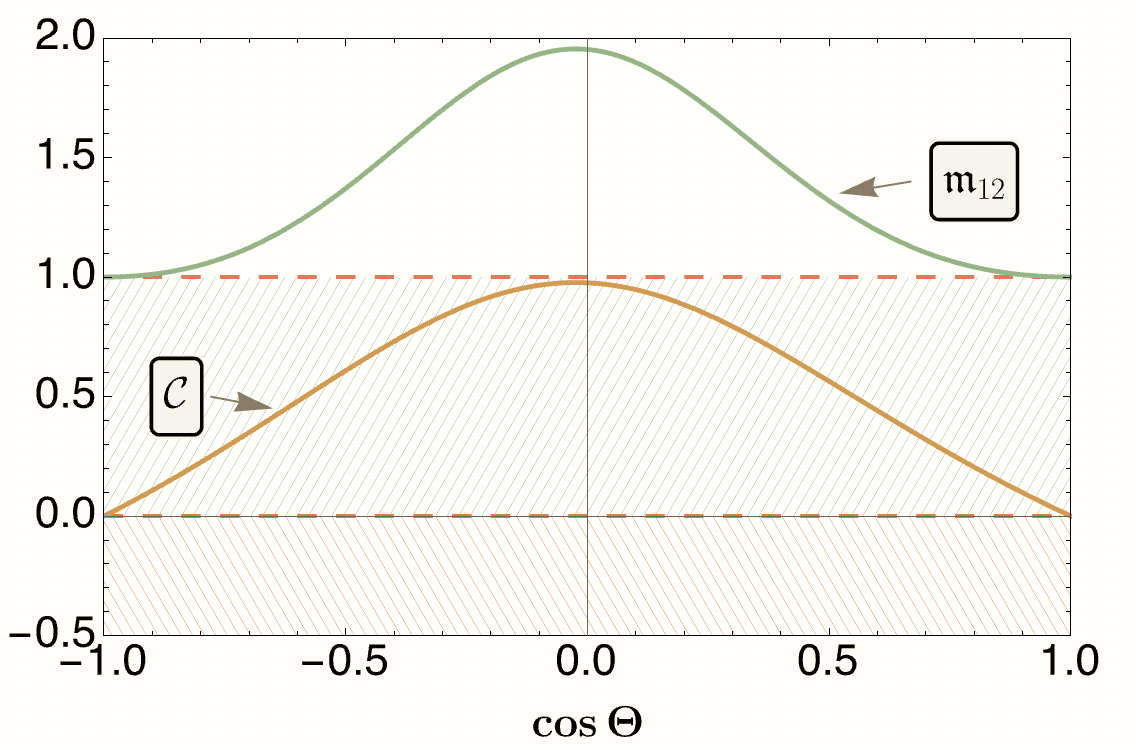}
\caption{\small   Concurrence and  $\mathfrak{m}_{12}$ for  $\tau \bar{\tau}$ pair production at SuperKEK. Figure revisited from \cite{Ehataht:2023zzt} (\href{https://creativecommons.org/licenses/by/4.0/}{CC BY 4.0}).}
\label{fig:superKEK} 
\end{center}
\end{figure}

For the process at SuperKEKB, a simple analytic formula can be computed~\cite{Ehataht:2023zzt}:
\begin{equation}
   \Rchsh{} = 1 + \left(\frac{\left(s - 4\, m_{\tau}^2\right) \sin^2\Theta}{4 \, m_{\tau}^2 \sin^2\Theta+ s \left(\cos^2\Theta + 1\right)}\right)^{2} \, ,
\end{equation}
The maximum value for both $\mathscr{C}[\rho]$ and $\Rchsh{}$ are reached for   scattering angles close to $\pi/2$, as shown in Fig.~\ref{fig:superKEK}.

\subsubsection{Monte Carlo simulations of events}
\label{sec:qubit:tau:simulations}
The production of $\tau \bar{\tau}$ pairs at SuperKEK appears very promising for the study of entanglement and Bell inequality violation because of the large number of events that are, in addition, very clean.

The polarization of the $\tau$-leptons can be extracted from the distribution in momenta of the final charged hadrons in the three decay channels: $\tau^- \to \pi^-\nu_\tau$, $\tau^- \to \pi^-\pi^0\nu_\tau$, and $\tau^- \to \pi^-\pi^+\pi^-\nu_\tau$.
The combination of these decay channels covers about $21\%$ of $\tau$ pair decays. 

In \cite{Ehataht:2023zzt}, a sample of $200$ million $e^+e^- \to \tau^+\tau^-$ Monte Carlo events was generated with the program \textsc{MadGraph5\-\_aMC@NLO}~\cite{Alwall:2014hca}, using leading-order matrix elements.
The program \textsc{PYTHIA}~\cite{Sjostrand:2014zea} was used for the modeling of parton showers, hadronization processes, and $\tau$ decays.
All the $\tau$ decay channels discussed above are included in the simulation.
The events are analyzed on Monte Carlo truth level and after taking realistic experimental resolutions into account.

Both entanglement and Bell inequality violation are predicted to be observable with a significance well in  excess of $5\sigma$, with a dataset comparable to that already recorded by Belle~II.


\subsection{Higgs boson decays in $\tau$-lepton pairs and two photons}
\label{sec:qubit:higgs-tau}

\begin{figure}[h!]
\begin{center}
\includegraphics[width=4.5in]{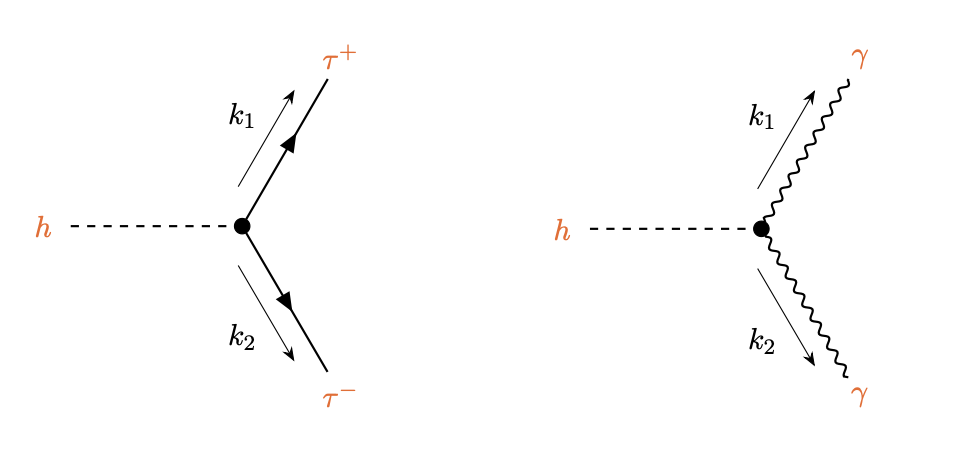}
\caption{\small \label{fig:Higgs-gamma} Feynman diagrams for the Higgs boson $h$ decay into two $\tau$-leptons (left) and into two photons (right). }
\end{center}
\end{figure}

The decay of the Higgs boson into a pair of fermions  or two photons (see, Fig.~\ref{fig:Higgs-gamma}), provides a physical process very similar to those utilized in atomic physics for studying entanglement. Because the final states originate from the decay a scalar particle, a pure state should be created for the spins. In this Section we  discuss first the qubits system provided by the Higgs boson decay into $\tau$-lepton pair, then the decay into two photons. In this last case, we  assume it will be possible in the future to determine the polarization of the photon.

\subsubsection{Entanglement and Bell inequalities in $h\to \tau\bar{\tau}$}
\label{sec:qubit:higgs-tau:entanglement}

The SM interaction Lagrangian for the decay of the Higgs boson  into a pair of $\tau$ leptons is given by 
\be
{\cal L}_{\text{\tiny SM}} = \frac{m_\tau}{v} \bar \tau \tau \, h \, ,
\ee
where $v$ is the vacuum expectation value of the Higgs field $h$.
On the basis of this interaction term, the elements of the matrix $C_{ij}$ entering the tau lepton-pair spin density matrix   can be easily computed and is given by \cite{Fabbrichesi:2022ovb}
\be
\CC=  \begin{pmatrix} 1&0&0\\
0&1&0 \\
0& 0 & -1
\end{pmatrix} \, ,
 \ee
where the $C$ matrix above is defined on the $\{\hat{n},\hat{r},\hat{k}\}$ spin basis as in \eq{matrixC}. The entanglement ${\cal C}[\rho]$ is maximal and equal to 1.
 The sum of the square of the two largest eigenvalues gives $\mathfrak{m}_{12}=2$, so that the 
Bell inequality (\ref{CHSH}) is predicted to be maximally violated.

\subsubsection{Monte Carlo simulations  and predictions}
\label{sec:qubit:higgs-tau:simulations}

The decay of the Higgs boson into $\tau$-lepton pairs has been analyzed in \cite{Altakach:2022ywa} and \cite{Ma:2023yvd}. Both studies investigate $HZ$ associated production at future $e^+e^-$ colliders; the process would be difficult to reconstruct experimentally in the resonant production of the Higgs boson in the $s$-channel production in hadron colliders.

The Monte Carlo simulations are performed by means of \textsc{MadGraph5\_aMC@NLO}~\cite{Alwall:2014hca}, using leading-order matrix elements.
The package \textsc{TauDecay}~\cite{Hagiwara:2012vz} is used for modeling  the $\tau$ decays. Only the  decays into a single pion are included. The momenta of the $\tau$-leptons are reconstructed by solving the kinematic equations holding for the (unknown) neutrino momenta. The kinematic reconstruction is possible up to a two-fold degeneracy.

In \cite{Altakach:2022ywa} it is found that one expects entanglement to be tested above 5$\sigma$ at both the International Linear Collider (ILC) and the Future Circular Collider (FCC-ee). The violation of Bell inequality is not expected to be observed at the ILC but it expected at the FCC-ee with a significance of about 3$\sigma$. It is found in \cite{Ma:2023yvd}  that the predicted significance is around $1\sigma$ for the Circular Electron Positron Collider (CEPC).


\subsubsection{Entanglement and Bell inequalities in $h\to \gamma\gamma$}
\label{sec:qubit:higgs-photons}

The entanglement of a system of two photon has been discussed in~\cite{Lyuboshitz:2007zzb} and, more recently,  in \cite{Fabbrichesi:2022ovb}.
 This system closely resembles those  in atomic physics, in which  the polarization of photons originating in atomic transitions are discussed.

The Higgs boson $h$ decays into two photons
\be
h\to \gamma(k_1)\,  \gamma(k_2)\, ,
\label{Hgg}
\ee
proceeds via an effective coupling $g_{\gamma\gamma h}$ provided in the SM by loop contributions. The effective Lagrangian in this case is given by
\be
{\cal L} = -\frac{1}{4} \, g_{\gamma\gamma h} \, h \, F^{\mu\nu}F_{\mu\nu}\, ,
\ee
where $F^{\mu\nu}$ is the field strength of the photon.

The corresponding polarized amplitude square is 
\be
{\cal M}(\lambda^{\prime}_1, \lambda^{\prime}_2){\cal M}(\lambda_1, \lambda_2)^\dagger = |g_{\gamma\gamma h}|^2 V^{\mu\nu}(k_1,k_2) V^{\rho\sigma}(k_1,k_2)
\left[\varepsilon^{\lambda_1}_\mu (k_1) \varepsilon^{\lambda^{\prime}_1 \ast}_\rho(k_1)\right] \left[\varepsilon^{\lambda_2}_\nu (k_2) \varepsilon^{\lambda^{\prime}_2 \ast}_\sigma (k_2)\right]\, ,
\ee
where $V^{\mu\nu}(k_1,k_2)=g^{\mu\nu}(k_1\cdot k_2)-k_1^{\nu}k_2^{\mu}$. Notice that, gauge invariance is guaranteed by the Ward Identities $k_1^{\mu}V_{\mu\nu}(k_1,k_2)=k_2^{\nu}V_{\mu\nu}(k_1,k_2)=0$.

Summing over the photon polarizations we obtain 
\be
|{\cal M}|^2=\frac{1}{2}\, |g_{h\gamma\gamma}|^2 m_h^4\, ,
\label{M2hS}
\ee
to which corresponds the width $\Gamma=g_{h\gamma\gamma}^2 m_h^3/(64\pi^2)$.

The  polarization density matrix in the case of the two-photons is readily obtained by  following the method of Section \ref{sec:third:toolbox}. After normalization over the unpolarized square amplitude in \eq{M2hS}, the correlation matrix $\CC$ is given by 
\cite{Fabbrichesi:2022ovb}
 \be
\CC =  \begin{pmatrix} 1&0&0\\
0&-1&0 \\
0& 0 & 1
\end{pmatrix} \, . \label{mC}
\ee
in the basis of the Stokes parameters $\{\xi_1,\xi_2,\xi_3\}$  defined in \eq{densitymat}. 

As we can see from the above result, for the matrix $\CC$ in \eq{mC}, the operator $\mathfrak{m}_{12}=2$  and the Bell inequalities are  maximally violated.

There is good motivation for wishing to perform such a test using a diphoton final state, however  it requires the detection of the polarization of the two photons.
The possibility of measuring photon polarizations depends on their energy. For high-energy photons, the dominant process is pair production as the photons traverse matter. There are two possible processes: the electron interacting with the nuclei ($A$) or the atom electrons:
\bea
\gamma + A &\to& A+ e^+ + e^- \nn \\
\gamma + e^- &\to& e^-+ e^+ + e^-  \, ,
\eea
with the latter dominating in the energy range we are interested in.

For a polarized photon,  the Bethe-Heitler  cross section for the \textit{Bremsstrahlung} production of electron pairs depends also on the azimuthal angles $\varphi_\pm$ of the produced electron and positron~\cite{May,Borsellino:1953zz} as
\be
\frac{\di s}{\di \varphi_+ \di \varphi_- } = \sigma_0 \Big[  \Sigma_{\text{un}} + \Sigma_{\text{pol}} \, P_\gamma \cos (\varphi_+-\varphi_-) \Big] \, ,\label{pairs}
\ee
where $P_\gamma$ is the linear polarization fraction of the incident photon, $ \Sigma_{\text{un}}$ and   $\Sigma_{\text{pol}} $  are the unpolarized and  polarized coefficients respectively, which depend on the kinematic variables. The explicit form of the cross section in \eq{pairs} can be found in~\cite{Boldyshev:1994bs}. The relevant  information on the azimuthal distribution   comes from  the dependence of the cross section on the
a-coplanarity of the outgoing electron and positron.
The measurement of the relative angle between these momenta gives information on the polarization of the photon.

Even though this possibility is not currently  implemented at the LHC, detectors able to perform such a measurement are already envisaged  for astrophysical $\gamma$ rays~\cite{Depaola:2009zz} and     an event generator to simulate the process already exists~\cite{Bernard:2013jea} and has been implemented within GEANT~\cite{GEANT4:2002zbu}
(for a recent review, see~\cite{Gros:2016dmp}). 

\newpage

\section{Qutrits: massive gauge bosons  and  vector mesons}
\label{sec:qutrits}

Systems of two  qutrits arise between the polarizations of pairs of massive gauge bosons  at the LHC  and between two vector mesons at  B-meson factories.

We discuss in Section \ref{sec:qutrits:gauge-bosons}  the SM production of two on-shell states $WW$ and $ZZ$ via the  electroweak processes induced at parton level by  quark-antiquark annihilation. Quantum entanglement and Bell inequality violations for these processes  have been analyzed in \cite{Barr:2022wyq,Ashby-Pickering:2022umy,Fabbrichesi:2023cev,Aoude:2023hxv}. In \cite{Fabbrichesi:2023cev,Aoude:2023hxv}  the potential for the same processes are also discussed at future colliders. 

In Section \ref{sec:qutrits:higgs}, we turn to  diboson production via resonant Higgs decays into $h\to WW^*$ and $h\to ZZ^*$, where $W^*,Z^*$ indicate the corresponding vector bosons as off-shell states. The field was initiated in \cite{Barr:2021zcp} in which entanglement and Bell inequality violation were studied in the decay of the Higgs boson into the two charged gauge bosons $W^+W^-$ by means of Monte Carlo simulations. It was followed by analyses of the same process in \cite{Aguilar-Saavedra:2022mpg,Bi:2023uop,Fabbri:2023ncz} and extended to the case of two neutral gauge bosons $ZZ$ first in \cite{Aguilar-Saavedra:2022wam,Ashby-Pickering:2022umy} and then in \cite{Bernal:2023ruk,Fabbrichesi:2023cev}. The result of these studies is that the most promising channel is $h\to ZZ$, because of the small background, and where the Bell inequality could be violated with a significance of more than 3$\sigma$ at the High-Lumi LHC.

In Section \ref{sec:qutrits:B} we consider the quantum entanglement and Bell inequality violation for the qutrits system  of two vector mesons arising from the neutral $B$ meson decays, which has been analyzed in \cite{Fabbrichesi:2023idl}.

\subsection{$B$-meson decays in two vector mesons}
\label{sec:qutrits:B}

The decays of the neutral $B$-mesons into two spin-1 mesons closely resemble those of the Higgs boson and the same tools can be put to work. 

There are three helicity amplitudes for the decay of a scalar, or pseudo-scalar,  into two massive spin-1 particles:
\be
h_{\lambda} = \langle V_{1}(\lambda) V_{2}(-\lambda)| {\cal H} |B\rangle\quad \text{with} \quad \lambda=(+,\, 0,\,-)\, ,
\label{helampl}
\ee
and ${\cal H}$ is the interaction Hamiltonian  giving rise to the decay.
For the spin quantization axis ($\hat z$) we can use the direction of the momenta of the decay products in the $B^0$ rest frame.
Helicities are here defined with respect to the $\hat z$ direction  in the rest frame of one of the two  spin-1 particles and $(+,\, 0,\,-)$ is a shorthand  for $ (+1,\, 0,\,-1)$.

The polarizations in the decay  are described by a quantum state that is pure for any values of the helicity amplitudes~\cite{Fabbrichesi:2023cev,Fabbrichesi:2023jep}. This state can be written as
\be
|\Psi \rangle = \frac{1}{\abs{\mathcal{M}}} \Big[  h_+\, |V_{1}(+)  V_{2}(-)\rangle 
 +  h_0 \, |V_{1}(0)  V_{2}(0)\rangle+  h_-\, |V_{1}(-)  V_{2}(+)\rangle \Big] \, ,\label{pure2}
\ee
with
\be
\abs{\mathcal{M}}^2= |h_0|^2 + |h_+|^2 + |h_-|^2 \, .
\ee
The relative weight of the transverse components $ |V_{1}(+)  V_{2}(-)\rangle $ and $ |V_{1}(-)  V_{2}(+)\rangle $ with respect to the longitudinal one $| V_{1}(0)  V_{2}(0 )\rangle$ is controlled  by the conservation of angular momentum. In general, only the helicity is conserved and the state in \eq{pure2} 
belongs to the $J_z=0$ component of the $S=0,1$ or $2$ states.

The polarization density matrix $\rho = |\Psi \rangle \langle \Psi |$ can be written in terms of the helicity amplitudes as
\be
\small
\rho=  \frac{1}{\abs{\mathcal{M}^2}}\, \begin{pmatrix} 
  0 & 0 & 0 & 0 & 0 & 0 & 0 & 0 & 0  \\
  0 & 0 & 0 & 0 & 0 & 0 & 0 & 0 & 0  \\
  0 & 0 &  h_+ h_+^* & 0 &  h_+  h_0^*& 0 &  h_+ h_-^*& 0 & 0  \\
  0 & 0 & 0 & 0 & 0 & 0 & 0 & 0 & 0  \\
  0 & 0 & h_0 h_+^* & 0 & h_0  h_0^*  & 0 & h_0 h_-^*& 0 & 0  \\
  0 & 0 & 0 & 0 & 0 & 0 & 0 & 0 & 0  \\
  0 & 0 &  h_- h _+^*& 0 &  h_- h_0^*& 0 &   h_-h_-^*& 0 & 0  \\
  0 & 0 & 0 & 0 & 0 & 0 & 0 & 0 & 0  \\
  0 & 0 & 0 & 0 & 0 & 0 & 0 & 0 & 0  \\
\end{pmatrix} \, ,
\label{rhoBVV}
\ee
on the basis given by the tensor product of the polarizations $(+,\, 0,\, -)$ of the produced spin-1 particles.

The polarizations of the spin-1 massive particles  can be reconstructed  using the momenta of the final charged mesons and  leptons in which they decay~\cite{Dighe:1995pd}. Usually, the experimental analysis provides the polarization amplitudes. These are mapped into the helicity amplitudes  by the correspondence
\be
\frac{h_{0}}{\abs{\mathcal{M}}}= A_{0} \,, \quad \frac{h_{+}}{\abs{\mathcal{M}}}  = \frac{ A_{\parallel}+A_{\perp}}{\sqrt{2}}\, , \quad
\frac{h_{-}}{\abs{\mathcal{M}}}  = \frac{ A_{\parallel} -A_{\perp}}{\sqrt{2}} \, .
\ee

The entanglement entropy and the Bell operator ${\cal I}_{3}$ can be readily be computed, the latter one after 
 the optimization procedure of \eq{uni_rot}.

 Data from the $B$-factories have been analyzed by the LHCb and Belle collaborations in terms of polarization amplitudes and provide 
 an abundant  source of processes in which it is possible to search for entanglement and test Bell inequality violation. The helicity measurements and the analyses have already been published and only the recasting in terms of entanglement markers and test of the Bell inequality need to be  validated by the experimental Collaborations.
 
The decay for which  the most precise polarization amplitudes are known is $B^0\to J/\psi\, \K$~\cite{LHCb:2013vga} for which, under the assumption that the density matrix takes the form \eq{rhoBVV}, it can be found~\cite{Fabbrichesi:2023idl}  that
\be
\mathscr{E} = 0.756 \pm  0.009\, \quad \text{and} \quad {\cal I}_3= 2.548 \pm 0.015\, ,
\ee
with a significance well in excess of 5$\sigma$ (numerically $36\sigma$)  for the violation of the Bell inequality ${\cal I}_3<2$.

To close the locality loophole---which exploits (see Section~\ref{sec:loopholes}) events not separated
by a space-like interval, as it is the case of the $J/\psi\, K^*$ decays---one must consider decays in which the produced particles are identical, as in the $B_s\to \phi\phi$ decay, and therefore their life-times are also the same. The actual decays take place with an exponential spread, with,   in the $\phi\phi$ case, more than 90\% of the events being separated by a space-like interval. 

For the decay $B_s\to \phi\phi$~\cite{LHCb:2023exl}, it is found~\cite{Fabbrichesi:2023idl}  that 
\be
\mathscr{E} = 0.734\pm 0.037\, \quad \text{and} \quad {\cal I}_3= 2.525\pm 0.064\, , 
\ee
with a significance of $8.2\sigma$ for the violation of the Bell inequality ${\cal I}_3<2$.

There is another reason why the decays of the $B$-mesons are interesting in testing for the presence of entanglement. It is possible to extract from the data~\cite{LHCb:2013vga,LHCb:2023exl} the strong phases arising from the final-state interactions in the $B$-meson decays and compute their contribution to the polarization amplitudes. We therefore know for the same process the amount of entanglement arising from the weak interactions, which are responsible for the decay, as well as that from the strong interactions in the subsequent re-scattering. The contribution to the  latter  can  be measured and shown to increase the overall  entanglement between the spins of the decay products.  

\vspace{0.5cm}
\subsection{Diboson production at LHC via quark-fusion}
\label{sec:qutrits:gauge-bosons}

The prospects for measurements in the production of $WW$ and of $ZZ$ gauge dibosons at the LHC have been analyzed in  \cite{Ashby-Pickering:2022umy,Fabbrichesi:2023cev,Aoude:2023hxv}.
These states can be produced via electroweak processes in a continuous range of diboson invariant masses. We show in the following how the polarization density matrix of this diboson system can be computed starting from the density matrices obtained for the involved parton contributions, presented in Fig.~\ref{fig:DYVV} for the processes at hand. We do not report here the results for 
$WZ$ production since, according to the analysis of \cite{Fabbrichesi:2023cev},   no significant excess above the null hypothesis for the Bell inequality violation has been found in the whole relevant kinematic region. While Bell violation is not expected, observation of entanglement would be possible~\cite{Ashby-Pickering:2022umy,Fabbrichesi:2023cev}.

The predicted correlation coefficients $h_{ab}$, $f_a$, and $g_a$ appearing in the decomposition of the polarization density matrix of two qutrits along the Gell-Mann matrix basis in Section \ref{sec:third:toolbox}, can be calculated as
a generalization of the corresponding coefficients of qubits
in \eq{eq:Cij-top}. In particular, for  $h_{ab}$ we get
\be
h_{ab} [\mVV,\Theta]= \frac{\sum_{q=u,d,s} L^{q\bar q} (\tau)
\Big(  \tilde{h}^{q \bar q}_{ab}[\mVV,\Theta] +
\tilde{h}^{q \bar q}_{ab}[\mVV,\Theta+\pi]\Big)}
{\sum_{q=u,d,s} L^{q\bar q} (\tau)\Big(A^{q \bar q}[\mVV,\Theta] +A^{q\bar q}[\mVV,\Theta+\pi]
  \Big)}\,
\label{eq:habDY}
\ee
where $\mVV$ stands for the invariant mass of the final diboson state and $\Theta$ the scattering angle in their CM frame. The abbreviations $A^{q \bar q} = |\mathcal{M}^{\;q \bar q}_{WW}|^2$ indicates the spin-summed square amplitude of the process, and $\tilde h_{ab} = A^{q \bar q} h_{ab} $. The $L^{q\bar q} (\tau)$ are the quark parton luminosity
functions defined in \eq{eq:lumPDF}. The sum of the terms with dependence by
$(\Theta+\pi)$ in \eq{eq:habDY} takes into account the fact that quarks or antiquarks can originate from both of the two proton beams of the LHC collider, and the two configurations have the same parton luminosity function.
For the sake of simplicity, when possible we leave implicit the dependence of the correlation coefficients $h_{ab}(\mVV,\Theta)$, $g_a(\mVV,\Theta)$ and $f_a(\mVV,\Theta)$ on the scattering angle $\Theta$ in the CM frame and on the invariant mass of the dibosons $\mVV$. 

Similar expressions hold for the remaining polarization correlation coefficients $f_a$ and $g_a$, of the Gell-Mann basis with $a\in\{1, \dots, 8\}$, obtained by replacing the $\tilde{h}_{ab}$ with corresponding quantities $\tilde{g}_a$ or $\tilde{f}_{a}$ ones.
We report in~\ref{sec:appendix:qutrits:qqZZ} the explicit expressions of $\tilde h_{ab}$, $\tilde f_{a}$, and $\tilde g_{a}$ functions only for the $ZZ$ production, while for all other processes these can be found in \cite{Fabbrichesi:2023cev}.

\subsubsection{Computing the observables: $p\, p \to W^+W^-$\label{sec:WW}}

\begin{figure}[h!]
  \begin{center}
  \includegraphics[width=5in]{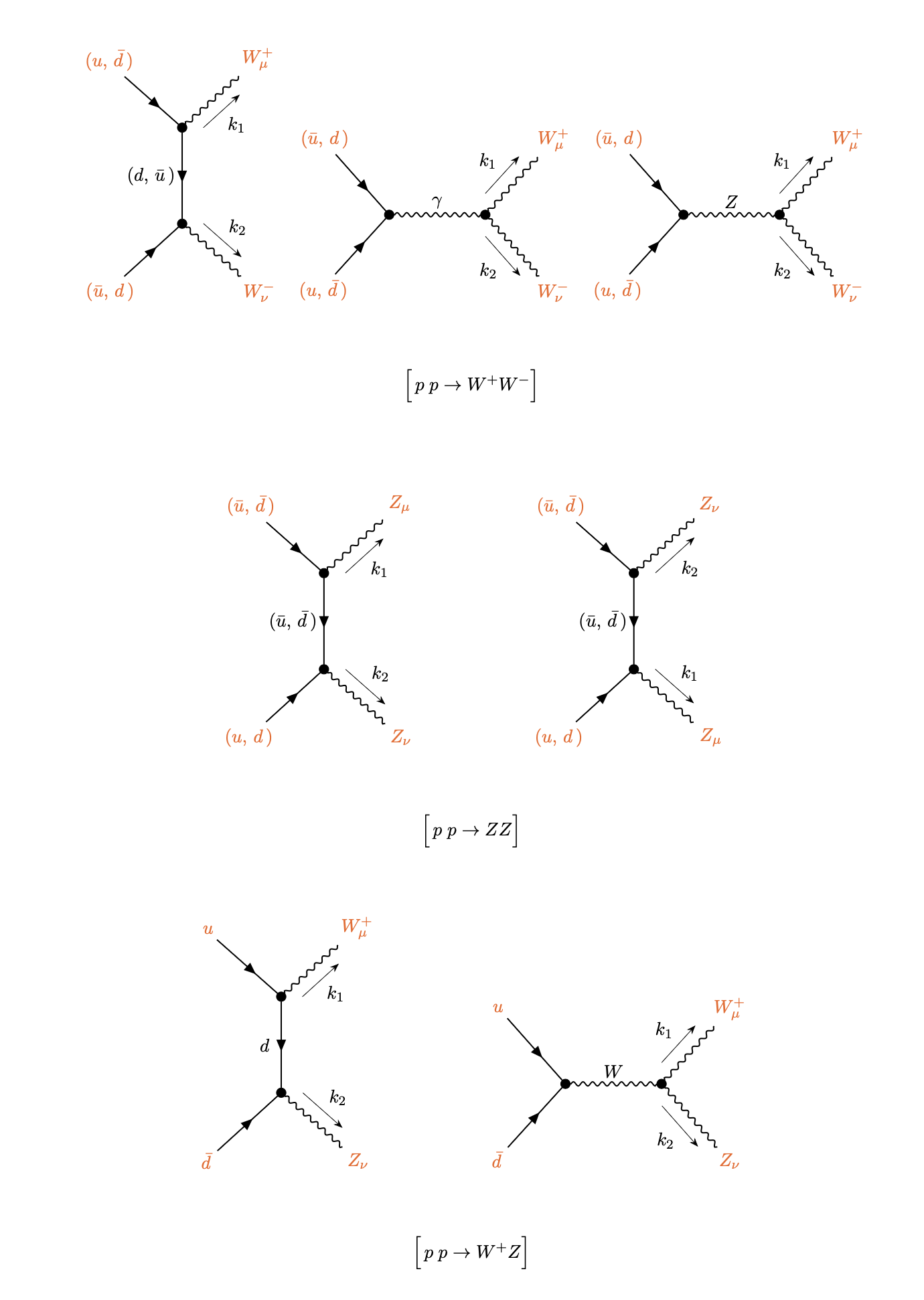}
  \caption{\small Feynman diagrams for the processes $p\; p\to W^+W^-$ (first row),  $p\; p\to Z Z$ (second row) and $p\; p\to W^+Z$ (third row) at the parton level for the first quark generation. We neglect Diagrams mediated by the Higgs boson are neglected (in  the limit of massless quarks). The arrows on the fermion lines indicate the momentum flow. 
  \label{fig:DYVV} 
  }
  \end{center}
  \end{figure}

The tree-level Feynman diagrams contributing to the parton level process
\be
\bar{q}(p_1) q(p_2) \to W^+(k_1,\lambda_1) W^-(k_2,\lambda_2)\, ,
\label{qqWW}
\ee
are shown in the top part of Fig.~\ref{fig:DYVV}.
The polarization vectors of $W^+$ and $W^-$ are $\varepsilon^{\mu}(k_1,\lambda_1)$ and $\varepsilon^{\nu}(k_2,\lambda_2)$, respectively.
The polarized amplitude for the process in \eq{qqWW}, for $u$ and $\bar u$ initial states, is  given by  \cite{Fabbrichesi:2023cev}
\bea
{\cal M}^{u\bar u}_{WW}(\lambdaA,\lambdaB)&=&-ie^2 \Big[\bar{v}(p_1) \Gamma^{\W\W}_{\mu\nu} u(p_2)\Big]
\varepsilon^{\mu}(k_1,\lambda_1)^{\star}\varepsilon^{\nu}(k_2,\lambda_2)^{\star}
\, ,
\label{MDYWW}
\eea
where the effective vertex $\Gamma^{WW}_{\alpha\beta}$ is
\bea
\Gamma^{\W\W}_{\mu\nu}&=&\frac{1}{s}
\left(\gamma^{\alpha} \bar{g}_V^q-\gamma^{\alpha}\gamma_5 \bar{g}_A^q\right)
V_{\alpha\nu\mu}(q,-k_2,-k_1)+\frac{1}{4 t \ssW^2}\gamma_{\nu}\left(\slashed{p}_2-\slashed{k}_1\right)\gamma_{\mu}(1-\gamma_5)\,,
\eea
with $\slashed{p}\equiv \gamma^\mu p_\mu$ and $\ssW= \sW$ and $e$ being the unit of electric charge. The effective couplings $\bar{g}^q_{V,A}$ are defined as
\be
\bar{g}_V^q=Q^q+\frac{g_V^q\chi}{\ssW^2}\, ,~~
\bar{g}_A^q=\frac{g_A^q\chi}{\ssW^2}\, ,
~~ \chi=\frac{s}{2(s-M_Z^2)}\, ,
\label{effgva}
\ee
where  $g_{V}^q=T_3^q-2Q^q\ssW^2$, $g_{A}^q=T_3^q$ and $T_3^q$ and $Q^q$ are the isospin and electric charge (in unit of $e$) of the quark $q$. The $\chi$ term in \eq{effgva}, which weights the contribution of the virtual $Z$ 
channel, is real since we neglect the $Z$ width contribution. The function $V_{\alpha\nu\mu}(k_1,k_2,k_3)$ is the usual Feynman rule for the trilinear vertex $V_{\alpha}(k_1)~W^+_{\nu}(k_2)~W^-_{\mu}(k_3)$, $V\in\{\gamma ,Z\}$ with all incoming momenta (see \cite{Fabbrichesi:2023cev} for its definition)
and the Mandelstam variables are defined as
\be
s=(p_1+p_2)^2, \quad t=(p_1-k_1)^2, \quad u=(p_1-k_2)^2\,\,.
\ee

From the amplitude in Eq.~(\ref{MDYWW}), summing over the spin of quarks one obtains the compact expression
\be
{\cal M}^{u\bar u}_{WW}(\lambdaA,\lambdaB) \Big[ {\cal M}^{u \bar u }_{WW}(\lambdaAp,\lambdaBp)\Big]^{\dag}\,=\,
\Tr\Big[\bar{\Gamma}^{\W\W}_{\mu\nu}\,\slashed{p}_1\,\Gamma^{\W\W}_{\mup\nup}\,\slashed{p}_2\Big]
 \mathscr{P}^{\mu\mup}_{\lambdaA\lambdaAp}(k_1)
 \mathscr{P}^{\nu\nup}_{\lambdaB\lambdaBp}(k_2)\, ,
 \label{M2WWpol}
\ee
where the symbol $\bar{\Gamma}_{\mu\nu}= \gamma_0 (\Gamma_{\mu\nu})^{\dag} \gamma_0$ and the projector $\mathscr{P}^{\mu\nu}_{\lambda\lambda^{\prime}}(k)$ is given in \eq{proj} with $M=M_W$.

The result for the $d\bar{d}\to W^+W^-$ process follows from
\eq{M2WWpol} through the substitutions 
\be
\gVb \to -\gVbD,\quad \gAb \to -\gAbD,\quad\betaW\to -\betaW
\, ,
\label{transfUD} 
\ee
with the angle $\Theta$ being defined as before  by the anti-quark and $W^+$ momenta. The contribution of strange quark initial states equals that of $d$ quarks in the considered massless limit.

Following the procedure explained in Section \ref{sec:third:toolbox}, from \eq{M2WWpol} (together with the corresponding ones for $d\bar d$ and $s\bar s$ processes) one can compute the unnormalized correlation coefficients $\tilde{f}_a$, $\tilde{g}_a$, and $\tilde h_{ab}$ of the density matrix for the process at and consequently, the expectation value of the operator ${\cal I}_3$  and  the observable $\cmb$.
The explicit expressions for $A^{q\bar q}$, $\tilde{h}_{ab}$ $\tilde{f}_{a}$, and $\tilde{g}_{a}$ as function of $\mWW$ and $\Theta$ can be found in the original work  \cite{Fabbrichesi:2023cev}.

As explained in Section~\ref{sec:Bell}, for the observable ${\cal I}_3$ one can find at each point in the kinematic space the unitary matrices $U$ and $V$ that maximize the violation of Bell inequalities.

\begin{figure}[h!]
  \begin{center}
  \includegraphics[width=3.5in]{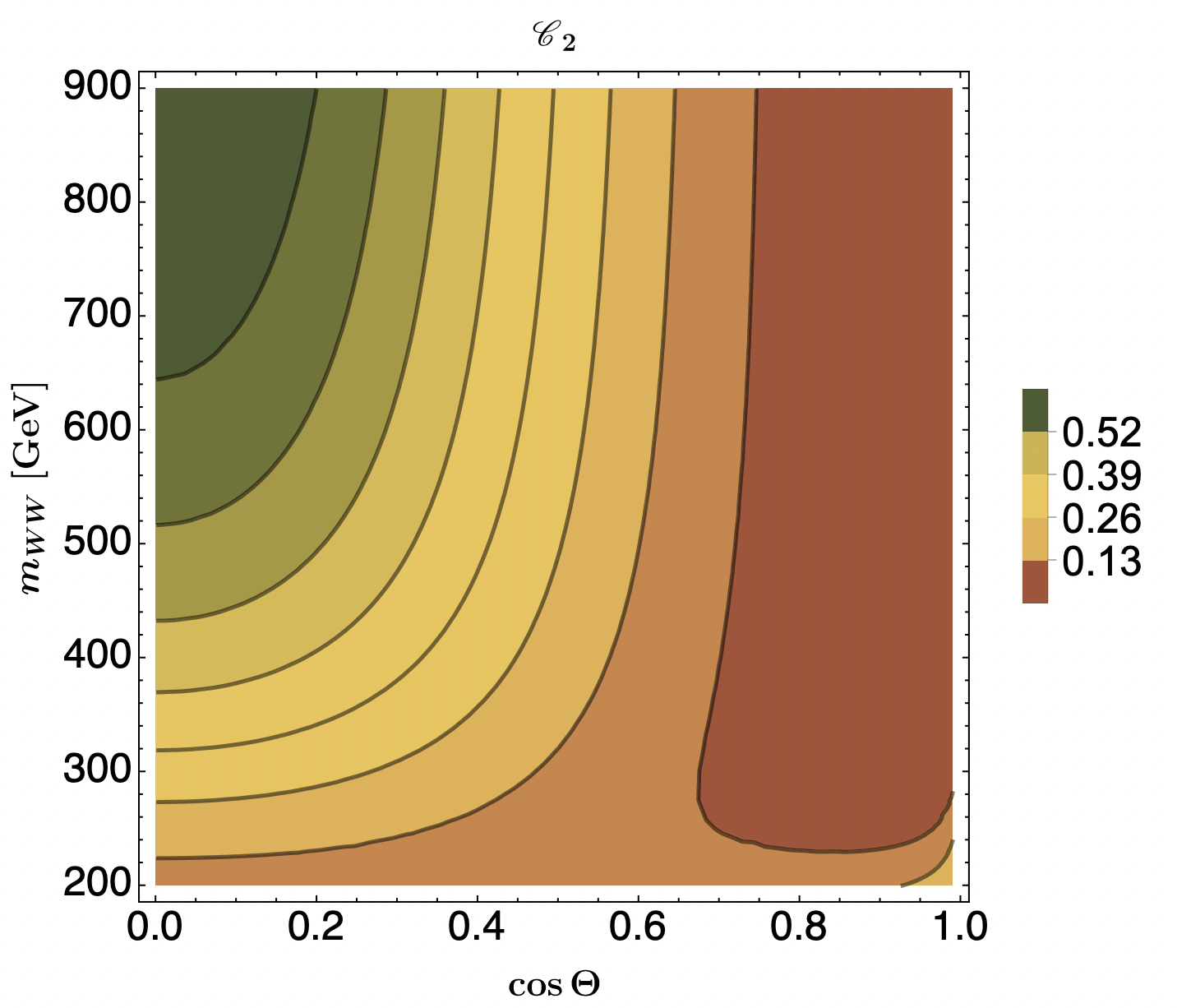}
  \includegraphics[width=3.25in]{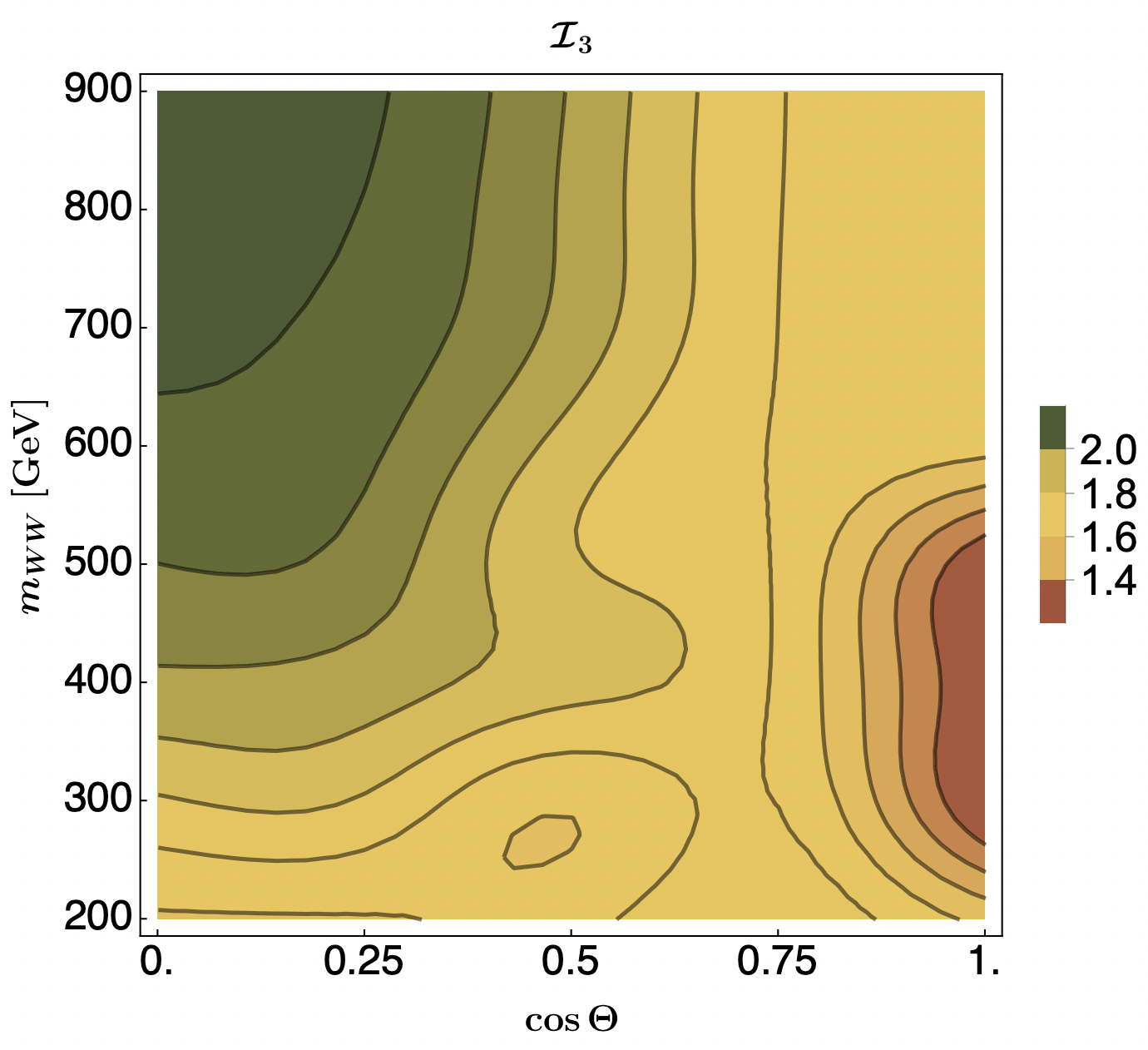}
  \caption{\small The observables $\cmb$ (left plot) and  ${\cal I}_3$  (right plot) for the process $p\;p\to W^+ W^-$ as functions of the invariant mass and scattering angle. Figures revisited from \cite{Fabbrichesi:2023cev} (\href{https://creativecommons.org/licenses/by/4.0/}{CC BY 4.0}). 
  \label{fig:WW} 
  }
  \end{center}
  \end{figure}

The results obtained in \cite{Fabbrichesi:2023cev} for the two observables of interest,  are reported in Fig.~\ref{fig:WW}, as functions of the two kinematic variables $\Theta$ and $\mWW$. Comparable results are obtained in~\cite{Ashby-Pickering:2022umy,Aoude:2023hxv}. From these results we can see that the violation of the Bell inequalities takes place only in a limited range of the kinematic variables, at higher $WW$ invariant mass and for scattering towards the transverse direction. The area in which ${\cal I}_3>2$ is indicated by the lighter-shaded area in  plot on the left of Fig.~\ref{fig:WW}. The explicit expression for the unitary $U$ and $V$ matrices (with accuracy at the percent level) that maximize the Bell observable in this particular kinematic region  can be found in \cite{Fabbrichesi:2023cev}.

The observable  $\cmb$   follows roughly the pattern of ${\cal I}_3$ and reaches the largest values in the upper-left quadrant, thus witnessing the presence of states more entangled than in the rest of the kinematic space. 
This feature can be made manifest by considering the density matrix of the process. For instance, by restricting to the region of maximum entanglement and Bell inequality violation, close to $m_{WW}=900$ GeV and $\cos \Theta=0$, the  polarization density matrix for the $W^+W^-$ states can be approximated
up to terms $O(10^{-3})$ by the following combination of pure state density matrices
\be
\rho= \alpha\, |\Psi_{+-}\rangle \langle \Psi_{+-}| + \beta\, |\Psi_{+-\,0}\rangle \langle \Psi_{+-\,0}| +
\gamma\, |00\rangle\langle 00| + \delta\, |\Psi_{0\,-}\rangle \langle \Psi_{0\,-}|\ ,
\label{density_WW}
\ee
with decreasing weights: $\alpha\simeq 0.72$, $\beta\simeq 0.18$, $\gamma\simeq 0.07$ and $\delta\simeq 0.02$;
the normalization condition $\alpha+\beta+\gamma+\delta=1$ is satisfied within the adopted approximation. The involved pure states are
\begin{eqnarray}
\nonumber
&&|\Psi_{+-}\rangle = \frac{1}{\sqrt{2}} \big( |++\rangle - |--\rangle \big)\ ,\\
&&|\Psi_{0\,-}\rangle = \frac{1}{\sqrt{2}} \big( |0\, -\rangle + |-0\rangle \big)\ ,\\
\nonumber
&&|\Psi_{+-\,0}\rangle = \frac{1}{\sqrt{3}} \big( |++\rangle - |--\rangle +|0\, 0\rangle \big)\ ,\\
\nonumber
\label{WW-states}
\end{eqnarray}
where $|a\, b\rangle= |a\rangle\otimes |b\rangle$ with $a,b\in \{+,\,0,\,-\}$ 
are the polarization states of the two $W$~gauge bosons at rest in the single spin-1 basis. As we can see,  the dominant contribution in (\ref{density_WW}) comes from the entangled pure state $|\Psi_{+-}\rangle$. This can justifies the high value of $\cmb$. However, by retaining all the  terms including the ones of $O(10^{-3})$,  the actual density matrix $\rho$ describes a mixture. This features explains why the corresponding value of $\cmb$, in this corner of the kinematic space, is large but far from maximal.

\subsubsection{Computing the observables: $p\, p \to ZZ$\label{sec:ZZ}}

The tree-level Feynman diagrams contributing to the process
\bea
\bar{q}(p_1) q(p_2) &\to& Z(k_1,\lambda_1) Z(k_2,\lambda_2)\, ,
\label{qqZZ}
\eea
at the parton level are shown in the middle row of Fig.~\ref{fig:DYVV}. We indicate the polarization vectors of the two $Z$ bosons with 
$\varepsilon^{\mu}(k_1,\lambda_1)$ and $\varepsilon^{\nu}(k_2,\lambda_2)$.

The polarized amplitude for the process in \eq{qqZZ} is given by
\bea
{\cal M}^{q \bar q}_{ZZ}(\lambdaA,\lambdaB)&=&-\frac{ie^2}{4\ccW^2\ssW^2} \Big[\bar{v}(p_1) \Gamma^{\Z\Z}_{\mu\nu} u(p_2)\Big]
\varepsilon^{\mu}(k_1,\lambda_1)^{\star}\varepsilon^{\nu}(k_2,\lambda_2)^{\star}\, ,
\label{MDYZZ}
\eea
where $\ccW=\cW$, 
\be
\Gamma^{\Z\Z}_{\mu\nu}\,=\,V^q_{\mu}
\frac{(\slashed{k}_1-\slashed{p}_1)}{u}\,V^q_{\nu}
+
V^q_{\nu}
\frac{(\slashed{k}_1-\slashed{p}_2)}{t}V^q_{\mu}\, ,
\label{GammaZZ}
\ee
and
\be
V^q_{\mu}\,=\,g_V^q\gamma_{\mu} -g_A^q \gamma_{\mu}\gamma_5
\label{VZ}
\ee
with the $g_{V,A}^q$ couplings defined as in \eq{effgva}.

Summing over the quark polarizations and colors we then obtain
\be
{\cal M}^{q \bar q}_{ZZ}(\lambdaA,\lambdaB) \Big[{\cal M}^{q \bar q }_{ZZ}(\lambdaAp,\lambdaBp)\Big]^{\dag}\,=\,
\Tr\Big[\bar{\Gamma}^{\Z\Z}_{\mu\nu}\,\slashed{p}_1\,\Gamma^{\Z\Z}_{\mup\nup}\,\slashed{p}_2\Big]
 \mathscr{P}^{\mu\mup}_{\lambdaA\lambdaAp}(k_1)
 \mathscr{P}^{\nu\nup}_{\lambdaB\lambdaBp}(k_2)\, ,
 \label{M2ZZpol}
\ee
where $\mathscr{P}^{\mu\nu}_{\lambda\lambda^{\prime}}(k)$ is given in \eq{proj} with $M=M_Z$ and the symbol $\bar{\Gamma}_{\mu\nu}$ is defined as in Section \ref{sec:WW}.

The corresponding $f_a,~g_a$ and $h_{ab}$ of the polarization density matrix, have been obtained in \cite{Fabbrichesi:2023cev} and can be derived by following the same procedure as explained in Section \ref{sec:WW}. We report their expressions in appendix B for completeness.

 \begin{figure}[h!]
\begin{center}
\includegraphics[width=3.5in]{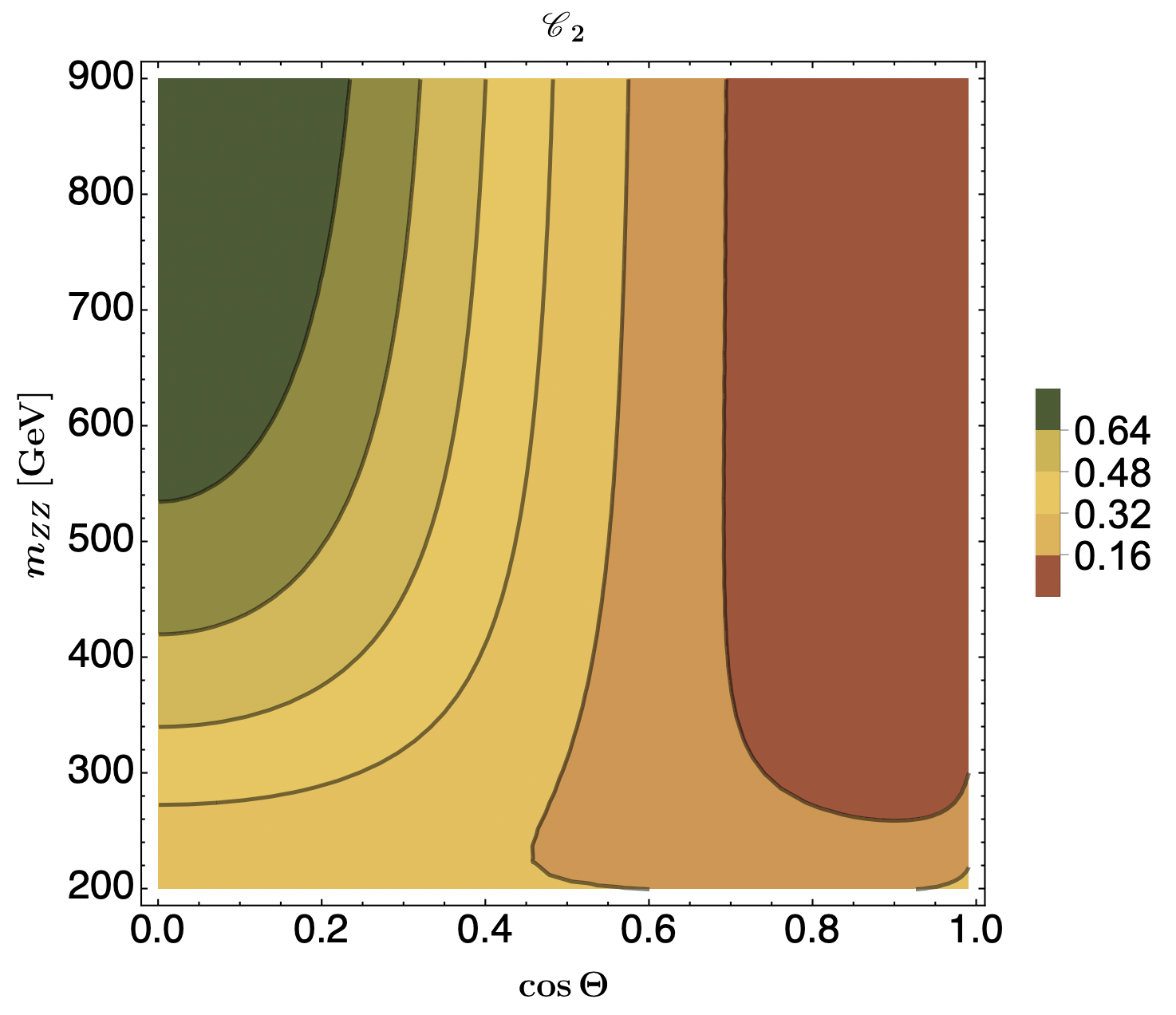}
\includegraphics[width=3.35in]{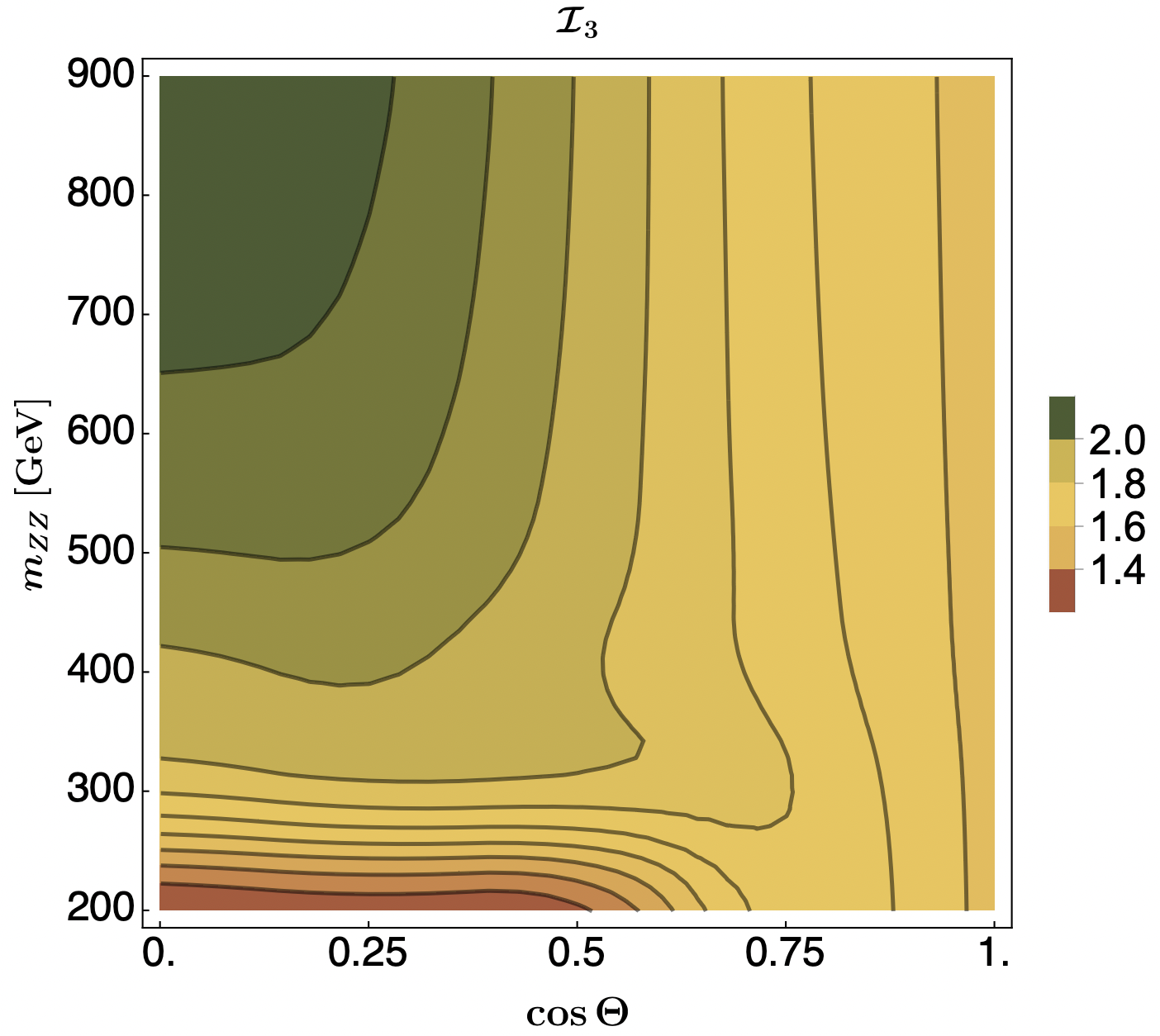}
\caption{\small The observables $\cmb$ (left plot) and ${\cal I}_3$ (right plot)  for the process $p\, p\to Z Z$ as functions of the invariant mass and scattering angle in the CM frame. Figures revisited from \cite{Fabbrichesi:2023cev} (\href{https://creativecommons.org/licenses/by/4.0/}{CC BY 4.0}). 
\label{fig:ZZ} 
}
\end{center}
\end{figure}

Fig.~\ref{fig:ZZ} shows the analytic results for the entanglement observables computed in \cite{Fabbrichesi:2023cev}. As we could see from these results, the violation of the Bell inequalities for the $ZZ$ production takes place only in a limited range of the kinematic variables.

The observable $\cmb$ follows the pattern of ${\cal I}_3$---as it does in the case of the $W^+W^-$ final states---and again reaches the largest values in the upper-left quadrant. In this region it witnesses the presence of states more entangled than in the rest of the kinematic space. 

\subsubsection{Monte Carlo simulations  and predictions}

Monte Carlo simulations of diboson production at the LHC has been performed in \cite{Ashby-Pickering:2022umy} and \cite{Aoude:2023hxv}. The \textsc{MadGraph5-\_aMC@NLO}~\cite{Alwall:2014hca} software is used including spin correlations and relativistic and Breit-Wigner effects. Events are generated at the leading order at CM energy of 13 TeV, and the 4-lepton final states considered.

Entanglement is proposed to be measured through the observable $\cmb$, \eq{C_2}, which provides a lower bound on the concurrence, and Bell inequality  by means of the expectation value ${\cal I}_3$ of a version of the Bell operator (\ref{B}), which is optimized along Cartesian planes. In agreement with the analytic results, entanglement is expected to be detected in the kinematic region of large scattering angles for invariant masses above 400 GeV for the $WW$ and $ZZ$ final states. For the tested observables Bell inequality violation is not predicted to reach a significant level even for a luminosity of 3 ab$^{-1}$ (Hi-Lumi) once the statistical uncertainty is taken into account.

\subsection{Higgs boson decays into $ WW^*$ and $ZZ^*$} \label{sec:qutrits:higgs}
The qutrits system of two massive gauge bosons is generated 
by the  decay of the SM Higgs boson
\be
h\to V(k_1,\lambda_1)\, V^*(k_2,\lambda_2)\, ,
\label{HVV}
\ee
with $V\in\{W,Z$\}, and $V^*$ regarded as an off-shell vector boson. We can treat the latter as an on-shell particle characterized by a fictitious mass 
\be
M_{V^*}= f M_V\, ,
\ee
which is the original mass $M_V$ reduced by a factor $f$, with $0<f<1$.
The Higgs boson is  produced at the LHC as a resonance in the $s$-channel.

The theoretically expected quantum entanglement and Bell inequality violation for the processes $h\to WW^*$ have been studied in \cite{Barr:2021zcp,Ashby-Pickering:2022umy}, and those for $h\to ZZ^*$  in \cite{Aguilar-Saavedra:2022wam,Ashby-Pickering:2022umy}.  Comparable results have been obtained in \cite{Fabbrichesi:2023cev} by using  analytical results for the polarization density matrix of the two gauge bosons in the helicity basis. We summarize here first the analytical results of \cite{Aguilar-Saavedra:2022wam,Fabbrichesi:2023cev} for the polarizations coefficients  and its implications for  quantum entanglement and Bell inequality violation observables. The corresponding results obtained by Monte Carlo simulation of events  are briefly discussed in the next Section~\ref{sec:qutrits:higgs:events}.

\subsubsection{Computing the observables}
\label{sec:qutrits:higgs:observables}
\begin{figure}[h!]
\begin{center}
\includegraphics[width=4.5in]{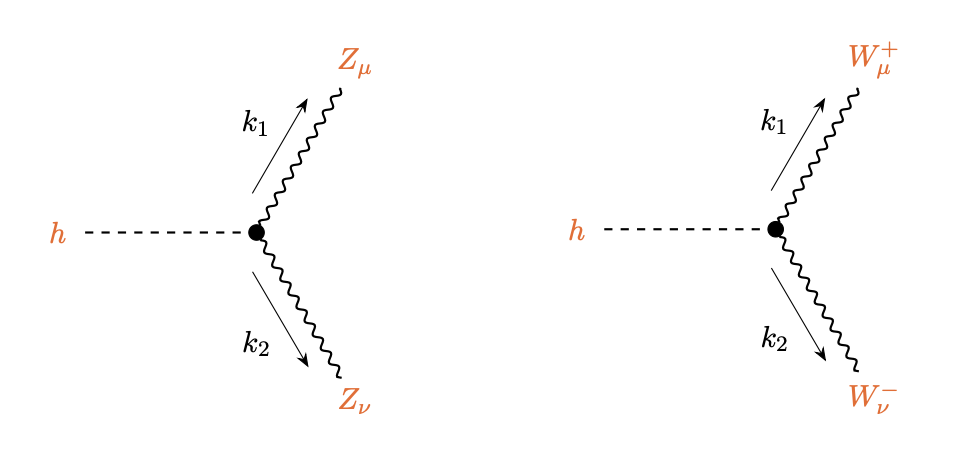}
\caption{\small Feynman diagrams for the decay of the Higgs boson into a pair of massive gauge bosons.
\label{fig:HtoVV} 
}
\end{center}
\end{figure}
The polarized amplitude for the Higgs boson decay in \eq{HVV}) is given by 
\bea
{\cal M} (\lambdaA,\lambdaB)=g \,  M_V\, \xi_V \, g_{\mu\nu}
\varepsilon^{\mu\star}(k_1,\lambdaA)\varepsilon^{\nu\star}(k_2,\lambdaB)
\, ,
\label{MHVV}
\eea
where $g$ is the weak coupling, $\xi_W=1$, and $\xi_Z=1/\cos{\theta_W}$ with $\theta_W$ the Weinberg angle. From the amplitude in \eq{MHVV} we obtain
\bea
{\cal M} (\lambdaA,\lambdaB) {\cal M}(\lambdaAp,\lambdaBp)^{\dag} &=&
g^2 \, M_V^2 \, \xi^2_V \, g_{\mu\nu}g_{\mup\nup}  \mathscr{P}^{\mu\mup}_{\lambdaA\lambdaAp}(k_1)
 \mathscr{P}^{\nu\nup}_{\lambdaB\lambdaBp}(k_2)\, .
\eea
where $\mathscr{P}^{\mu\nu}_{\lambda\lambda^{\prime}}(k)$ is given in \eq{proj} with $M=M_V$ or $M=M^*_V$ for the on-shell and off-shell boson, respectively.

Following the procedure explained in section~\ref{sec:toolbox} for a CM energy $\sqrt s=m_h$, one can obtain the coefficients $f_a$, $g_a$, and $h_{ab}$ ($a,b\in\{1, \dots,8\}$). These coefficients have been computed in \cite{Fabbrichesi:2023cev} and their  expression can be found in Appendix
\ref{sec:appendix:qutrits:HZZ}.
No dependence is expected of these coefficients on the scattering angle $\Theta$ because we are considering the decay of the scalar Higgs boson at rest. 

The main theoretical uncertainty affecting the correlation coefficients in \eq{fghHiggs} is due to the missing next-to-leading electroweak corrections to the tree-level values.
In \cite{Fabbrichesi:2023cev} it was estimated that the error induced by these missing corrections yields at most a few percent of uncertainty on the main entanglement observables, in the relevant kinematic regions in which one of the two electroweak gauge boson are on-shell. This expectation is based on the fact that these corrections give a 1-2\% effect on the total width ~\cite{Boselli:2015aha}.
Corrections for these effects can and should be applied when making actual experimental measurements. 

The polarization density matrix $\rho$ for the two vector bosons emitted in the decay of the Higgs boson is calculated to be~\cite{Fabbrichesi:2023cev}
\be
\rho = 2 \begin{pmatrix} 
  0 & 0 & 0 & 0 & 0 & 0 & 0 & 0 & 0  \\
  0 & 0 & 0 & 0 & 0 & 0 & 0 & 0 & 0  \\
  0 & 0 &  h_{44} & 0 &  h_{16} & 0 & h_{44} & 0 & 0  \\
  0 & 0 & 0 & 0 & 0 & 0 & 0 & 0 & 0  \\
  0 & 0 &  h_{16} & 0 & 2\, h_{33} & 0 & h_{16} & 0 & 0  \\
  0 & 0 & 0 & 0 & 0 & 0 & 0 & 0 & 0  \\
  0 & 0 &  h_{44} & 0 &  h_{16} & 0 &  h_{44}& 0 & 0  \\
  0 & 0 & 0 & 0 & 0 & 0 & 0 & 0 & 0  \\
  0 & 0 & 0 & 0 & 0 & 0 & 0 & 0 & 0  \\
\end{pmatrix} \, ,
\label{rhoH}
\ee
with the condition $\Tr[\rho_H]=1$ following from the relation $4(h_{33}+h_{44})=1$. There are therefore only two independent coefficients under these assumptions. 

In \cite{Aguilar-Saavedra:2022wam} the spin density matrix of the system is written in terms of tensor components $T^L_M$ (see Section~\ref{sec:toolbox}) and the  correlation coefficients entering the density matrix  are indicated as $C_{L_1,M_1,L_2,M_2}$ (see \eq{eq:rho-tensor2}). These coefficients are related to those in  \eq{rhoH} by the correspondence
\be
\frac{1}{6} C_{2,2,2,-2}= h_{44}  \quad \text{and}  \quad \frac{1}{6} C_{2,1,2,-1} =h_{16}  \, . \label{Ctensors}
\ee
Assuming that the state is pure, there would be entanglement if and only if the two components in \eq{Ctensors} are different from zero.

Although some $f_a$ and $g_a$ are non-vanishing, the dependence of $\rho_H$ on these quantities cancels in the final expression. Furthermore, due to the following identity among the correlation coefficients 
$ h_{44} = 2 \left(h_{16}^2 + 2 h_{44}^2\right)$
the above polarization density matrix is idempotent
\bea
\rho^2=\rho\, ,
\label{rhorel}
\eea
as expected from the assumption that the final $VV^*$ state is a pure state. The density matrix in \eq{rhoH} can then be written as ~\cite{Aguilar-Saavedra:2022wam}
\be
\rho  = |\Psi \rangle \langle \Psi | \, ,
\ee
where (in the basis $|\lambda\, \lambdap\rangle = |\lambda\rangle\otimes|\lambdap\rangle$ with $\lambda,\lambdap\in\{+,0,-\}$)  
\be
|\Psi  \rangle = \frac{1}{\sqrt{2 +  \varkappa^2}} \left[  |{\small +-}\rangle -  \varkappa \,|{\small 0\, 0}\rangle + |{\small -+}\rangle  \right] \label{pure}
\ee
with
\be
 \varkappa = 1+ \frac{m_h^2 - (1+f)^2 M_V^2}{2 f M^2_V} 
\ee
and $ \varkappa=1$ corresponding to the production of two gauge bosons at rest.

\begin{figure}[h!]
\begin{center}
\includegraphics[width=3in]{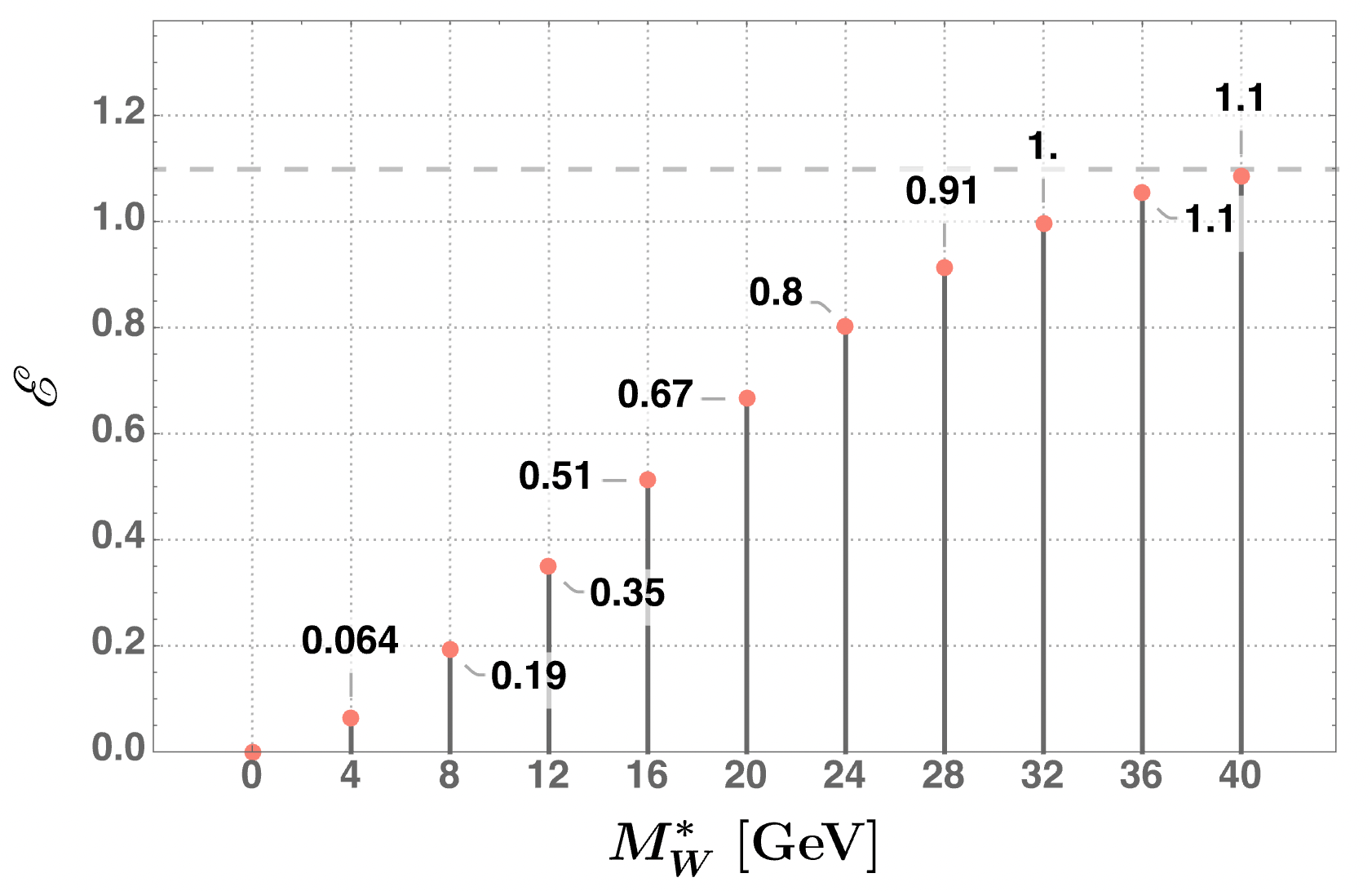}
\includegraphics[width=3in]{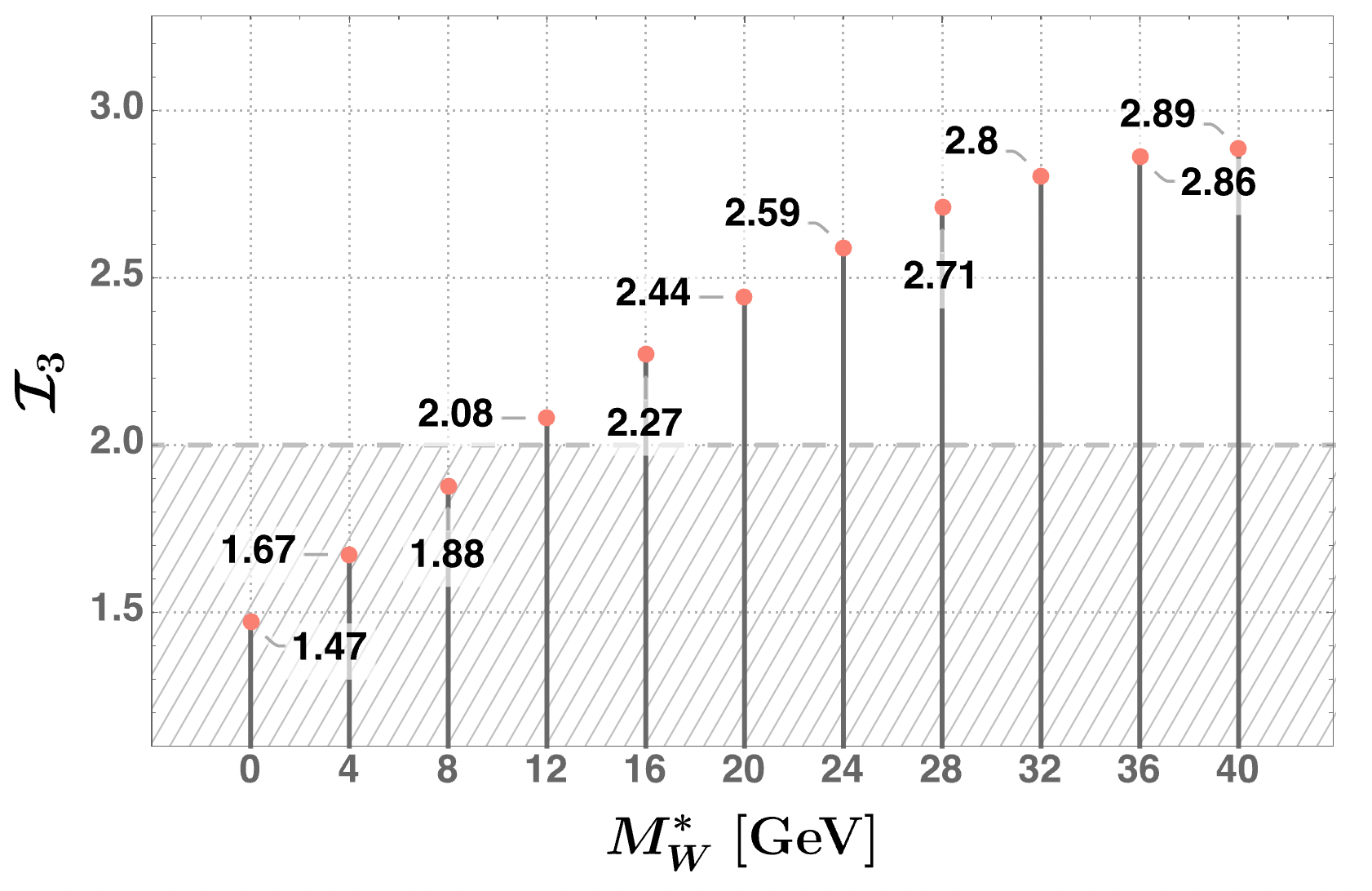}
\caption{\small Predictions of the entropy of entanglement ${\cal E}$ (left plot) and the Bell operator expectation value ${\cal I}_3$ (right plot) for the pair production of $W$ bosons in Higgs boson decays as functions of the virtual $W^{*}$ mass in the range $0<M_{W^*}<40$~GeV~\cite{Fabbrichesi:2023cev}. The dashed horizontal line in the right-hand side plot marks the Bell-inequality violation condition ${\cal I}_3>2$. The dashed line in the left-hand side plot denotes the maximum value 
of $\ln 3$. Figures revisited from \cite{Fabbrichesi:2023cev} (\href{https://creativecommons.org/licenses/by/4.0/}{CC BY 4.0}).
\label{fig:HWW}
}
\end{center}
\end{figure}

\begin{figure}[h!]
\begin{center}
\includegraphics[width=3in]{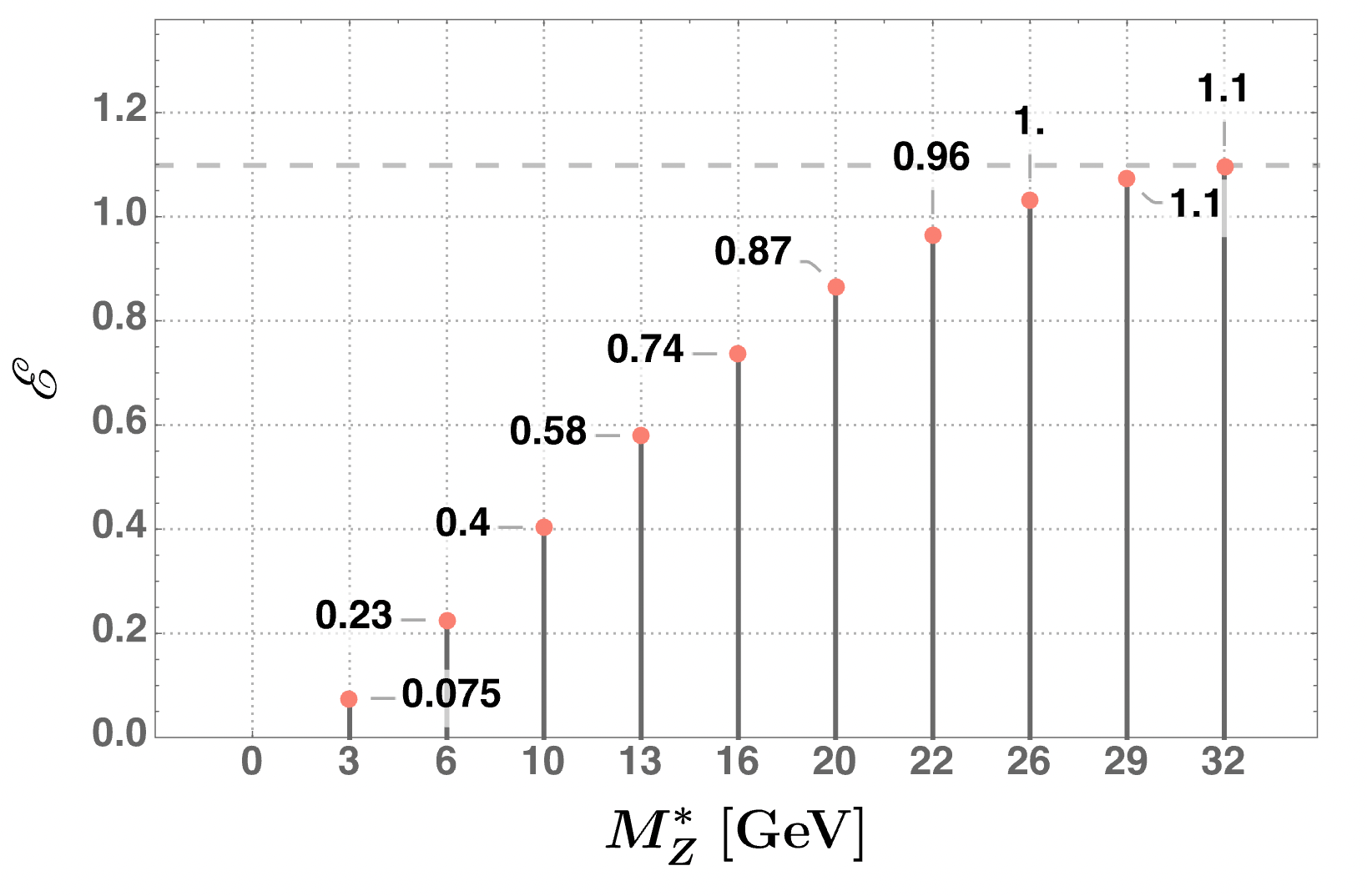}
\includegraphics[width=3in]{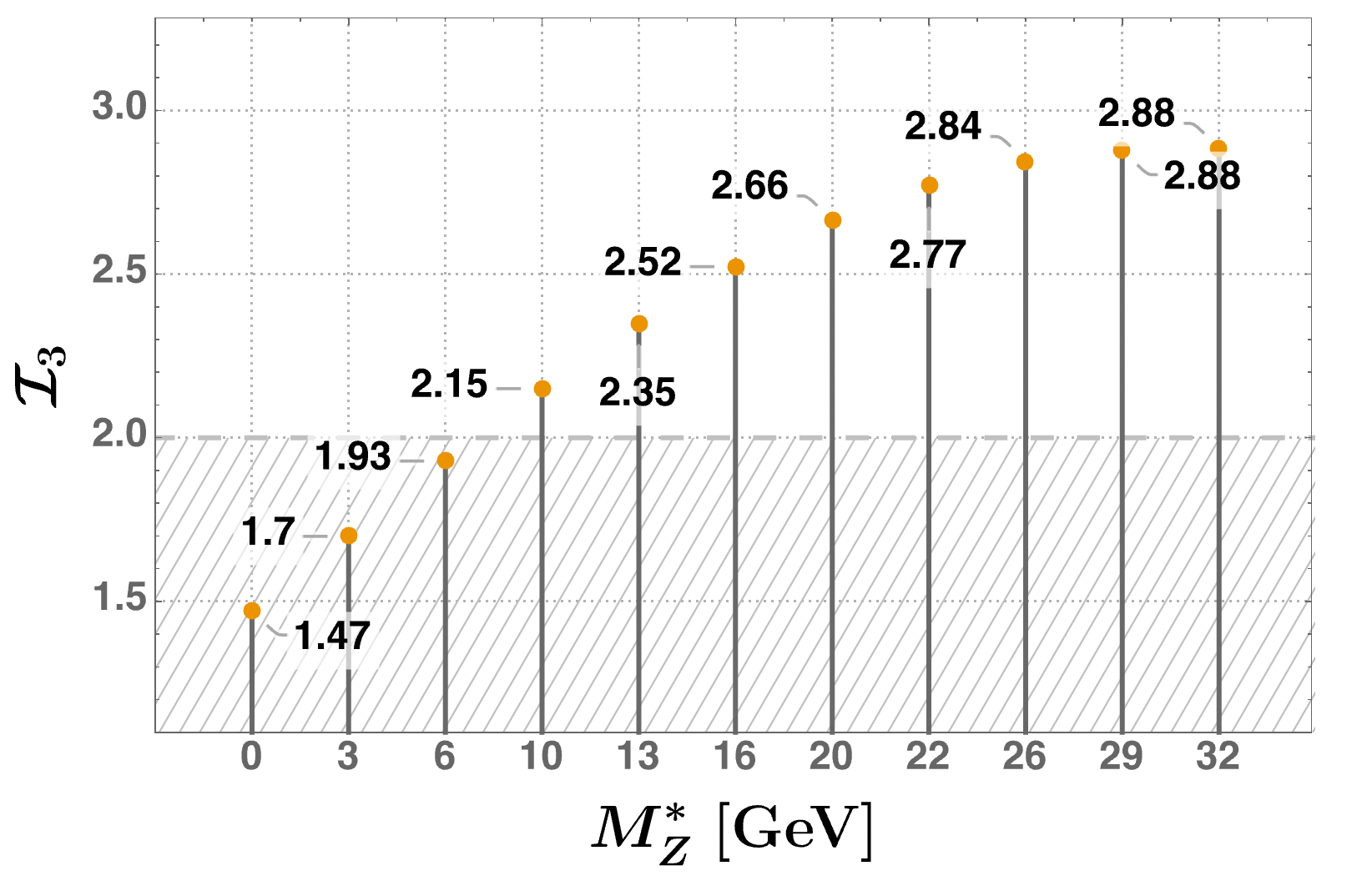}
\caption{\small Predictions of the entropy of entanglement ${\cal E}$ (left plot) and the Bell operator expectation value ${\cal I}_3$ (right plot) for the pair production of $Z$ bosons in Higgs boson decays as functions of the virtual $Z^{*}$  mass in the range $0<M_{Z^*}<32$~GeV~\cite{Fabbrichesi:2023cev}. The dashed line in the left-hand side plot denotes the maximum value 
of $\ln 3$. Figures revisited from \cite{Fabbrichesi:2023cev} (\href{https://creativecommons.org/licenses/by/4.0/}{CC BY 4.0}). \label{fig:HZZ} 
}
\end{center}
\end{figure}

If one makes the assumption that the diboson system is described by a pure state, then one can measure its entanglement through the  entropy of entanglement defined in \eq{entropy}. This quantity is plotted in Figs.~\ref{fig:HWW} and \ref{fig:HZZ} as a function of the mass of virtual $W$ or $Z$ boson \cite{Fabbrichesi:2023cev}. As we can see from these calculations, the entropy of entanglement is expected to reach its maximum at the kinematic threshold, signaling a maximally entangled state. The dependence of the polarization entanglement on the  mass of the virtual state  is due  the contribution of the longitudinal polarization, parametrized by the coefficient $\varkappa$ in \eq{pure}. Indeed, this contribution starts out bigger and decreases  to 1  at the threshold. The value of 1 corresponds to a pure singlet state and thus to the maximum in the entanglement of the state. 

\begin{figure}[h!]
\begin{center}
\includegraphics[width=4in]{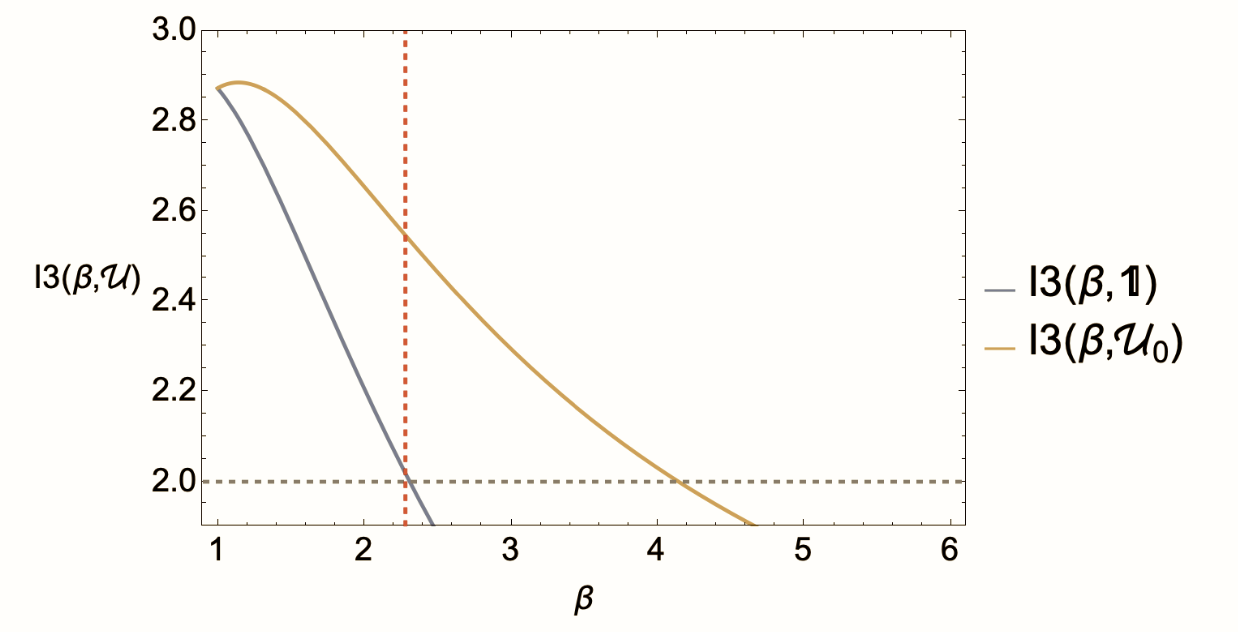}
\end{center}
\caption{\label{optimal_operator_ZZ}
Expectation values of unoptimised (lower curve) and unitary-operator optimised (upper curve) Bell operators for $H\rightarrow ZZ$ decays as a function of the relative coefficient $\beta$ ($\varkappa$ in \eq{pure}) of the longitudinal polarisation of the state. Figure from~\cite{Aguilar-Saavedra:2022wam} (\href{https://creativecommons.org/licenses/by/4.0/}{CC BY 4.0}).
}
\end{figure}

The maximization of the ${\cal I}_{3}$ observable, which depends in this case only on the $M_V^{*}$ mass, is obtained  through the unitary rotation in \eq{uni_rot} of the $\mathscr{B}$ matrix in \eq{B}, that maximizes the value of the corresponding expectation value.
This maximization must be performed point by point as the density matrix varies with $M_V^{*}$. The unitary matrices that maximizes the ${\cal I}_{3}$ observable in the last bins (in which $M_{W^*}=40$ GeV and $M_{Z^*}=32$ GeV) for the $h\to WW^*$ and $h\to ZZ^*$ decays are given in \cite{Fabbrichesi:2023cev}.

The plots on the left-hand side in Figs.~\ref{fig:HWW} and~\ref{fig:HZZ}  nicely show that the value of the entropy of entanglement \eq{entropy} is expected to decrease as the pure state in \eq{rhoH}  becomes less and less entangled, for decreasing values of $M^*_V$.
The advantages obtained from using the optimal Bell operator can be seen in Figure~\ref{optimal_operator_ZZ}. The relative coefficient $\beta$ of the longitudinal component to the polarisation increases is unity when the when the $Z$ bosons are produced at rest in their zero-momentum frame (ZMF), and increases at larger relative momenta.  

The same  Figs.~\ref{fig:HWW} and \ref{fig:HZZ}  show the calculations  for the  Bell operator expectation value ${\cal I}_{3}$ (right panels)  in the $h\to WW^*$ and $h\to ZZ^*$ decays. The plots are for different values of the virtual gauge boson masses $M_{{W}^{*}}$ and $M_{{Z}^{*}}$, respectively. 
The Bell inequality violation in $WW^*$ and
$ZZ^*$ final states starts above 12 GeV and 10 GeV for $M_{W^*}$  and $M_{Z^*}$ invariant masses respectively,  reaching its maximum allowed value of order ${\cal I}_{3} \sim 2.9$  at the largest invariant mass of the corresponding off-shell gauge boson.

\subsubsection{Monte Carlo simulations  and predictions}
\label{sec:qutrits:higgs:events}
The simulation for the process $h\to W W^*$ has been performed in~\cite{Barr:2021zcp} in which most of the tools for the analysis of qutrit systems (as discussed  in Section \ref{sec:toolbox}) were introduced as well. The \textsc{MadGraph5\_aMC@NLO}~\cite{Alwall:2014hca} software is used including spin correlations and relativistic and Breit-Wigner effects. The Bell operator ${\cal I}_3$ is optimized along Cartesian planes. Only the fully leptonic decays are considered.  There are two neutrinos in the final state and the reconstruction of the rest frame of each gauge boson necessarily introduces a potentially large uncertainty. Various scenarios about the overall uncertainty are discussed (by attributing a smearing in the value of the lepton momenta) and the significance for the Bell inequality violation shown to vary from about 5$\sigma$ (for the most optimistic momenta reconstruction) to 1$\sigma$ (for a less sanguine one) at the luminosity of 140 fb$^{-1}$ at the LHC. An analysis is also presented in~\cite{Aguilar-Saavedra:2022mpg}.

The same decay is discussed in \cite{Fabbri:2023ncz} by looking at 
the semi-leptonic decay $h \to jj \ell \nu_\ell$ (rather than the fully leptonic one).  The momentum from the $s$-jet (identified via the $c$-tagging of the companion jet) is used to measure the polarization of one of the two $W$-bosons. It has been shown that the efficiency
of the jet tagging and the decreased uncertainty in the single neutrino momentum may improve the polarization reconstruction.

All these analyses must be taken with a grain of salt since the final state $WW^*$ is hidden inside a large background that makes generally hard to select the events of the signal. 

The process $h\to ZZ^*$ has been simulated and analyzed using tensor~\cite{Aguilar-Saavedra:2022wam} and Gell-Mann~\cite{Ashby-Pickering:2022umy} bases. There are no neutrinos in the final state and the rest frame of the gauge bosons can be reconstructed with precision. The basis that maximizes  the Bell operator is explicitly written out in~\cite{Aguilar-Saavedra:2022wam}. The \textsc{MadGraph5\_aMC@NLO}~\cite{Alwall:2014hca} software is used to generate the events. It is found that, for a luminosity of 3 ab$^{-1}$ (Hi-lumi at the  LHC),  the significance for the violation of the Bell inequality  can be as large as 4.5$\sigma$. 
This process is actually the most promising  to test the Bell inequality in weak boson decays because of the clean reconstruction and low background. 

\subsection{Vector-boson fusion}

Processes in which  vector-boson fusion takes place, as in
\be
W^+W^- \to W^+W^-\, , \quad Z\gamma \to W^+W^- \quad \text{or} \quad \gamma \gamma \to W^+W^-
\ee
have been analyzed in \cite{Morales:2023gow} by means of the computation of the corresponding tree level amplitudes within the SM. It is interesting that this family of scattering process contains final states with two quibits (photons), one qubit and one qutrit (photon and massive gauge bosons) and two qutrits (massive gauge bosons). 

As before for other process, the amount of entanglement depends on  phase space. More or less all channels share a comparable amount of entanglement but for the $ZZ\to ZZ$, whose entanglement is suppressed. 

The violation of Bell inequality can be tested in vector-boson fusion by measuring the expectation value of the appropriated Bell operator in regions of the phase space that are identified and listed in \cite{Morales:2023gow}.

\newpage
	\section{Possible loopholes in testing Bell inequalities at colliders}\label{sec:loopholes}

As pointed out in the Section~\ref{history} of the Introduction, soon after  the first test of a Bell inequality was performed, ways to escape the consequences were put forward. Since then, these attempts have been grouped together  under the  label of `loopholes'.

The existence of a \textbf{loophole} in the test of a Bell inequality shows how to avoid the exclusion of deterministic, local theories even in the presence of an experimentally verified  violation of the inequality. A violation of the inequality that is free of loopholes excludes these theories and confirms quantum mechanics. If the test is open to one or more loopholes, the possibility of a description in terms of local, deterministic models is, in principle, still possible. 

 The discussion of  loopholes has taken place  so far mostly in the framework of experiments in optics and atomic physics. It is important to bear in mind that (almost) all possible  loopholes have been closed in low-energy tests with photons~\cite{Giustina:2015yza,Hensen:2015ccp}
and in atomic physics~\cite{PhysRevLett.119.010402}. This means that devising a local hidden variable model -- be it deterministic or stochastic -- exploiting any or all of these loophole is nowadays  a formidable if not indeed impossible task.

The extension to collider physics of any discussion about possible loopholes  is delicate and still little explored~\cite{Severi:2021cnj,Ehataht:2023zzt}. The implications for collider experiments of Bell-violation measurements and considerations of the possible dependencies on hidden variables was recently considered in the philosophy of physics literature~\cite{Timpson23}.

We note that the existence of a loophole does not mean that a test of the Bell inequality is useless or meaningless. The test and the loophole are two distinct entities and the existence of a loophole only implies that there exists, in principle, a way to bypass the ruling out of locally local hidden variable models. At the same time, the hypothetical model, required to exploit the loophole, is necessarily made rather complicated and unnatural by its  accounting for the violation. Indeed, all these models have to satisfies  so involved a series of requirements that they are very difficult to conceive and very few of them have even been actually defined.\footnote{Bohm's pilot wave theory~\cite{Bohm:1951xw,Bohm:1951xx}, perhaps the best known example of a hidden-variable model,  yields explicitly nonlocal dynamics for the hidden variables.}

Contrary to  experiments at low energies, those at colliders were not designed to test Bell violation and  therefore seem more prone to loopholes and other shortcomings. Nevertheless, as we discuss below, most loopholes appear to be closed already by the current most common settings of collider detectors.

The potential loopholes that could be  present in any test of  Bell inequality are:

\begin{itemize}
\item[-] {\bf Detection loophole} \cite{Pearle:1970zt}: If the efficiency in detecting the entangled states is not 100\%, the undetected states could, had they been taken into account, restore the inequality;
\item[-] {\bf Locality loophole} \cite{Bell:1964}: Bell locality, even if satisfied, could be bypassed if it is possible for the entangled states to communicate by means of a local interaction;
\item[-] {\bf Coincidence loophole} \cite{Larsson:2003efo}: The  states are misidentified and do not belong to the entangled pair;
\item[-] {\bf Freedom of choice loophole} \cite{Bohm:1957zz}: The lack of freedom in choosing the measurement to be performed alters the outcome;
\item[-] {\bf Super-determinism loophole} \cite{Larsson_2014}: if the initial conditions fully predict all successive developments, possible experiments included, Bell locality is always satisfied.
\end{itemize}

How do these loopholes affect a test of  Bell inequality at colliders?
\begin{itemize}
\item The detection loophole is  always present at colliders where only a small fraction of the final states are actually recorded. Here one must appeal to the assumption of having a fair sampling of these events. This is what is routinely assumed in high-energy physics since also a  measurement of a cross section or branching ratio would be open to the same loophole. 
Given such an assumption, due to the high efficiency of the detectors at colliders, as far as the measure of the momenta of charged particles,  the detection loophole might be closed. For qubits the loophole would be closed if the efficiency were more than about 80\%~\cite{Clauser:1969ny} and this requirement is even lower for states belonging to larger Hilbert spaces~\cite{Vertesi:2010zvq}. By comparison, the efficiency of the LHCb detector is more than 90\%~\cite{Dordei:2017rtt} for kaon, pion and muon identification. However analysis selection efficiencies would also need to be considered.

\item The locality loophole is potentially present for states made of particles  that end up decaying with a relative time-like interval, either because they decayed at different times or because they do not move apart fast enough. It could be particularly serious in the case of charged particles for which the electromagnetic interaction can be used in bypassing the test. Fig.~\ref{fig:loophole1} shows the kinematics exploited by the locality loophole. To close the locality loophole it is desirable to consider decays in which the produced particles are identical,  and therefore their life-times are also the same. Even in this case, the actual decays take place with an exponential spread. To take this into account, one must verify that the majority of the events do take place  separated by a space-like interval and/or weed out those that do not.

\begin{figure}[h!]
\begin{center}
\includegraphics[width=3in]{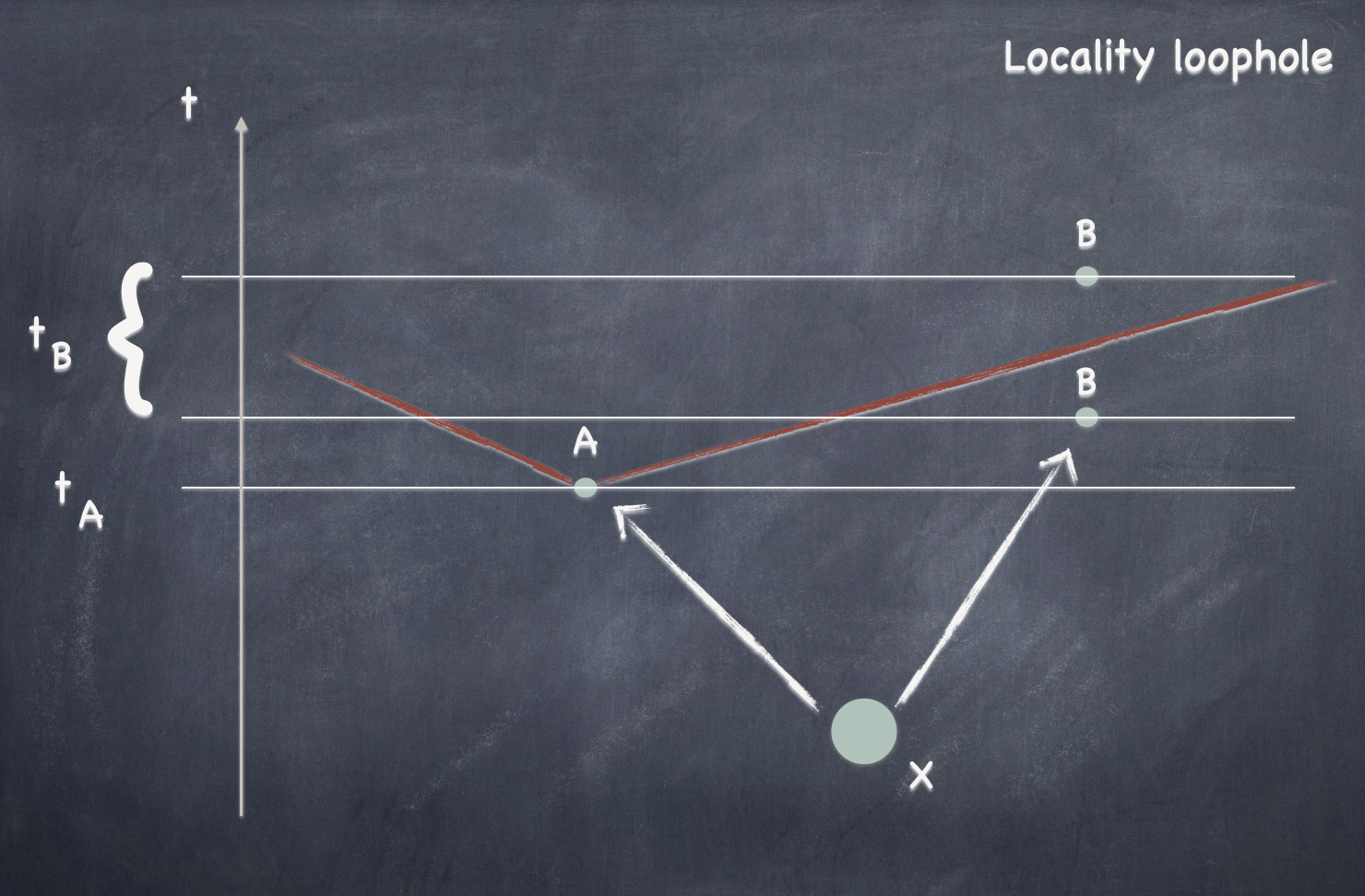}
\caption{\small Kinematics of the locality loophole. Particle B can decay within the future cone of particle A either because of a longer lifetime or because the random spread in its decay time.
\label{fig:loophole1} 
}
\end{center}
\end{figure}

Fig.~\ref{fig:loophole2} shows a typical distribution of decays events as a function of their relative distances. The relative velocity $v$ with which the  pair flies apart is sufficiently large to create, at the times $t_1$ and $t_2$ of decay, a space-like separation iff
\be 
\frac{|t_1-t_2|\, c}{(t_1+t_2)\, v} < 1\, .
\ee
The separation prevents local interactions (as those arising through the exchange of photons between charged particles) and ensures that the locality loophole is closed~\cite{Bell:1987hh}. The selection of these events could be implemented with a suitable cut on the relative momentum of the two particles. If the amount of available data is large and the  fraction of pairs rejected  by the cut is small, this refinement would not affect the significance of the Bell test under consideration.

\begin{figure}[h!]
\begin{center}
\includegraphics[width=4in]{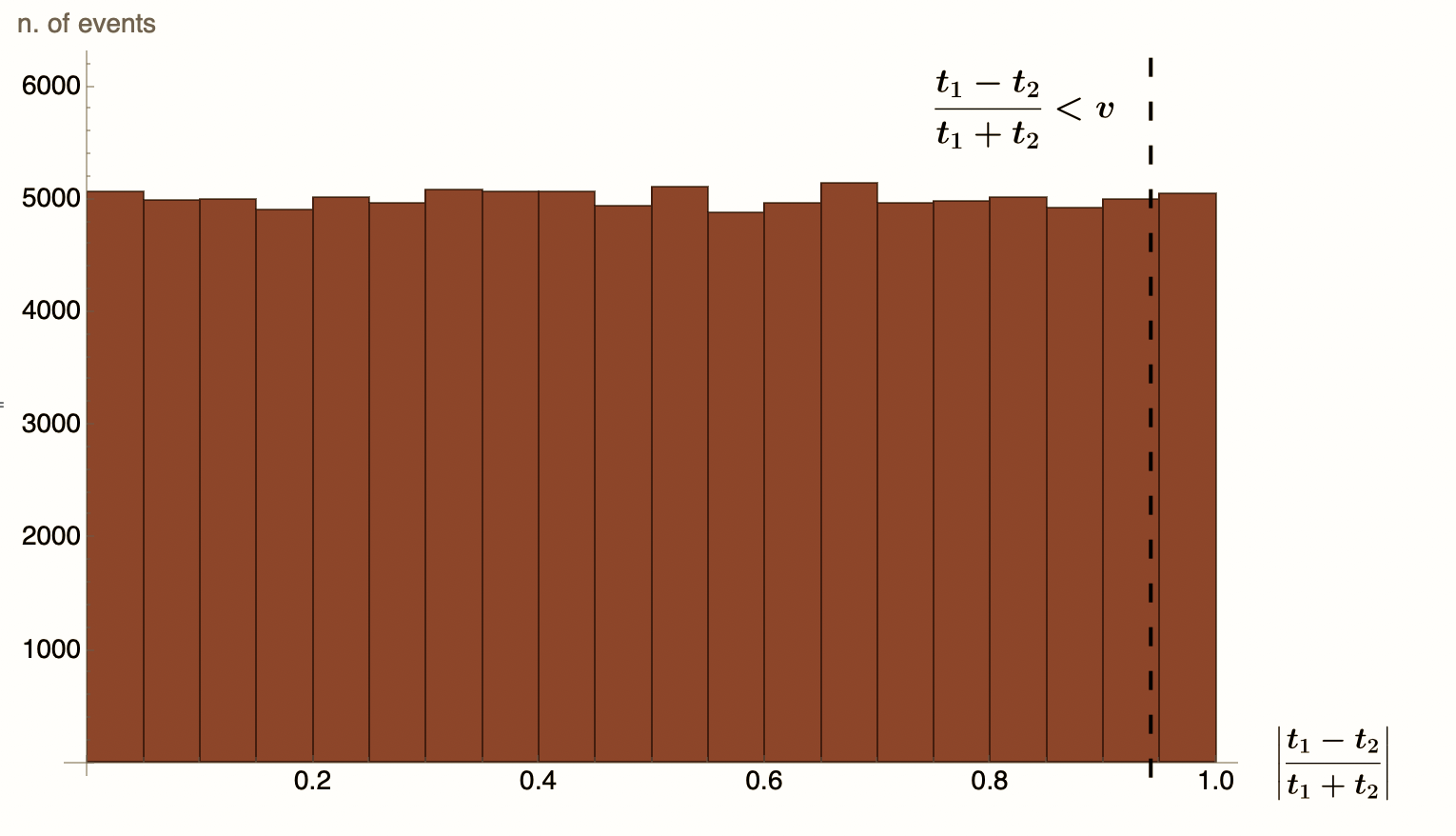}
\caption{\small Example about the fraction of events separated by a space-like interval (95\% in this histogram).  Histogram of the number of events as a function of the ratio $|t_1-t_2|/(t_1+t_2)$ between the difference and the sum of the decay times of the two taus. The events have been generated by $10^5$ pseudo-experiments in which the decay times are randomly varied within an exponential distribution. The black-dashed vertical line distinguishes events separated by a time-like interval (to the right of the line) from those that are space-like separated (to the left of the line). Figure revisited from \cite{Ehataht:2023zzt} (\href{https://creativecommons.org/licenses/by/4.0/}{CC BY 4.0}) 
\label{fig:loophole2}}
\end{center}
\end{figure}

\item The coincidence loophole does not seems to be problematic at colliders. Such a misidentification is always accounted for in the quoted uncertainty in the results of the experiments. 

\item The freedom-of-choice loophole relates to the possible dependence of  
is---depending on whom you ask---either the hardest or the simplest to close at a collider setting. At low-energy experiments the loophole is addressed by coupling the polarization measurement to a (pseudo)random choice that is made after the entangled states have been produced, and with a space-like separation at the point of `choice'. This is not possible at colliders where the detector is fixed by its construction design. Though this seems to be a show stopper, we have advanced an alternative point of view:  the polarization measurement is made inside the detector by the particles themselves as they decay into the final state; because the decay is a quantum process, it is the ultimate random process and one could argue that therefore the freedom of choice is implemented. It can be argued that the objection  that the quantum theory one would like to put to the test is employed in closing the loophole can be extended also to the (pseudo)random choice in the loophole-free low-energy setting. Be that as it may, as our brief discussion reveals, the physics surrounding this loophole is not settled yet and needs further discussion.

\item The `super-determinism' loophole is related; a dependence of the measurement outcomes on information in the overlapping past light-cones of the respective measurements can break the assumed form of the probability distribution~\eq{Bell-locality}. This loophole cannot be closed at colliders, nor can it be be closed in atomic-physics experiments or, indeed, at all. 
\end{itemize}

The discussion of the role of loopholes in the violation of Bell inequality at high energies is still at its first steps~\cite{Timpson23}. It is fair to say that models exploiting these loopholes to save local hidden variable theories -- either deterministic or stochastic -- will become even harder to define once the violation will be extended at colliders and in the presence of strong and electroweak forces because they will have to account for both the low- and the high-energy experiments. We are not aware of any definite model claiming to be able to achieve this. 

\newpage
\section{Probing new particles and fields with entanglement}\label{sec:newphysics}
	
The sensitivity of entanglement on the specific form of the couplings between the states produced at colliders makes it a promising observable to be used in gaining sensitivity to new physics -- particles, fields and interactions -- beyond the SM. The overall advantage in sensitivity with respect to a more usual observable like the cross section is tempered by the added uncertainty necessarily present in the determination of the polarizations. Yet the use of entanglement can contribute to better constrain  interactions and models beyond the SM, as the examples reviewed below show. This is a field still in its infancy but we expect quantum state tomography to become part of the routine tools  in the physical analysis of the experimental events.

\subsection{Top quark}

 The SM Effective Field Theory (SMEFT) expansion parametrizes  possible new particles and fields, characterized by  heavy new states, in terms of operators that are obtained by integrating out these new states. Modifications of the entanglement of the spins of top-quark pairs in this framework has been studied in \cite{Aoude:2022imd}. In this approach  the effective Lagrangian is given by
\be
{\cal L}_{\text{\tiny SMEFT}}= {\cal L}_{\text{\tiny SM}} + \frac{1}{\Lambda^2} \sum_i c_i {\cal O}_{i}\,, \label{L_smeft}
\ee
in which, at the leading order in QCD, all $CP$-even operators of dimension six are included. There is 1 operator with zero fermions, 2 operators with two fermions and 14 with four fermions (see \cite{Aoude:2022imd} for their explicit form). In \eq{L_smeft}, $\Lambda$ is the scale of the effective  theory (roughly the mass of the heavy states) and $c_i$ the Wilson coefficients of the corresponding operators. 

The concurrence is modified by the presence in the differential cross section of terms linear in $c_i/\Lambda$ (arising from the interference between the SM and the dimension six operators) and by  terms quadratic in $c_i/\Lambda$ (arising from the square of   the dimension six operators).
The qualitative result of the analysis is that, at threshold, the linear interference terms  modify the concurrence very little while the quadratic terms reduce it. Both classes of terms reduce the concurrence in the high-energy regime by a sizable amount.

The impact of higher-order terms in the SMFET expansion have been studied
in \cite{Severi:2022qjy}. While NLO k-factors do not radically change the predictions,  some NLO corrections are shown that are not captured by LO scale variations. 

\subsubsection{Gluon magnetic-like dipole moment}

To show how  entanglement can provide constraints on higher-order operators, let us focus on a single one, the gluon magnetic-like dipole operator, as  discussed in~\cite{Fabbrichesi:2022ovb}, which gives rise to the effective Lagrangian 
\be
{\cal L}_{\text{\tiny  dipole}}=\frac{c_{\,tG}}{\Lambda^2} \big( {\cal O}_{tG} +{\cal O}_{tG}^\dag \big) \quad \text{with}  \quad
 {\cal O}_{tG} =g_s \left(\bar Q_L \, \sigma^{\mu \nu} \, T^a\, t_R \right) \tilde{H}  G_{\mu\nu}^a \, . \label{L_dip}
\ee
In \eq{L_dip} above,  $Q_L$ and $t_R$ stands for the $SU(2)_L$ left-handed doublet of top-bottom quarks and right-handed top quark fields respectively, while $\tilde{H}$ is as usual the dual of the $SU(2)_L$  doublet Higgs field, with SM vacuum expectation value $v$ given by $\langle 0| \tilde{H} |0\rangle =(v/\sqrt{2},0)$.

The  magnetic-like dipole moment  is given by 
\be
\mu = - \frac{\sqrt{2} m_t v}{\Lambda^2} c_{\,tG}\ .
\ee

The addition of an effective magnetic dipole moment term to the SM Lagrangian, gives rise in general to further mixture
contributions, thus weakening the  entanglement of the $t \bar{t}$ spin state produced by the SM interaction.
It is  this loss of entanglement  both in the $q \bar{q}$ and $gg$ production channels
that allows the  bound on the magnitude of the extra, effective parameter $\mu$, to be obtained.

By running a simple Monte Carlo, the authors of~\cite{Fabbrichesi:2022ovb} find that---in the kinematic region $m_{t\bar t} > 900$ GeV and $2 \Theta/\pi > 0.85$,  where the relative difference between the SM and the new physics  is largest and equal to about 3\%---a separation of 2.3$\sigma$ is  possible down to  the value of $\mu=0.003$ with the data of run 2 at the LHC.  This result is in agreement  with what found in~\cite{Aoude:2022imd} (with $c_{tG} =-0.1$ for $\Lambda=1 \,\text{TeV}$) and compares favorably with current determinations~\cite{Lillie:2007hd,CMS:2019kzp} which find  a bound around  $\mu=0.02$.

\subsection{$\tau$ lepton}

Quantum state tomography has been used in the study of the properties of the $\tau$ lepton and its coupling to quarks and the Higgs boson.

\subsubsection{Contact interactions}

Contact interactions in $\tau$-pair entanglement were discussed in \cite{Fabbrichesi:2022ovb}. 
The most general contact operators for the production of $\tau$-leptons from quarks can be written, in chiral components, as
 \begin{align}
     {\cal L}_{\text{\tiny  cc}} =& -\frac{4\pi }{\Lambda^2} \eta_{LL}  ( \bar q_L \gamma^\alpha q_L) \,(\bar \tau_L \gamma_\alpha \tau_L) 
 -\frac{4\pi}{\Lambda^2} \eta_{RR} (\bar q_R \gamma^\alpha q_R)\, (\bar \tau_R \gamma_\alpha \tau_R)
 \\ \nonumber &
 -\frac{4\pi}{\Lambda^2}  \eta_{LR} ( \bar q_L \gamma^\alpha q_L) \,(\bar \tau_R \gamma_\alpha \tau_R )
 -\frac{4\pi}{\Lambda^2} \eta_{RL} (\bar q_R \gamma^\alpha q_R)\, (\bar \tau_L \gamma_\alpha \tau_L) \label{CI} \, .
 \end{align}
 
It is the  change in the entanglement content of the $\tau$-pair spin state induced by the presence
of the contact term contribution, both in the $u \bar{u}$ and $d \bar{d}$ production channels,
that makes possible  obtaining bounds on the magnitude of the new physics scale $\Lambda$.

 The entanglement becomes larger in the kinematic regions where either the photon or the $Z$-boson diagram dominates. Because the new-physics terms increase as the energy in the CM, these  regions---being as they are at relatively low-energies---are not favorable for distinguishing between SM and new higher-scale physics. It is at higher energies, just below 1 TeV that the two can best be compared. At these energies, the amount of entanglement is modest but very sensitive to the addition of new terms in the amplitude. The authors of~\cite{Fabbrichesi:2022ovb} therefore consider the kinematic region $m_{\tau\bar\tau}>800$  as a compromise between having enough events and having new-physics effects sizable.
 
For  $m_{\tau\bar\tau}>800$ GeV and scattering angles close to $\pi/2$,  the relative difference  between SM and the new physics (with $\Lambda=25$ TeV) is largest and equal to about 70\%.  Such a large effect shows that  the contact interaction and its scrambling of the two $\tau$-lepton polarizations is a very effective way of changing the concurrence of their spins. The SM hypothesis can be rejected with a significance of 2.7 for a contact interaction with a scale $\Lambda=25$ TeV  at Hi-Lumi LHC. This result compares favorably with current determinations of four-fermion operators~\cite{ATLAS:2017eqx,CMS:2018ucw} (see also, the dedicated Section in \cite{ParticleDataGroup:2022pth}).

\subsubsection{CP properties of the coupling to the Higgs boson}
 The CP nature of the Higgs boson coupling to the $\tau$ leptons has been proposed to be constrained by means of entanglement in \cite{Altakach:2022ywa}. The authors consider the associated production $Zh$ at $e^+e^-$ colliders and look in the subsequent decay $h\to \tau^+\tau^-$  at the generic interaction Lagragian
\be
{\cal L}_h = - \frac{m_\tau}{v} \kappa \, h \, \bar \tau \Big( \cos \delta + i \gamma_5 \sin \delta \Big) \tau \, .
\ee
Quantum state tomography of the decay is proposed via the computation of the correlation matrix, which is  given by
\be
C_{ij} =  \begin{pmatrix} \cos 2 \delta & \sin 2 \delta& 0 \\
-\sin 2 \delta&  \cos 2 \delta &0\\ 0 & 0 &-1\end{pmatrix}\,. 
\ee

Monte Carlo events are generated with the program \textsc{MadGraph5\_aMC@NLO}~\cite{Alwall:2014hca}, using leading-order matrix elements for two benchmark colliders: the ILC and FCC-ee. The one-prong decays $\tau^+ \to \pi^+ \nu_\tau$ and  $\tau^- \to \pi^- \bar \nu_\tau$ are used. The kinematic constraints of the process are used to reconstruct the neutrino momenta and  find those of the $\tau$-leptons, in the rest frame of which the entries  $C_{ij}$ are computed. Since the concurrence is maximal regardless of the $CP$ phase~\cite{Fabbrichesi:2022ovb}, the determination of the $CP$ phase would be obtained by a direct fit of the entries in the $C_{ij}$ matrix.

The simulations suggest that a zero value of the phase $\delta$ could be constrained at the 9\% CL to the intervals: 
\bea
&[-10.89^\circ,\, 9.21^\circ] &\quad \text{(ILC)} \nn \\ 
&[-7.36^\circ,\, 7.31^\circ] &\quad \text{(FCC-ee)}
\eea
for the two benchmark considered. A sensitivity (at 1$\sigma$) of roughly $7.5^\circ$ is found for the ILC and  $5^\circ$  for the
FCC-ee. These values are comparable to those found by more traditional methods (see, for instance ~\cite{Berge:2013jra}).

\subsubsection{Electromagnetic couplings and compositness}

The electromagnetic couplings of the $\tau$ leptons are constrained by means of entanglement in \cite{Fabbrichesi:2024xtq}. The effective vertex used to model these interactions is 
\be
-ie\,\bar \tau \,\Gamma^\mu(q^2) \,\tau \,A_\mu(q)
=
-ie \,\bar \tau \left[ \gamma^\mu F_1(q^2)  +  \frac{i \sigma^{\mu\nu}q_\nu}{2 m_\tau} F_2(q^2) 
+  \frac{\sigma^{\mu\nu} \gamma_5 q_\nu}{2 m_\tau} F_3(q^2)  \right]   \tau\, A_\mu(q) \, , \label{int}
\ee
and it defines the  magnetic and electric dipole moments as
\be
a_\tau = F_2(0) \quad \mbox{and} \quad d_\tau = \frac{e}{2 m_\tau} F_3(0)\,.
\ee
The potential compositeness of the $\tau$ lepton can be investigated by means of the mean squared electromagnetic radius
\be
\langle \vec r^{\;2} \rangle = -  6 \left .\frac{d }{d \vec q^{\;2}}  \left[  
 F_{1}(q^{2}) + \frac{q^{2}}{4 m_{\tau}^{2}} F_{2}(q^{2}) \right] \right|_{q^2=0} \,.\label{rm}
\ee

\begin{table}[h!]
\centering
  \begin{tabular}{*{2}{p{0.4\textwidth}}}
  \toprule
      PDG (2022) & Quantum observables \\
  \midrule
 $-1.9 \times 10^{-17} \leq d_{\tau} \leq 6.1 \times 10^{-18} $ e cm& $  |d_{\tau}| \leq 1.7 \times 10^{-17} $ e cm    \\
 $ -5.2 \times 10^{-2} \leq a_{\tau} \leq 1.3 \times 10^{-2}$ &  $ | a_{\tau} |\leq 6.3 \times 10^{-4} $    \\
 $\qquad \quad \Lambda_{\rm C.I.}\geq 7.9\; \text{TeV}$ & $ \sqrt{\langle \vec r^{\;2} \rangle} <  5.1 \times 10^{-3} \; \text{fm.}\implies  \Lambda_{\rm C.I.}\geq 2.6\; \text{TeV}$    \\
 \bottomrule
\end{tabular}
  \caption{Marginalization of 95\% joint confidence intervals on the magnatic and electric dipole of the $\tau$ lepton obtained with quantum observables for a benchmark luminosity of 1 ab$^{-1}$ at Belle II. The current experimental limits are reported in the first column. The scale $\Lambda_{\rm C.I.}$ suppresses the four-fermion contact interaction related to the $\tau$ lepton electromagnetic radius.}
  \label{tab:taucouplings} 
  \end{table}
  
To constrain these quantities the authors employ a $\chi^2$ test targeting deviations of the concurrence, cross section and antisymmetric part of the $\tau$-pair polarization density matrix from the corresponding SM values. The uncertainties associated with the quantum operators were obtained via a Monte Carlo simulation~\cite{Ehataht:2023zzt}, whereas the one affecting the cross section was obtained by rescaling the error quoted in Ref.~\cite{Banerjee:2007is} to the benchmark luminosity used in the study. The limit obtained with this methodology are reported in Tab.~\ref{tab:taucouplings}, together with the corresponding current experimental bounds.

\subsection{Diboson production}

The possibilities for using concurrence bounds, purity, and Bell inequalities to gain sensitivity to new particles and fields using the qutrit bipartite  system representing two massive gauge bosons are discussed---both analytically and in  Monte Carlo simulations---for lepton and hadron colliders in~\cite{Aoude:2023hxv}. The SM results agree with 
\cite{Fabbrichesi:2023cev}. In addition, it is shown that spin observables can serve as  probes for heavy new physics as parametrized by higher dimensional operators in the SMEFT expansion. In particular, it is found that these observables offer increased sensitivity to operators whose contributions do not interfere with the SM amplitudes at the level of differential cross sections. As expected, lepton colliders have better sensitivity than hadron colliders because in the latter the quantum state of the system is the incoherent sum of different partonic channels and therefore tends to be mixed.

Production of  $ZZ$ pairs is the least interesting process when it comes to sensitivity to heavy new particles and fields, as the phenomenology is completely determined by only two possibly anomalous couplings (the right-handed and the left-handed coupling to the $Z$ boson) and the dimension-6 operators do not introduce new Lorentz structures.  On the other hand, $WW$ and $WZ$ production show a rather large sensitivity to heavy new-physics effects in the spin density matrix already with operators of  dimension six with significant changes expected in the entanglement pattern across phase space.

\subsection{Higgs boson coupling to $W^\pm$ and $Z$}
The power of entanglement and quantum observables to constrain non-standard interactions between Higgs and massive gauge bosons has been discussed in~\cite{Fabbrichesi:2023jep,Bernal:2023ruk}. These anomalous couplings have been previously studied by means of dedicated observables~\cite{Soni:1993jc,Chang:1993jy,Skjold:1993jd,Buszello:2002uu,Choi:2002jk,Gao:2010qx,Christensen:2010pf,Desai:2011yj,Bolognesi:2012mm,Dwivedi:2016xwm,Anderson:2013afp}, within effective field theories~\cite{Artoisenet:2013puc,Boselli:2017pef,Brivio:2019myy} and by means of helicity amplitudes~\cite{Hellmund:1981kq,Shim:1995ax,Mahlon:1998jd,Bern:2011ie,Stirling:2012zt,Maina:2020rgd,Maina:2021xpe,Rao:2020hel}.

The most general interaction Lagrangian involving the Higgs boson $h$ and the gauge bosons $W^\pm$ and $Z$ allowed by Lorentz invariance is given by
\begin{align}
{\cal L}_{hVV}  = & g\, M_{W} W^{+}_{\mu}W^{-\mu} h + \frac{g}{2 \cos\theta_{W}} M_{Z} Z_{\mu}Z^{\mu} h\nn \\
& -\frac{g}{M_{W} }\Bigg[ \frac{a_{W}}{2} W^{+}_{\mu\nu} W^{-\mu\nu} +
\frac{\widetilde a_{W}}{2} W^{+}_{\mu\nu}\widetilde W^{-\mu\nu}  +\,\frac{a_{Z}}{4} Z_{\mu\nu} Z^{\mu\nu}  + 
\frac{\widetilde a_{Z}}{4} Z_{\mu\nu}\widetilde Z^{\mu\nu}\Bigg]  h\, ,\label{eq:Lhvv}
\end{align}
where $V^{\mu\nu}$ is the field strength tensor of the gauge boson $V=W$ or $Z$ and the corresponding dual tensor is defined as $\widetilde V^{\mu\nu}=\frac12 \epsilon^{\mu\nu\rho\sigma}V_{\rho\sigma}$. The anomalous couplings $a_V$ allow for a momentum dependent interaction vertex whether the couplings $\widetilde a_{V}$ signal the presence of a pseudoscalar component, which could result in the violation of the CP symmetry through the interference with the SM contribution. The latter is obtained for $a_V=\widetilde a_{V}=0$. 


Following the conventions of Sec.~\ref{sec:qutrits:higgs},  off-shell states are denoted with $V^*$, $V=W, Z$. From the Lagrangian in Eq.~\eqref{eq:Lhvv} it is possible obtain the following amplitude for the $h\to V(k_1,\lambda_1)\, V^*(k_2,\lambda_2)$ process 
\begin{equation}
{\cal M}(\lambdaA,\lambdaB)=\mathcal{A}_{\mu\nu}
\varepsilon^{\mu\star}(k_1,\lambdaA)\varepsilon^{\nu\star}(k_2,\lambdaB)
\, ,
\label{eq:MHVV}
\end{equation}
where
\begin{equation}
\mathcal{A}^{\mu\nu}=g\, M_V\xi_V \, g^{\mu\nu} -\frac{g}{M_W}
\Big[a_V \left(k_1^{\nu} k_2^{\mu}-g^{\mu\nu} k_1\cdot k_2\right)+
\tilde{a}_V \epsilon^{\mu\nu\alpha\beta}k_{1\alpha} k_{2\beta}\Big]\,.   
\end{equation}
and the parameter $\xi_V$ takes values $\xi_W=1$ and $\xi_Z=1/(\cos\theta_W)$, with $\theta_W$ being the Weinberg angle. The spin-summed amplitude square is then
\begin{align}
\label{eq:M2hvv}
    \abs{\mathcal{M}}^2  = & \frac{\xi_V^2\,g^2}{4 f^2 M_V^2} \Bigg\{ 
     \Big[ 1  + 2\, f^2 \left(\LtVV + \LVV \right)\Big] m_h^4 -2 \Big[1  + f^2 \Big(1 + 2 \LtVV + 2 \LVV  - 6 \LV \Big)
\nonumber\\
&+   2\,  f^4 \Big(\LtVV + \LVV \Big)\Big] m_h^2 M_V^2+\Big[1  + 
  2 f^6 \Big(\LtVV + \LVV \Big)
\nonumber\\
&+    2 f^2 \Big(5 + \LtVV + \LVV  - 6 \LV \Big) +    f^4 \Big(1 - 4 \LtVV + 8 \LVV - 
      12 \LV\Big)\Big] M_V^4\Bigg\}\,,
\end{align}
where, as before, $f= M_{V^*}/M_V$ quantifies how much the particle $V^*$ is off-shell.

The procedure in Sec.~\ref{sec:third:toolbox} results in a density matrix having the same structure as that of Eq.~\eqref{rhoBVV}, 
with helicity amplitudes~\eqref{helampl} now given by 
\begin{align}
 h_0 = & g\,\xi_V\left[\,a_V\,f\,M_V\,(x^2-1)-\left(M_V+a_V \frac{k_1\cdot k_2}{M_V}\right) x\right] \,, \label{eq:hehvv1}\\
 h_{\pm} = & g\,\xi_V\left[ \left(M_V+a_V \frac{k_1\cdot k_2}{M_V}\right)\mp i\,\tilde{a}_V\,f\,M_V \sqrt{x^2-1}\right]\,,\label{eq:hehvv2}
\end{align}
where $x = m_h^2/(2 f M_{V}^{2}) - (f^2+1)/(2f)$. The coefficients $f_a$, $g_a$ and $h_{ab}$ entering the alternative decomposition of the density matrix on the basis formed by the tensor products of the Gell-Mann matrices and the identity matrix are listed in appendix~\ref{appendix:HVV}. The density matrix continues to describe a pure state also in presence of anomalous coupling; an explicit expression can be obtained from Eq.~\eqref{pure} by means of Eqs.~\eqref{eq:hehvv1} and~\eqref{eq:hehvv2}.  

To constrain the anomalous couplings in the Lagrangian~\eqref{eq:Lhvv}, the authors of \cite{Fabbrichesi:2023jep}  employ two observables made easily accessible by quantum state tomography:
\begin{itemize}[{\bf-}]
\item The entanglement between the polarizations of the massive gauge bosons emitted in the decay under consideration, given for a pure state by the entropy of entanglement defined in Eq.~\eqref{entropy}. The anomalous coupling $a_V$ enters the observable linearly, whereas the dependence on $\LtV$ is only quadratic and, therefore, suppressed in the expected range of values. 
\item An observable tailored to single out the anti-symmetric part of the density matrix
\be
\mathscr{C}_{odd}=\frac{1}{2}\, \sum_{\substack{a,b\\ a< b}} \Big| h_{ab} -h_{ba} \Big| \, , \label{CPodd} 
\ee
corresponding to kinematics variables that involve the triple products of momenta and polarizations, for instance $\vec k \cdot \Big(\vec \varepsilon_{\hat n} \times \vec \varepsilon_{\hat r} \Big) \,$ where $\vec k$ is the momentum of one of the particles while $\vec \varepsilon_{\hat n}$ and  $\vec \varepsilon_{\hat r}$ are the projections of the polarizations along two directions orthogonal to the momentum. The observable $\mathscr{C}_{odd}$ depends linearly on the anomalous coupling $\LtV$, while the effects of $a_V$ are suppressed as the parameter enters the expression only multiplied by $\LtV$.
\end{itemize}

The values of the anomalous couplings can be constrained by a $\chi^2$ test set for a 95\% joint CL. The relevant uncertainties can be computed by taking the error affecting the Higgs boson mass measured from the $p \,p \to h\to W^+\ell^-\bar \nu_\ell$~\cite{ATLAS:2022ooq} and $p \,p \to h\to  Z\ell^+\ell^-$~\cite{ATLAS:2020rej} processes as a proxy for the uncertainty in the reconstruction of the resonant Higgs boson rest frame, crucial for the determination of gauge boson polarizations. The error is consequently propagated to the observables via a Monte Carlo simulation where $m_h$ is varied within the experimental limits. Tab.~\ref{tab:thcouplings} shows the marginalized 95\% joint confidence intervals obtained for the anomalous couplings. 
\begin{table}[t]
\centering
\begin{tabular}{*{2}{p{0.15\textwidth}}}
  \toprule
       LHC run2&
       LHC Hi-Lumi  \\
  \midrule
   $ |a_{W}| \leq 0.033$    & $|a_{W}|\leq 0.0070 $ \\
     $|\widetilde a_{W}|\leq 0.031$   & $|\widetilde a_{W}|\leq 0.0068$  \\
   $ |a_{Z}| \leq 0.0019 $   & $|a_{Z}|\leq 0.00040$ \\
     $|\widetilde a_{Z}|\leq 0.0039$   & $|\widetilde a_{Z}|\leq 0.00086$  \\
  \bottomrule%
\end{tabular}
\caption{Marginalized 95\% joint confidence intervals for the anomalous couplings obtained from the LHC data neglecting the backgrounds. The operators used in the $\chi^2$ test are the entropy of entanglement and $\mathscr{C}_{odd}$.}
\label{tab:thcouplings}
\end{table}
The proposed strategy outperforms, in power, alternative strategies employing polarization observables not related to entanglement~\cite{Rao:2020hel} and goes beyond the projected reach of even future lepton collider searches exploiting classical spin correlations and cross sections~\cite{Han:2000mi,Craig:2015wwr,Sharma:2022epc}.

Although the proposed observables seem optimal to constrain the anomalous couplings, a careful assessment of the power of the method must include the effect of  backgrounds originating, for instance, from the gauge boson and quark electroweak fusions. A first effect of these processes is that of impairing the purity of the bipartite qutrit final state, thereby complicating the quantification of entanglement which now must rely on the concurrence~\eqref{C_rho} or on its lower bound~\eqref{eq:cmb-qutrit}. According to current estimates, the $W$ plus jets background affecting the $h\to WW^*$ channel overcomes the signal, whereas a signal-to-background ratio of $0.8$ can be achieved for the $ZZ^*$ channel in the kinematic region of interest~\cite{CMS:2021ugl}. In the latter case, the inclusion of background processes does not significantly worsen the results in Tab.~\ref{tab:thcouplings}.

\newpage
\section{Other processes and ideas}
\label{sec:ideas}
The study of entanglement in particle physics is just at its beginnings and new ideas and applications are coming to light and being explored. We give a short summary of some of them in this Section.

\subsection{Three-body decays}\label{sec:3body}

The extension to three-body decays of the computations of entanglement is natural and potentially fruitful in the physics of colliders. The authors of~\cite{Sakurai:2023nsc}  explain how the concurrence can be generalized to measure tripartite systems. Two kinds of concurrence can be defined for  a three qubit state $|\Psi\rangle$: one 
\be
{\cal C}_{ij} = {\cal C}\,[\rho_{ij}]=
\Tr_k\, |\Psi\rangle \langle \Psi | 
\ee
in which one of the three sub-states is traced out, and one
\be
{\cal C}_{i(kj)} = \sqrt{2 (1 - \Tr \rho_{kj}^2)}
\ee
in which the concurrence of the sub-part $i$ is measured with respect to the other two.

The properties and peculiarities of the three-body system can be analyzed by means of the \textbf{monogamy} inequality~\cite{Coffman:1999jd,PhysRevLett.96.220503}
\be
{\cal C}_{i(kj)}^2 \leq {\cal C}_{ij}^2+{\cal C}_{ik}^2
\ee
and the genuine multiparticle entanglement quantified by the \textbf{concurrence triangle} given by~\cite{Jin:2022kxb}
\be
F_3= \frac{4}{\sqrt{3}} \Big[ Q\, (Q-{\cal C}_{1(23)}) \, (Q-{\cal C}_{2(13)})\, (Q-{\cal C}_{3(12)})  \Big]^{1/2}
\ee
with $Q=[{\cal C}_{1(23)}+{\cal C}_{2(13)}+{\cal C}_{3(12)}]/2$;
$F_3$ takes values from 0 and 1. 

Monogamy  and the concurrence triangle are discussed in \cite{Sakurai:2023nsc} for various kinds of possible  interactions (scalar, pseudoscalar, vector and axivector) in a three-body decay process.
The general properties of multipartite systems are discussed in~\cite{Bernal:2023xzp}, which introduces the concept of the concurrence vector.

\subsection{Post-decay entanglement}\label{sec:post-decay}

	An idea first discussed in \cite{Bernabeu:2019gjs} for  kaon system, has been extended in \cite{Aguilar-Saavedra:2023hss,Aguilar-Saavedra:2023lwb} to the generic case of the decay into two particles, one of which is projected into an eigenstate by a Stern-Gerlach-type  experiment.

 The procedure is applied to top-quark pairs produced at the LHC to show~\cite{Aguilar-Saavedra:2023hss}, by means of a Monte Carlo simulation, that it is possible to measure entanglement between one top-quark and the $W$
gauge boson originating from the decay of the other top-quark. If implemented, such a measure would be the first  showing entanglement between a fermion and a boson.

\subsection{Maximum entanglement}\label{sec:max} 

A direct computation of many QED processes shows  that the entanglement between the polarizations of the particles in the final state is maximum for certain scattering angles. This behavior comes about because of the structure of the interactions in the processes considered. 

This result has inspired a line of research in which \textbf{maximum entanglement} is taken  as a principle and used in an attempt to determine some of the SM interactions and parameters~\cite{Cervera-Lierta:2017tdt}. For example,  the application of this principle  to the determination of the Weinberg angle in tree-level scattering of leptons
leads to the value  $\sin \theta_W = 1/2$. It comes from  the cancellation of the vector-like coupling in the electroweak current. Off by about 10\% of the actual value though this is, it is an interesting result which may be hinting to some underlining interplay between quantum mechanics and particle physics.

\subsection{Minimum entanglement}\label{sec:min}

The idea of connecting \textbf{minimal entanglement} to emergent symmetries in hadron physics and low-energy QCD has been initiated in \cite{PhysRevLett.122.102001}, in which 
the Wigner $SU(4)$ symmetry for two flavors and an $SU(16)$ symmetry for three flavors is conjectured to arise from  dynamical entanglement suppression of the strong interactions in the infrared.

Further discussion of entanglement suppression in hadron physics are presented in~\cite{Low:2021ufv,Beane:2021zvo,Liu:2022grf,Bai:2022hfv,Liu:2023bnr} and applied to a model for the SM Higgs boson based on entanglement suppression of the $SO(8)$ symmetry in a scalar model with two Higgs bosons which are flavor doublets~\cite{Carena:2023vjc}

\subsection{Quantum process tomography and beyond-quantum tests}\label{sec:x}

As well as understanding the spin structure of the final state, we have reason to be interested in the mapping that takes an incoming initial state, characterised by some spin density matrix $\rho_{\rm in}$ to some final spin density matrix $\rho_{\rm out}$ -- what is known as the quantum process. This map $\Phi: \mathbb{C}^{m\times m}\rightarrow \mathbb{C}^{n\times n}$ needs to satisfies some requirements in order to be physically acceptable (for instance complete positivity, see~\cite{Nielsen:2012yss}) and, due to quantum state-channel duality, can also be represented by a larger matrix, the Choi matrix~\cite{CHOI1975285}. 
The formalism allows us to advance Feynman's proposal~\cite{Feynman:1981tf} of using quantum systems (quantum computers) to efficiently simulate quantum dynamics (scattering processes). A dictionary mapping between the language of quantum computers and of particle physics processes was developed in \cite{Altomonte:2023mug}, as well as simulating an example process -- the spins of an $e^+e^-\rightarrow t\bar{t}$ scattering process on an IBM quantum computer.\footnote{The broader use of quantum computing methods in high-energy physics was recently reviewed, for instance, in~\cite{Humble:2022vtm,Bauer:2022hpo,Catterall:2022wjq,Rodrigo:2024say}.}
The authors of \cite{Eckstein:2021pgm} advocate measuring experimentally the Choi matrix for subatomic processes since such tests could indicate sensitivity to unexplored physics and even probe `post-quantum' theories that do not necessarily have unitary evolution.

	\newpage
	\section{Outlook}\label{sec:sum}

 The detection of entanglement at colliders might have seemed, at first blush, a rather far fetched proposition. High-energy collisions have all sorts of  multiple vertex interactions and superposition of processes weighted by the respective distribution probabilities. How can quantum coherence survive through all that?
 
Unlikely though it might have seemed at first, the study of entanglement at colliders turned out to be not only possible but a new and promising field whose very existence is enriching for particle physics. Many works have recently been published in a very short span of time as different processes have been investigated and an increasing number of results harvested. We hope to have produced a useful survey of those released up to the beginning of the year 2024.

 After these developments, the experiments are now weighing in. It has begun with the analyses of $B$-meson decays at the LHCb and Belle-II~\cite{Fabbrichesi:2023idl}  and the  detection of entanglement at the LHC~\cite{ATLAS:2023fsd,CMS:2024hgo} and we expect more results will be forthcoming for  $\tau$-lepton pairs final states at Belle II--- whose experiments have by far  the best statistics. Most likely, these will be followed by analyses for top-quark pairs and diboson final states from Higgs boson decays from the data of run 1 and 2 at the LHC, which are already under way, and  will be extended into the Hi-Lumi runs as well.
The results of all these experiments will provide the basis for the next round of theoretical enquires toward perhaps a more detailed view of the processes discussed in Section~\ref{sec:qubits}, \ref{sec:qutrits} and \ref{sec:newphysics}   or new directions, some of which have been briefly discussed in Section~\ref{sec:ideas}.
  
    We believe that the possible experimental program of investigation on the structure of quantum mechanics at the existing and future colliders is very broad, and continues to be developed. The implications of these measurements are only just  starting to be investigated.  It is refreshing for our generation of collider physicists to recall that, regardless of whether additional new particles are found, there is a great deal of highly interesting and challenging physics out there for us to investigate.

	\section*{Acknowledgements}
	
AJB is funded through STFC grants ST/R002444/1 and 
ST/S000933/1, by the University of Oxford and by Merton College, Oxford. LM is supported by the Estonian Research Council grants PRG803, RVTT3 and by the CoE program grant TK202 \textit{Fundamental Universe}.

AJB is grateful to Christopher Timpson for helpful comments on Section~\ref{sec:loopholes}. MF thanks Michele Pinamonti for  useful discussions.  We thank  the organizers of the workshop \textit{Quantum Observables for Collider Physics}, November 2023, funded by the  Galileo Galilei Institute for Theoretical physics of the \textit{Istituto Nazionale di Fisica Nucleare}, for hosting several of the authors during the preparation of this manuscript.

	
 \newpage

\begin{appendices}
	\section{Qubits}
        \label{appendix:qubits}

        This Appendix contains some kinematic definitions utilized in Section~\ref{sec:qubits} and \ref{sec:qutrits} and the explicit expressions for the SM functions $\tilde A$ and $\tilde C_{ij}$, $\tilde B_{i}$ entering
        the  coefficients $C_{ij}$ and $B_i$, respectively, for the top-quark and $\tau$-lepton  pair production, as discussed in Section~\ref{sec:qubits}.

\subsection{Kinematics} 
\label{appendix:kinematics}

Let us consider the generic production of fermion pair via quark anti-quark annihilation
        \be
        q(q_1) + \bar q(q_2) \to f(k_1) + \bar f(k_2)\, .
\label{eq:qqff}
\ee
The momenta $k_1$ and $k_2$, corresponding to the final fermion and anti-fermion, and  $q_1$ and $q_2$ of the entering quark and anti-quark, respectively,  can be written  in the CM system as~\cite{Fabbrichesi:2022ovb}
\begin{align}
k_1 &=  \left( \frac{m_f}{\sqrt{1-\beta_f^2}},\, \frac{m_f \beta_f \sin \Theta}{\sqrt{1-\beta_f^2}},\, 0, \, \frac{m_f \beta_f \cos \Theta}{\sqrt{1-\beta_f^2}}\right) \nn \\
k_2&=  \left( \frac{m_f}{\sqrt{1-\beta_f^2}},\,- \frac{m_f \beta_f \sin \Theta}{\sqrt{1-\beta_f^2}},\, 0, \, -\frac{ m_f \beta_f \cos \Theta}{\sqrt{1-\beta_f^2}}\right)  \nn \\
q_1 &= \left( \frac{m_f}{\sqrt{1-\beta_f^2}},\, 0,\, 0, \, \frac{m_f}{\sqrt{1-\beta_f^2}}\right) \nn \\
q_2&=  \left( \frac{m_f}{\sqrt{1-\beta_f^2}},\,0,\, 0, \, -\frac{ m_f }{\sqrt{1-\beta_f^2}}\right) \, ,
\label{eq:momenta}
\end{align}
where $m_f$ is the mass of the final fermions and 
\be
\beta_f = \sqrt{1 -  4 \frac{m_f^2}{m_{f\bar f}^2}}\, , \label{eq:beta_f}
\ee
where $m_{f\bar f}$ is the fermion pair invariant mass, with $\Theta$ the angle between the initial and final fermion momenta in the CM frame.

Throughout the review, we adopt the orthonormal basis in \eq{basis} introduced in~\cite{Bernreuther:2001rq} in order to describe the spin correlations.

The elements $\CC_{ij}$ of the correlation matrices  are obtained on the various components of the chosen basis by means of the polarizations vectors $s_i^{\mu}$ appearing in \eqs{eq:projU}{eq:projV} ~\cite{Fabbrichesi:2022ovb}
\begin{align}
s_1^k &=  \left( \frac{\beta_f}{\sqrt{1-\beta_f^2}},\, \frac{\sin \Theta}{\sqrt{1-\beta_f^2}},\, 0, \, \frac{\cos \Theta}{\sqrt{1-\beta_f^2}}\right) \nn\\
s_2^k &=  \left(- \frac{\beta_f}{\sqrt{1-\beta_f^2}},\, \frac{\sin \Theta}{\sqrt{1-\beta_f^2}},\, 0, \, \frac{\cos \Theta}{\sqrt{1-\beta_f^2}}\right) \nn \\
s_1^r &=  \zeta_2^r = \left(0,\, -\cos\Theta ,\, 0, \, \sin\Theta \right) \nn \\
s_1^n &=  s_2^n= \left(0,\, 0 ,\, 1, \, 0 \right)
\label{eq:spins}
\end{align}
where the indices 1 and 2 stand for the final fermion and anti-fermion respectively. 
        
\subsection{Top-quark pairs}
\label{appendix:top}

Here are the  complete expressions~\cite{Bernreuther:1993hq,Uwer:2004vp} for the coefficients $\tilde{A}^{q\bar{q}}$, $\tilde{B}^{qq}_{i}$, and $\tilde{C}^{qq}_{ij}$ entering in \eq{eq:Cij-top} for the $t\bar t$ pair production via $q\bar{q}$ and $g g$ scattering in the SM:
\begin{subequations}
\begin{align}
  \tilde{A}^{gg} &= F_{gg} \Big[1 + 2\beta_t^2\sin^2\Theta
  -\beta_t^4\left(1+\sin^4\Theta\right)\Big],\\
  \tilde{C}^{gg}_{nn} &= - F_{gg} \Big[1-2\beta_t^2
  +\beta_t^4\left(1+\sin^4 \Theta\right)\Big],\\
  \tilde{C}^{gg}_{rr} &=  - F_{gg} \Big[1
  - \beta_t^2 \left(2-\beta_t^2\right)\left(1+ \sin^4\Theta \right)  \Big],\\
    \tilde{C}^{gg}_{kk} &= - F_{gg} \Big[1 - \beta_t^2\frac{\sin^2 2\Theta}{2} -\beta_t^4\left(1+ \sin^4\Theta \right) \Big],\\
    \tilde{C}^{gg}_{kr} &=  \tilde{C}^{gg}_{rk}= F_{gg} \, \beta_t^2\sqrt{1-\beta_t^2} \sin 2\Theta\sin^2\Theta\, 
    \\   \tilde{B}^{gg}_k&=\; \tilde{B}^{gg}_r=\tilde{B}^{gg}_n=0\, ,
\end{align}
\end{subequations}
with $\displaystyle F_{gg} =  \frac{N_c^2\left(1+\beta_t^2\cos^2\Theta\right)-2}{64N_c\left(1-\beta_t^2 \cos^2\Theta\right)^2}$\, and
\begin{subequations}
\begin{align}
    \tilde{A}^{q\bar{q}} & = F_{q\bar{q}} \Big(2-\beta_t^2\sin^2\Theta\Big),\\
    \tilde{C}^{q\bar{q}}_{nn} &= -F_{q\bar{q}}\, \beta_t^2 \sin^2\Theta,\\
    \tilde{C}^{q\bar{q}}_{rr} &= F_{q\bar{q}} \Big(2-\beta_t^2 \Big)\sin^2\Theta,\\
    \tilde{C}^{q\bar{q}}_{kk} &= F_{q\bar{q}} \Big(2\cos^2\Theta+\beta_t^2 \sin^2\Theta \Big),\\
    \tilde{C}^{q\bar{q}}_{kr} &=\tilde{C}^{q\bar{q}}_{rk}= F_{q\bar{q}}\, \sqrt{1-\beta_t^2}\sin 2\Theta,
    \\   \tilde{B}^{gg}_k& =\; \tilde{B}^{gg}_r=\tilde{B}^{gg}_n=0\, ,
\end{align}
\end{subequations}
with $\displaystyle F_{q\bar{q}} = \frac{1}{2N^2_c}$.

\subsection{$\tau$-lepton pairs}\label{sec:appendix:qubits:tau}

Here are the  complete expressions \cite{Fabbrichesi:2022ovb} for the coefficients   $\tilde{A}^{q\bar{q}}$, $\tilde{B}^{qq}_{i}$, and $\tilde{C}^{qq}_{ij}$   entering  in \eq{eq:Cij-tau}) for the $\tau^+\tau^-$ pair production via $q\bar{q}$ scattering in the SM:
\begin{subequations}
  \begin{align}
  \tilde{A}^{q\bar{q}} &= F_{q\bar{q}}\bigg\{
  Q_q^2 Q_\tau^2 \Big[2-\beta_\tau^2 \sin^2\Theta \Big] + 2 Q_q Q_\tau
  \Rechi \Big[ 2 \beta_\tau g_A^q g_A^\tau\cos\Theta + g_V^q g_V^\tau\left( 2  - \beta_\tau^2 \sin^2\Theta\right) \Big]
  \nn \\
  & + \Abschi^2 \bigg[ \Big(g_V^{q 2}+ g_A^{q 2}\Big)\Big(2g_V^{\tau 2}+
      2\beta_\tau^2 g_A^{\tau 2}-\beta_\tau^2  \left(g_V^{\tau  2}+ g_A^{\tau  2}\right)\sin^2\Theta
      \Big)
  + 8 \beta_\tau  g_V^q g_V^\tau g_A^q g_A^\tau  \cos \Theta
  \bigg]\bigg\}\, , 
\end{align}
\end{subequations}
\begin{subequations}
    \begin{align}
\tilde{C}^{q\bar{q}}_{nn} &= -F_{q\bar{q}} \beta_\tau^2 \sin^2\Theta
\bigg\{ Q_q^2 Q_\tau^2 + 2 Q_q Q_\tau\Rechi g_V^{q} g_V^{\tau} -\Abschi^2\Big(g_V^{q 2}+g_A^{q 2}\Big) 
\Big(g_A^{\tau 2}-g_V^{\tau 2}\Big)\bigg\}\, , \\
\tilde{C}^{q\bar{q}}_{rr} &= -F_{q\bar{q}} \sin^2\Theta \,\bigg\{ \left(\beta_\tau^2-2\right)     Q_q^2 Q_\tau^2  + 2 Q_q Q_\tau \Rechi g_V^q g_V^\tau \left(\beta_\tau^2-2\right) \nn \\
& +
    \Abschi^2  \Big[\beta_\tau^2 \left(g_A^{\tau 2}+g_V^{\tau 2}\right) - 2g_V^{\tau 2}\Big] \left(g_V^{q 2}+ g_A^{q 2}\right) \bigg\}\, ,
    \\
\tilde{C}^{q\bar{q}}_{kk} &= F_{q\bar{q}}\bigg\{ Q_q^2 Q_\tau^2 \Big[ \left(\beta_\tau^2-2\right)  \sin^2\Theta + 2 \Big] \nn \\
& + 2 Q_q Q_\tau \Rechi\Big[2 \beta_\tau  g_A^q g_A^\tau  \cos \Theta  + g_V^q g_V^\tau \big( (\beta_\tau^2 -2) \sin^2\Theta + 2 \big) \Big]
\nn \\
& + \Abschi^2 \Big[8 \beta_\tau g_A^q g_A^\tau g_V^q g_V^\tau \cos \Theta
  +\Big(g_V^{q 2}+g_A^{q 2}\Big)\Big(
  2g_V^{\tau 2} \cos^2\Theta
  - \beta_\tau^2 \left(g_A^{\tau 2}-g_V^{\tau 2}\right)\sin^2\Theta
  +2\beta_\tau^2g_A^{\tau 2}\Big)
\Big]\bigg\}\, , \\
    \tilde{C}^{q\bar{q}}_{kr} &= \tilde{C}^{q\bar{q}}_{rk} = 2 F_{q\bar{q}}  \sin \Theta \sqrt{1-\beta_\tau^2} \Bigg\{ Q_q^2 Q_\tau^2 \cos\Theta + 
     Q_q Q_\tau   \,\Rechi \Big[ \beta_\tau g_A^q g_A^\tau + 2 
     g_V^q g_V^\tau\cos \Theta \Big] \nn \\ 
     & + \Abschi^2 \Big[ 2\beta_\tau  g_A^q g_A^\tau g_V^q g_V^\tau
     +  g_V^{\tau 2} \left(g_V^{q 2}+ g_A^{q 2}\right) \cos \Theta  \Big] \Bigg\}\, , \nn \\
     \tilde{C}^{q\bar{q}}_{rn} &=  \tilde{C}^{q\bar{q}}_{nr}=
     \tilde{C}^{q\bar{q}}_{kn}=\tilde{C}^{q\bar{q}}_{nk}=0\, ,
     \\ \nn \\
     \tilde{B}^{q\bar{q}}_{k} &= -2F_{q\bar{q}}\bigg\{   
     Q_qQ_\tau\Rechi\Big[\beta_\tau g_A^{\tau}g_V^{q}\left(1+\cos^2\Theta\right)
       +2 g_A^{q}g_V^{\tau}\cos\Theta\Big] 
     \nn \\
&+ \Abschi^2 \left[2 g_A^{q}g_V^{q}\Big( \beta_\tau^2g_A^{\tau 2} +g_V^{\tau 2}\Big)\cos\Theta+
       \beta_\tau g_A^{\tau}g_V^{\tau}\left(g_V^{q 2}+g_A^{q 2}\right)\left(1+\cos^2\Theta\right)\right]
     \bigg\}\, ,
    \\ 
    \tilde{B}^{q\bar{q}}_{r} &= -2F_{q\bar{q}}\sin\Theta\sqrt{1-\beta_\tau^2}\bigg\{Q_qQ_\tau\Rechi\Big[\beta_\tau  g_A^{\tau}g_V^{q}\cos\Theta
      +2g_A^{q}g_V^{\tau}\Big] \nn \\
    &+ \Abschi |^2 g_V^{\tau}\Big[\beta_\tau g_A^{\tau}
      \left(g_V^{q 2}+g_A^{q 2}\right)\cos\Theta +2 g_A^q g_V^{q}g_V^{\tau}\Big]     \bigg\}\, ,
    \\
      \tilde{B}^{q\bar{q}}_{n} &= 0\, ,
\end{align}
\end{subequations} 
with $\displaystyle F_{q\bar{q}} =  \frac{1}{16}$, $Q_{q,\tau}$ the electric charges, $\beta_\tau$ the 
$\tau^{\pm}$ velocity in their CM frame,
\be
g_V^i = T^i_3 - 2 \,Q_i \sin^2 \theta_W\, , \quad g_A^i = T^i_3\, ,
\ee
 and
\be
\Re\big[\chi(q^2)\big] =  \frac{q^2(q^2-m_Z^2)}{\sin^2{\theta_W}\cos^2{\theta_W} \left[(q^2-m_Z^2)^2 + q^4 \Gamma_Z^2/m_Z^2\right]}\quad\, ,
\ee
\be
\left|\chi(q^2)\right|^2 =  \frac{q^4}{\sin^4{\theta_W}\cos^4{\theta_W}\left[(q^2-m_Z^2)^2 + q^4\Gamma_Z^2/m_Z^2\right]}\, ,
\ee
where $\theta_W$ is the Weinberg angle, $m_Z$ and $\Gamma_Z$ the mass and total width of the $Z$ boson respectively, and $q^2=(q_1+q_2)^2$.

\newpage
\section{Qutrits}\label{appendix:qutrits}
This Appendix contains  some basic definitions for the spin and Gell-Mann matrices and   the explicit form of the Wigner functions, which are utitlized  in Section~\ref{sec:qutrits}. 

\subsection{Spin and Gell-Mann matrices}\label{appendix:Gell-Mann}

The spin-1 representation of the three $SU(2)$ generators $S_i$, $i\in\{1,2,3\}$, used throughout the text is 
\be
S_1 =  \frac{1}{\sqrt{2}} \begin{pmatrix} 0& 1 & 0 \\1& 0 & 1 \\ 0 & 1 &0 \end{pmatrix}\,, 
\qquad  
S_2 =  \frac{1}{\sqrt{2}} \begin{pmatrix} 0& -i & 0 \\i& 0 & -i \\ 0 & i &0 \end{pmatrix}\,, 
\qquad S_3 = \begin{pmatrix} 1& 0 & 0 \\0& 0 & 0 \\ 0 & 0 &-1 \end{pmatrix} \, .
\ee
They can be expressed in terms of the Gell-Mann matrices $T^a$ as
\be
\label{eq:s3}
S_1=\frac{1}{\sqrt{2}} \Big( T^1 + T^6 \Big)\, , \quad S_2=\frac{1}{\sqrt{2}} \Big( T^2 + T^7 \Big)\,,\quad 
S_3=\frac{1}{2} \, T^3 + \frac{\sqrt{3}}{2} \,T^8  \, . 
\ee
In similar fashion, the matrices $S_{ij}$ in \eq{Sij} are given, in terms of the Gell-Mann matrices, as
\begin{align}
S_{31} &=S_{13}= \frac{1}{\sqrt{2}} \, \Big( T^1 - T^6 \Big)\, ,  \nn \\
S_{12} &=S_{21}= T^5 \, , \nn \\
S_{23} &=S_{32}= \frac{1}{\sqrt{2}} \, \Big( T^2 - T^7 \Big)  \nn \\
S_{11} &=   \frac{1}{2 \sqrt{3}} \, T^8 +T^4 - \frac{1}{2} \,T^3 \, , \nn \\
S_{22} &=  \frac{1}{2 \sqrt{3}} \, T^8 -  T^4 - \frac{1}{2} \, T^3  \, , \nn \\
S_{33} &=     T^3 -  \frac{1}{\sqrt{3}} \,T^8 \, . 
\end{align}

The Gell-Mann matrices $T^a$ are: 
\bea
T^1 = \begin{pmatrix} 0& 1 & 0 \\1& 0 & 0 \\ 0 & 0 &0 \end{pmatrix}\,, & \quad  T^2 = \begin{pmatrix} 0& -i & 0 \\i& 0 & 0 \\ 0 & 0 &0 \end{pmatrix}\,, & \quad T^3 = \begin{pmatrix} 1& 0 & 0 \\0& -1 & 0 \\ 0 & 0 &0 \end{pmatrix}\,, \nn \\
T^4 = \begin{pmatrix} 0& 0 & 1 \\0& 0 & 0 \\ 1 & 0 &0 \end{pmatrix}\,, &\quad T^5 =\begin{pmatrix} 0& 0 & -i \\0& 0 & 0 \\ i & 0 &0 \end{pmatrix}\,, & \quad T^6 =\begin{pmatrix} 0& 0 & 0 \\0& 0 & 1 \\ 0 & 1&0 \end{pmatrix}\,, \nn \\
T^7 =\begin{pmatrix} 0& 0 & 0 \\0& 0 & -i \\ 0 & i &0 \end{pmatrix}\,, &\quad T^8 = \dfrac{1}{\sqrt{3}} \begin{pmatrix} 1& 0 & 0 \\0& 1 & 0 \\ 0 & 0 &-2 \end{pmatrix} \, .
\eea
\subsection{The Wigner functions $\mathfrak{q}^n_\pm$  and $\mathfrak{p}^n_\pm$  and the matrix   $\mathfrak{a}^{n}_{m}$\label{sec:Aqp}}

In this Appendix we follow~\cite{Ashby-Pickering:2022umy}.
The  $\mathfrak{q}^n_\pm$ functions introduced in Section~\ref{sec:qp} are given by the following expressions
\begin{align}
 \mathfrak{q}^1_\pm &=  \dfrac{1}{\sqrt{2}} \, \sin \theta^\pm \Big( \cos \theta^\pm \pm 1\Big) \cos \phi^\pm \,,\nn\\
 \mathfrak{q}^2_\pm&=   \dfrac{1}{\sqrt{2}} \, \sin \theta^\pm \Big( \cos \theta^\pm \pm 1 \Big) \sin \phi^\pm  \,,\nn\\
 \mathfrak{q}^3_\pm&= \dfrac{1}{8} \,  \Big( 1 \pm 4 \,   \cos \theta^\pm  + 3 \cos 2 \theta^\pm  \Big) \,,\nn\\
 \mathfrak{q}^4_\pm&=  \dfrac{1}{2} \, \sin^{2} \theta^\pm \cos 2 \,\phi^\pm\,, \nn\\
 \mathfrak{q}^5_\pm&=  \dfrac{1}{2} \, \sin^{2} \theta^\pm \sin 2 \,\phi^\pm  \,,\nn\\
 \mathfrak{q}^6_\pm&=  \dfrac{1}{\sqrt{2}} \, \sin \theta^\pm \Big(- \cos \theta^\pm \pm 1\Big) \cos \phi^\pm \,,\nn\\
 \mathfrak{q}^7_\pm&=  \dfrac{1}{\sqrt{2}} \, \sin \theta^\pm \Big( - \cos \theta^\pm \pm 1 \Big) \sin \phi^\pm \,, \nn\\
 \mathfrak{q}^8_\pm&=  \dfrac{1}{8\sqrt{3}} \,  \Big(- 1 \pm  12   \cos \theta^\pm  -3 \cos 2 \theta^\pm  \Big)  \, , \label{Q}
\end{align}
in terms of the spherical coordinates of the two decaying particle rest frames. 

The $\mathfrak{p}^n_\pm$ functions utilized in Section~\ref{sec:qp}  are given  by the following expressions:
\begin{align}
 \mathfrak{p}^1_\pm &=  \sqrt{2} \, \sin \theta^\pm \Big( 5\, \cos \theta^\pm \pm 1\Big) \cos \phi^\pm \,,\nn\\
 \mathfrak{p}^2_\pm&=  \sqrt{2} \, \sin \theta^\pm \Big( 5\, \cos \theta^\pm \pm 1 \Big) \sin \phi^\pm  \,,\nn\\
 \mathfrak{p}^3_\pm&= \dfrac{1}{4} \,  \Big( 5 \pm 4 \,   \cos \theta^\pm  + 15\,  \cos 2 \theta^\pm  \Big) \,,\nn\\
 \mathfrak{p}^4_\pm&=  5 \, \sin^{2} \theta^\pm \cos 2 \,\phi^\pm \,,\nn\\
 \mathfrak{p}^5_\pm&= 5 \, \sin^{2} \theta^\pm \sin 2 \,\phi^\pm  \,,\nn\\
 \mathfrak{p}^6_\pm&= \sqrt{2} \, \sin \theta^\pm \Big(- 5\, \cos \theta^\pm \pm 1\Big) \cos \phi^\pm \,,\nn\\
 \mathfrak{p}^7_\pm&=  \sqrt{2}\, \sin \theta^\pm \Big( -5 \cos \theta^\pm \pm 1 \Big) \sin \phi^\pm  \,,\nn\\
 \mathfrak{p}^8_\pm&=  \dfrac{1}{4\sqrt{3}} \,  \Big(- 5 \pm  12   \cos \theta^\pm  -15 \cos 2 \theta^\pm  \Big)  \, . \label{P}
\end{align}

The matrix $\mathfrak{a}^{n}_{m}$ used in Section~\ref{sec:qp} is the following
\be
\mathfrak{a}^{n}_{m}= \dfrac{1}{g_L^2-g_R^2} \begin{pmatrix} g_{R}^2&0&0&0&0&g_L^2&0&0\\
0& g_{R}^2&0&0&0&0&g_L^2&0\\
0&0&  g_{R}^2-\frac{1}{2}\, g_L^2&0&0&0&0&\frac{\sqrt{3}}{2}\, g_L^2\\
0&0&  0&g_{R}^2- g_L^2&0&0&0&0\\
0&0&  0&0& g_{R}^2- g_L^2&0&0&0\\
g_L^2&0&  0&0&0& g_{R}^2&0&0\\
0& g_L^2&0&  0&0&0& g_{R}^2&0\\
0&0&\frac{\sqrt{3}}{2}\, g_L^2&0&0&0&0&\frac{1}{2}\, g_L^2-g_R^2
\end{pmatrix} \label{Anm}\, .
\ee
The coefficients in \eq{Anm} are $g_{L} =-1/2 + \sin^2 \theta_{W}\simeq -0.2766$  and $g_{R}= \sin^2 \theta_{W}\simeq 0.2234$.


\subsection{Polarization density matrix for $q\,\bar{q}\to ZZ$}\label{sec:appendix:qutrits:qqZZ}
The coefficients $\Aqq [\Theta,\mVV]$, $\ftqq_a[\Theta,\mVV],\, \gtqq_a[\Theta,\mVV]$, and $\htqq_{ab}[\Theta,\mVV]$,
appearing in the polarization density matrix for
$q\, \bar{q}\to ZZ$,  that has been computed in \cite{Fabbrichesi:2023cev}  (and here amended of few typographical errors). The angle $\Theta$ is the scattering angle in the CM frame from the anti-quark and one of the $Z$-boson momenta. The convention adopted is that the $Z$ is in this case the one with momentum parallel to the $\hat{k}$ unit vector of  the spin right-handed basis in \eq{basis}.  Results below are given for a generic quark $q$. 

\bea
A^{q \bar q}= |\xbar{{\cal M}}^{\;q \bar q}_{ZZ}|^{2}=
\frac{8\fZZ (\gAAAA + 6 \gAA \gVV + \gVVVV)
}{\DZ}\Big\{2 -
       \betaZ^2 \big[\betaZ^4 + (9 - 10 \betaZ^2 + \betaZ^4) \Ct^2 + 
         4 \betaZ^2 \Ct^4-3\big]\Big\}\, ,
\label{M2ZZ}
\eea
where
\be
\fZZ =  \frac{8\alpha^2\pi^2N_c}
      {\DZ\ccW^4 \ssW^4}\, ,
      \quad \text{and} \quad
      \DZ =  1 +\betaZ^4 + 2\betaZ^2 (1 - 2 \Ct^2)\, ,
\label{fZZ}
\ee
with $\betaZ=\sqrt{1-4M_Z^2/\mZZ^2}$. The angle $\Theta$ is here defined as the angle between the anti-quark momentum and the 3-momentum of one of the two Z in the CM frame, where the orientation of the latter coincides with that of the $\hat{k}$ unit vector of the basis in \eq{basis}.
Throughout the following expressions we use $\Ct\equiv \cos{\Theta}$, $\St\equiv \sin{\Theta}$.

The non-vanishing elements $\htqq_{ab}$ ($\htqq_{ba}=\htqq_{ab}$),
are given by
\vspace{0.5cm}
\begin{align}
\htqq_{11}[\Theta,\mZZ]&=
\fZZ(1 - \betaZ^2)
\Big\{
(1 + \Ct^2) (\gAAAA + 6 \gAA \gVV + \gVVVV)+ 8 \Ct \gA \gV (\gAA + \gVV) 
   \Big\}
\nonumber\\
\nonumber\\
\htqq_{15}[\Theta,\mZZ]&=
\fZZ\sqrt{2}\sqrt{1-\betaZ^2}\St
\Big\{
\Ct (\gAAAA + 6 \gAA \gVV + \gVVVV)+4 \gA \gV (\gAA + \gVV) 
\Big\}
\nonumber\\
\nonumber\\
\htqq_{16}[\Theta,\mZZ]&=
\fZZ(1-\betaZ^2)\St^2
\Big\{\gAAAA + 6 \gAA \gVV + \gVVVV
\Big\}
\nonumber\\
\nonumber\\
\htqq_{22}[\Theta,\mZZ]&=
\frac{\fZZ (1-\betaZ^2)}{\DZ}
\Bigg\{
-8 \Ct \big[3 + 2 \betaZ^2 - \betaZ^4 - 4 \Ct^2\big] \gA \gV (\gAA + \gVV)
\nonumber\\
&+\Big[(1 + \betaZ^2)^2 - (7 + 10 \betaZ^2 - \betaZ^4) \Ct^2 + 
  4 (2 + \betaZ^2) \Ct^4\Big] (\gVVVV + 6 \gAA \gVV +\gAAAA)
\Bigg\}
\nonumber\\
\nonumber\\
\htqq_{23}[\Theta,\mZZ]&=
\frac{\fZZ 2\sqrt{2}\sqrt{1-\betaZ^2}\St}{\DZ}
\Bigg\{ \Big[\Ct (1 + \betaZ^2 + (\betaZ^2-3) \Ct^2)\Big]
(\gAAAA + 6 \gAA \gVV + \gVVVV)\nonumber\\
&+\Big[2 (1 + \betaZ^2)^2 - 2 (5 - 2 \betaZ^2 + \betaZ^4) \Ct^2\Big]
\gA \gV (\gAA + \gVV)
\Bigg\}
\end{align}
\begin{align}
\htqq_{24}[\Theta,\mZZ]&=
\frac{\fZZ\sqrt{2}\sqrt{1-\betaZ^2}\St}{\DZ}
\Bigg\{ \Big[ (3 - \betaZ^2) (1 + \betaZ^2) \Ct  -
 4 \Ct^3 \Big](\gAAAA + 6 \gAA \gVV + \gVVVV)\nonumber\\
&+\Big[4 (1 + \betaZ^2)^2 -  8 (1 + \betaZ^4) \Ct^2\Big]
\gA \gV (\gAA + \gVV)
\Bigg\}
\nonumber\\
\nonumber\\
\htqq_{27}[\Theta,\mZZ]&=-
\frac{\fZZ (1-\betaZ^2)\St^2}{\DZ}
\Bigg\{\Big[(1 + \betaZ^2)^2 + 4 (\betaZ^2-2) \Ct^2\Big]
(\gAAAA + 6 \gAA \gVV + \gVVVV) 
\Bigg\}
\nonumber\\
\nonumber\\
\htqq_{28}[\Theta,\mZZ]&=
\frac{\fZZ 2\sqrt{2}\sqrt{1-\betaZ^2}\St}{\sqrt{3}\DZ}
\Bigg\{
\Big[ 2 (1 + \betaZ^2)^2 (1 + \Ct^2)-8 (1 + \betaZ^2) \Ct^2\Big]
\gA \gV (\gAA + \gVV)
\nonumber\\
&+ \Big[2 (1 - 3 \betaZ^2) \Ct^3 + (1 + \betaZ^2)
  (3 \betaZ^2 + \Ct^2-2) \Ct \Big]
(\gAAAA + 6 \gAA \gVV + \gVVVV)
\Bigg\}
\nonumber\\
\nonumber\\
\htqq_{33}[\Theta,\mZZ]&=
\frac{\fZZ}{\DZ}
\Bigg\{8 \Ct \Big[2 + \betaZ^2
  + \betaZ^6 + (-3 + 2 \betaZ^2 - 3 \betaZ^4) \Ct^2\Big]\gA \gV (\gAA + \gVV)
\nonumber\\
&+\Big[(\betaZ + 
  \betaZ^3)^2 + (7 - 5 \betaZ^2 - 3 \betaZ^4 + \betaZ^6) \Ct^2
\nonumber\\
  &- (9 - 
      10 \betaZ^2 + 5 \betaZ^4) \Ct^4\Big] (\gAAAA + 6 \gAA \gVV + \gVVVV)
\Bigg\}
\nonumber\\
\nonumber\\
\htqq_{34}[\Theta,\mZZ]&=
\frac{\fZZ(1-\betaZ^2)\St^2}{\DZ}
\Bigg\{\Big[2 (3 + \betaZ^2)\Ct^2-(1 + \betaZ^2)^2  \Big]
(\gAAAA + 6 \gAA \gVV + \gVVVV)
\nonumber\\
&+8 \Ct (1 +\betaZ^2)\gA \gV (\gAA + \gVV)
\Bigg\}
\nonumber\\
\nonumber\\
\htqq_{37}[\Theta,\mZZ]&=-
\frac{\fZZ \sqrt{2}(1-\betaZ^2)^{3/2}\Ct\St}{\DZ}
\Bigg\{
3 (1 + \betaZ^2 - 2 \Ct^2)(\gAAAA + 6 \gAA \gVV + \gVVVV)
\nonumber\\
&+4 (1 - \betaZ^2) \Ct \gA \gV (\gAA + \gVV)
\Bigg\}
\nonumber\\
\nonumber\\
\htqq_{38}[\Theta,\mZZ]&=
\frac{\fZZ}{\sqrt{3}\DZ}
\Bigg\{\Big[2 + 3 \betaZ^2 - 
  \betaZ^6 - (9 - 9 \betaZ^2 - \betaZ^4 + \betaZ^6) \Ct^2
\nonumber\\
&+ (9 - 18 \betaZ^2 + 
5 \betaZ^4) \Ct^4\Big] (\gAAAA + 6 \gAA \gVV + \gVVVV)
\nonumber\\
&+8 \Ct \Big[2 + \betaZ^2 + \betaZ^6 - (3 - 2 \betaZ^2 + 3 \betaZ^4)
\Ct^2\Big]
\gA \gV (\gAA + \gVV)
\Bigg\}
\end{align}

\begin{align}
\htqq_{44}[\Theta,\mZZ]&=
\frac{2\fZZ\St^2}{\DZ}
\Bigg\{\Big[2 (1 + \betaZ^4) \Ct^2-(1 + \betaZ^2)^2\Big]
(\gAAAA + 6 \gAA \gVV + \gVVVV)
\nonumber\\
\nonumber\\
\htqq_{47}[\Theta,\mZZ]&=
\frac{\fZZ \sqrt{2}\sqrt{1-\betaZ^2}\St}{\DZ}
\Bigg\{\Ct\Big[(\betaZ^2-3) (1 + \betaZ^2)  +4 \Ct^2\Big]
(\gAAAA + 6 \gAA \gVV + \gVVVV)
\nonumber\\
&+4\Big[(1 + \betaZ^2)^2-2 (1 + \betaZ^4) \Ct^2\Big]
\gA \gV (\gAA + \gVV)
\Bigg\}
\nonumber\\
\nonumber\\
\htqq_{48}[\Theta,\mZZ]&=
\frac{\fZZ(1-\betaZ^2)\St^2}{\sqrt{3}\DZ}
\Bigg\{
\Big[(1 + \betaZ^2)^2 - 2 (3 + \betaZ^2) \Ct^2\Big]
(\gAAAA + 6 \gAA \gVV + \gVVVV)
\nonumber\\
&+24 (1 + \betaZ^2) \Ct\gA \gV (\gAA + \gVV)
\Bigg\}
\nonumber\\
\nonumber\\
\htqq_{55}[\Theta,\mZZ]&=
\fZZ 2\St^2 \Big[\gAAAA + 6 \gAA \gVV + \gVVVV\Big]
\nonumber\\
\nonumber\\
\htqq_{56}[\Theta,\mZZ]&=-
\fZZ\sqrt{2}\sqrt{1-\betaZ^2}\St
\Big\{
\Ct(\gAAAA + 6 \gAA \gVV + \gVVVV)-4 \gA \gV (\gAA + \gVV)
\Big\}
\nonumber\\
\nonumber\\
\htqq_{66}[\Theta,\mZZ]&=
\fZZ(1-\betaZ^2)\Big\{
(1 + \Ct^2)(\gAAAA + 6 \gAA \gVV + \gVVVV)
-8 \Ct \gA \gV (\gAA + \gVV)
\Big\}
\nonumber\\
\nonumber\\
\htqq_{77}[\Theta,\mZZ]&=
\frac{\fZZ(1-\betaZ^2)}{\DZ}
\Bigg\{
8 \Ct \Big[3 + 2 \betaZ^2 - \betaZ^4 - 4 \Ct^2\Big]
 \gA \gV (\gAA + \gVV)
\nonumber\\
&+\Big[(1 + \betaZ^2)^2 - (7 + 10 \betaZ^2 - \betaZ^4) \Ct^2 +
  4 (2 + \betaZ^2) \Ct^4\Big](\gAAAA + 6 \gAA \gVV + \gVVVV)
\Bigg\}
\nonumber\\
\nonumber\\
\htqq_{78}[\Theta,\mZZ]&=
\frac{\fZZ\sqrt{2}\sqrt{1-\betaZ^2}\St}{\sqrt{3}\DZ}
\Bigg\{
\Ct\Big[1 + 4 \betaZ^2 + 3 \betaZ^4 -2 (3 + \betaZ^2) \Ct^2\Big]
(\gAAAA + 6 \gAA \gVV + \gVVVV)
\nonumber\\
&+4\Big[(9 - 2 \beta^2 + \betaZ^4) \Ct^2-2 (1 + \betaZ^2)^2 \Big]
\gA \gV (\gAA + \gVV)
\Bigg\}
\nonumber\\
\nonumber\\
\htqq_{88}[\Theta,\mZZ]&=
\frac{\fZZ}{3\DZ}
\Bigg\{
\Big[(1 + \betaZ^2)^2 (4 + \betaZ^2) + (3 + 3 \betaZ^2 - 7 \betaZ^4 + 
  \betaZ^6) \Ct^2
\nonumber\\
  &- (9 + 6 \betaZ^2 + 5 \betaZ^4) \Ct^4\Big]
(\gAAAA + 6 \gAA \gVV + \gVVVV)
\nonumber\\
&-24 \Ct \Big[2 + \betaZ^2 + \betaZ^6 - (3 - 2 \betaZ^2 + 3 \betaZ^4)
  \Ct^2\Big]
\gA \gV (\gAA + \gVV)
\Bigg\}
\end{align}

\newpage
The non-vanishing elements $\ftqq_{a}$ are given by
\vspace{0.5cm}
\begin{align}
\ftqq_{2}[\Theta,\mZZ]&=
\frac{\fZZ 2\sqrt{2}\sqrt{1-\betaZ^2}\St}{3\DZ}
\Bigg\{\Ct \Big[2 \betaZ^2 + 3 \betaZ^4 -4 \betaZ^2 \Ct^2-1\Big]
 (\gAAAA + 6 \gAA \gVV + \gVVVV)
\nonumber\\
&+4\Big[(1 + \betaZ^2)^2  + 4 \betaZ^2 (\beta^2-2) \Ct^2\Big]
\gA \gV (\gAA + \gVV)
\Bigg\}
\nonumber\\
\nonumber\\
\ftqq_{3}[\Theta,\mZZ]&=
\frac{\fZZ}{3\DZ}
\Bigg\{\Big[(1 + \betaZ^2)^3 + (15 \betaZ^2 - 13 \betaZ^4 + \betaZ^6-3) \Ct^2
\nonumber\\
&+   4 \betaZ^2 (\betaZ^2-3) \Ct^4\Big]
(\gAAAA + 6 \gAA \gVV + \gVVVV)
\nonumber\\
&+8 \Ct \Big[1 + 3 \betaZ^4 - \betaZ^6 + \betaZ^2 (5 - 8 \Ct^2)\Big]
\gA \gV (\gAA + \gVV)
\Bigg\}
\nonumber\\
\nonumber\\
\ftqq_{4}[\Theta,\mZZ]&=
\frac{\fZZ 2(1-\betaZ^2)\St^2}{3\DZ}
\Bigg\{\Big[1 + \betaZ^4 + \betaZ^2 (2 + 4 \Ct^2)\Big]
(\gAAAA + 6 \gAA \gVV + \gVVVV)
\Bigg\}
\nonumber\\
\nonumber\\
\ftqq_{7}[\Theta,\mZZ]&=
\frac{\fZZ 2\sqrt{2}\sqrt{1-\betaZ^2}\St}{3\DZ}
\Bigg\{\Ct\Big[1 - 2 \betaZ^2 - 3 \betaZ^4 +4 \betaZ^2 \Ct^2\Big]
(\gAAAA + 6 \gAA \gVV + \gVVVV)
\nonumber\\
&+4\Big[(1 + \betaZ^2)^2 +4 \betaZ^2 (\betaZ^2-2) \Ct^2\Big]
\gA \gV (\gAA + \gVV)
\Bigg\}
\nonumber\\
\nonumber\\
\ftqq_{8}[\Theta,\mZZ]&=-
\frac{\fZZ}{3\sqrt{3}\DZ}
\Bigg\{\Big[(1 + \betaZ^2)^3 + (15 \betaZ^2 - 13 \betaZ^4 + \betaZ^6-3) \Ct^2
\nonumber\\
  &+ 4 \betaZ^2 (\betaZ^2-3) \Ct^4\Big]
(\gAAAA + 6 \gAA \gVV + \gVVVV)
\nonumber\\
&+24 \Ct \Big[ \betaZ^6 + \betaZ^2 (8 \Ct^2-5)-1 - 3 \betaZ^4 \Big]
\gA \gV (\gAA + \gVV)
  \Bigg\}\, .
\end{align}
The elements $\gtqq_{a}$ are identical: $\gtqq_{a}=\ftqq_{a}$.
\subsection{Polarization density matrix for $h\to ZZ^*$}\label{sec:appendix:qutrits:HZZ}
Here we write  the coefficients $g_{a}$, $f_a$, and $h_{ab}$ ($a,b\in\{1,\dots,8\}$) 
appearing in the polarization density matrix for the Higgs boson decay
$h\to ZZ^*$, as well as the unpolarized squared amplitude,  as in~\cite{Fabbrichesi:2023cev}.

The non-vanishing $f_{a}$ elements are
\bea
&&f_{3}=\frac{1}{6}\, \frac{-m_h^4 + 2 (1 + f^2) m_h^2 M_Z^2 - (1 - f^2)^2 M_Z^4}{m_h^4 - 2 (1 + f^2) m_h^2 M_Z^2 +(1 + 10 f^2 + f^4) M_Z^4}
\, ,
\nonumber\\
\nonumber\\
&&f_{8}=-\frac{1}{\sqrt{3}}f_3\, ,
\eea
where $g_{a}=f_a$ for $a\in\{1,\dots,8\}$. The non-vanishing $h_{ab}$ elements are 
\bea
&&h_{16}=h_{61}=h_{27}=h_{72}=\frac{f M_Z^2 \Big[-m_h^2 + (1 + f^2) M_Z^2\Big]}
    {m_h^4 - 2 (1 + f^2) m_h^2 M_Z^2 + (1 + 10 f^2 + f^4) M_Z^4}\, ,
    \nonumber\\
    \nonumber\\
&&h_{33}=\frac{1}{4}\, \frac{\Big[m_h^2 - (1 + f^2) M_Z^2\Big]^2}{m_h^4 - 2 (1 + f^2) m_h^2 M_Z^2 + (1 + 10 f^2 + f^4) M_Z^4}\, ,
    \nonumber\\
    \nonumber\\
&&h_{38}=h_{83}=-\frac{1}{4\sqrt{3}}
    \nonumber\\
    \nonumber\\
&&h_{44}=h_{55}=\frac{2 f^2 M_Z^4} {m_h^4 - 2 (1 + f^2) m_h^2 M_Z^2 + (1 + 10 f^2 + f^4) M_Z^4}\, ,
\nonumber\\
\nonumber\\
&& h_{88}=\frac{1}{12}\, \frac{m_h^4 - 2 (1 + f^2) m_h^2 M_Z^2 + (1 - 14 f^2 + f^4) M_Z^4}
    {m_h^4 - 2 (1 + f^2) m_h^2 M_Z^2 + (1 + 10 f^2 + f^4) M_Z^4}\, . \label{fghHiggs}
\eea
The unpolarized square amplitude $|{\cal M}|^2$ of the process is instead
\bea
|{\cal M}|^2=\frac{g^2}{4 \cos\theta_W^2 f^2 M_Z^2}
\Big[ m_h^4 - 2 (1 + f^2) m_h^2 M_Z^2 + (1 + 10 f^2 + f^4) M_Z^4\Big]\, .
\eea

\subsection{Polarization density matrix for $h\to WW^*$ and $h\to ZZ^*$ in presence of anomalous couplings}
\label{appendix:HVV}
Here  write  the expression for the coefficients $g_{a}$, $f_a$, and $h_{ab}$ ($a,b\in\{1,\dots,8\}$)
appearing in the polarization density matrix for the Higgs boson decay
$h\to VV^*$, $V=W$ or $Z$, in presence of anomalous couplings.

The square amplitude summed over the gauge boson spin is 
\begin{align}
    \abs{\mathcal{M}}^2 & = \frac{\xi_V^2\,g^2}{4 f^2 M_V^2} \Bigg\{ 
     \Big[ 1  + 2\, f^2 \left(\LtVV + \LVV \right)\Big] m_h^4 -2 \Big[1  + f^2 \Big(1 + 2 \LtVV + 2 \LVV  - 6 \LV \Big)
\nonumber\\
&+   2\,  f^4 \Big(\LtVV + \LVV \Big)\Big] m_h^2 M_V^2+\Big[1  + 
  2 f^6 \Big(\LtVV + \LVV \Big)
\nonumber\\
&+    2 f^2 \Big(5 + \LtVV + \LVV  - 6 \LV \Big) +    f^4 \Big(1 - 4 \LtVV + 8 \LVV - 
      12 \LV\Big)\Big] M_V^4\Bigg\}\,,
\end{align}
where $\xi_W=1$ and $\xi_Z = \cos\theta_W^{-1}$. We find $f_a=g_a$ $\forall a\in\{1,\dots,8\}$ and the non-vanishing coefficients $\tilde f_a = \abs{\mathcal{M}}^2 f_a$ and $\tilde g_a = \abs{\mathcal{M}}^2 g_a$ are: 
\begin{align}
    \tilde f_3 = \tilde g_3 &= -\frac{\xi_V^2 g^2 \left(1-f^2 \left(\LV^2+\LtV^2\right)\right) \left(-2 \left(f^2+1\right) m_h^2 m_V^2+\left(f^2-1\right)^2 m_V^4+m_h^4\right)}{24
   f^2 m_V^2}\,,\nonumber\\
   \tilde f_8 = \tilde g_8  &=-\frac{1}{\sqrt 3} \tilde f_3 \,.
\end{align}
The non-vanishing elements of the matrix $\tilde  h_{ab} = \abs{\mathcal{M}}^2  h_{ab}$ for a gauge boson $V=W,Z$ instead are:
\begin{align}
     \tilde h_{16} &= \frac{g^2 \xi_V^2 \left(m_V^2 \left((1-2 \LV) f^2+1\right)-m_h^2\right) \left(\LV m_h^2-m_V^2 \left(\LV \left(f^2+1\right)-2\right)\right)}{8 f m_V^2} \,,
     \nonumber\\
    \tilde h_{17} &= \frac{ g^2 \xi_V^2 \LtV \sqrt{-2 \left(f^2+1\right) m_h^2 m_V^2+\left(f^2-1\right)^2 m_V^4+m_h^4} \left(m_V^2 \left((1-2 \LV) f^2+1\right)-m_h^2\right)}{8 f m_V^2}\,,
     \nonumber\\
     \tilde h_{26} &= - \tilde h_{17} = - \tilde h_{62} =
     \tilde h_{71}\,,
     \nonumber\\
     \tilde h_{27} &=  \tilde h_{16} =  \tilde h_{61} =  \tilde h_{72}\,,
     \nonumber\\
     \tilde h_{33} &= \frac{g^2 \xi_V^2 \left(m_h^2-m_V^2 \left((1-2 \LV) f^2+1\right)\right)^2}{16 f^2 m_V^2}\,,
    \nonumber \\
     \tilde h_{38} &= -\frac{\abs{\mathcal{M}^2}}{4\sqrt{3}}\,,
    \nonumber \\\nonumber
    \tilde h_{44} &= \frac{g^2 \xi^2_V}{8 m_V^2} \left[2 m_h^2 m_V^2 \left(-\left(\LV^2 \left(f^2+1\right)\right)+2 \LV+\LtV^2 \left(f^2+1\right)\right) \right.\\\nn &\quad\left.+m_V^4 \left(\LV^2 \left(f^2+1\right)^2-4 \LV \left(f^2+1\right)-\LtV^2 f^4+2 \LtV^2 f^2-\LtV^2+4\right)+m_h^4 \left(\LV^2-\LtV^2\right)\right]\,,
    \nonumber \\ 
    \tilde h_{45} &= \frac{\LtV g^2 \xi_V^2 \sqrt{-2 \left(f^2+1\right) m_h^2 m_V^2+\left(f^2-1\right)^2 m_V^4+m_h^4} \left(\LV m_h^2-m_V^2 \left(\LV \left(f^2+1\right)-2\right)\right)}{4 m_V^2}\,,
   \nonumber \\
    \tilde h_{54} &= -  \tilde h_{45} \,,
   \nonumber \\
    \tilde h_{55}  &=   \tilde h_{44} \,,
   \nonumber \\
    \tilde h_{83} &=  \tilde h_{38} \,,
   \nonumber \\
    \tilde h_{88} &=  \frac{g^2 \xi_V^2}{48 f^2 m_V^2} \Bigg\{m_h^4 \Big[-4 f^2 \left(\LV^2+ \LtV^2\right)+1\Big]-2 m_h^2 m_V^2 \Big[-4 f^4 \left(\LV^2 + \LtV^2\right)  
    \\ \nn &\quad
    +f^2 \left(-4 \LV^2+6 \LV-4 \LtV^2+1\right)+1\Big]+m_V^4 \Big[-4 f^6 \left( \LV^2+ \LtV^2\right)
    \\ \nn&\quad+f^4 \left(-4 \LV^2+12 \LV+8 \LtV^2+1\right) -2 f^2 \left(2 \LV^2-6 \LV+2 \LtV^2+7\right)+1\Big]\Bigg\}\,.
\end{align}
\end{appendices}

	\bibliography{review_PPNP}
	



\end{document}